\documentclass[11pt]{article}
\pdfoutput=1 
\usepackage{./jheppub}

\usepackage[T1]{fontenc}

\usepackage{graphicx}
\usepackage{amsfonts,amsmath,amssymb}
\usepackage{verbatim}
\usepackage{array}
\usepackage{mathrsfs}
\usepackage{mathrsfs}
\usepackage{yfonts}
\usepackage{dsfont}
\usepackage{bbm}
\usepackage{colonequals}
\usepackage{amscd}

\usepackage{float}
\usepackage{afterpage}

\newcommand{\x}[1]{\textcolor{cyan}{[\bf #1 ]}}

\usepackage{subcaption}

\usepackage{wrapfig}

\usepackage{suffix,mathtools,cancel,bbm}

\usepackage{multirow}

\usepackage{enumerate}

	\usepackage{tikz, tikz-3dplot, pgfplots}
	\usepackage{tkz-graph}
	\usetikzlibrary[positioning,patterns] 
\newcommand{\bn}{\begin{enumerate}}
\newcommand{\en}{\end{enumerate}}

\def\CN{{\cal N}}
\def\CO{{\cal O}}

\def\z{\zeta}

\newcommand{\beq}{\begin{equation}}
\newcommand{\eeq}{\end{equation}}

\newcommand\nn{\nonumber}

\newcommand{\cA}{\mathcal{A}}
\newcommand{\cB}{\mathcal{B}}
\newcommand{\cC}{\mathcal{C}}
\newcommand{\cD}{\mathcal{D}}

\newcommand{\cF}{\mathcal{F}}
\newcommand{\cG}{\mathcal{G}}
\newcommand{\cH}{\mathcal{H}}

\newcommand{\cK}{\mathcal{K}}

\newcommand{\cN}{\mathcal{N}}
\newcommand{\cO}{\mathcal{O}}
\newcommand{\cP}{\mathcal{P}}
\newcommand{\cQ}{\mathcal{Q}}

\newcommand{\cS}{\mathcal{S}}

\numberwithin{equation}{section}

\def\NO#1#2{{\rm NO}(#1,#2)}

\def\bea{\begin{eqnarray}}
\def\eea{\end{eqnarray}}

\DeclarePairedDelimiterX\MeijerM[3]{\lparen}{\rparen}%
{\begin{smallmatrix}#1 \\ #2\end{smallmatrix}\delimsize\vert\,#3}

\newcommand\MeijerG[8][]{%
  G^{\,#2,#3}_{#4,#5}\MeijerM[#1]{#6}{#7}{#8}}

\WithSuffix\newcommand\MeijerG*[7]{%
  G^{\,#1,#2}_{#3,#4}\MeijerM*{#5}{#6}{#7}}

\def\cH{\mathcal{H}}

\def\cN{\mathcal{N}}

\def \beg#1{\begin{#1}} 

\def \bea{\beg{eqnarray}}
\def \eea{\end{eqnarray}}
\def \ee{\end{equation}}

\def \restr#1#2{{\left.\kern-\nulldelimiterspace#1\vphantom{\big|}\right|_{#2}}}

\def \nn{\nonumber}

\def \PP{\mathcal{P}}

\newcommand{\ba}[1]{\begin{align} #1 \end{align} }

\newcommand{\bs}[1]{\begin{split} #1 \end{split} }

\usepackage{colortbl}
\definecolor{mygray}{gray}{0.93}

\title{\boldmath Aspects of Irregular Punctures via Holography
}

\author[a]{Ibrahima Bah,}
\author[b]{Federico Bonetti,}
\author[c]{Emily Nardoni,}
\author[a]{and Thomas Waddleton} 
\affiliation[a]{Department of Physics and Astronomy, Johns Hopkins University, 3400 North Charles Street, Baltimore, MD 21218, USA}
\affiliation[b]{Mathematical Institute, University of Oxford, Woodstock Road, Oxford, OX2 6GG, UK}
\affiliation[c]{Kavli Institute for the Physics and Mathematics of the Universe (WPI),
The University of Tokyo Institutes for Advanced Study,
The University of Tokyo, Kashiwa, Chiba 277-8583, Japan}

\emailAdd{iboubah@jhu.edu, federico.bonetti@maths.ox.ac.uk, emily.nardoni@ipmu.jp, twaddle1@jhu.edu}

\abstract{
We present new families of $AdS_5$ solutions in M-theory 
preserving 4d ${\mathcal N = 2}$ supersymmetry.
We perform a systematic analysis 
of holographic observables for these  solutions,
providing evidence for an interpretation in terms
of 4d superconformal field theories (SCFTs) of
Argyres-Douglas type,
realized in class $\mathcal S$
via a sphere with one irregular, and one regular puncture.
The gravity solutions
exhibit
internal M5-brane sources that correspond to the
irregular puncture. For a family of  solutions, we identify
explicitly the   class $\mathcal S$
puncture data
and perform a detailed match, including
Higgs branch operators.
 For   other families
we comment on proposed field theory duals, based on
irregular punctures labeled by nested Young tableaux.
}

\usepackage{etoolbox}
\appto\appendix{\addtocontents{toc}{\protect\setcounter{tocdepth}{1}}}

\appto\listoffigures{\addtocontents{lof}{\protect\setcounter{tocdepth}{1}}}
\appto\listoftables{\addtocontents{lot}{\protect\setcounter{tocdepth}{1}}}

\begin{document} 

\setcounter{tocdepth}{2}

\maketitle
\flushbottom



\section{Introduction and summary} \label{sec:introduction}

\subsection{Holographic duals of Argyres-Douglas SCFTs}


Infinite families of non-trivial 4d superconformal
field theories (SCFTs) can be realized by
reducing a 6d (2,0) SCFT on a Riemann surface, 
implementing a partial topological twist to preserve
supersymmetry. 
This idea dates back to the original class $\cS$
constructions of 4d $\cN = 2$ SCFTs \cite{Gaiotto:2009we,Gaiotto:2009hg}, 
as well as their   
generalizations to 4d
 $\cN =1$ theories \cite{Maruyoshi:2009uk,Benini:2009mz,Bah:2011je,Bah:2011vv,Bah:2012dg}.

One of the key ingredients in the class $\cS$ constructions is  
a rich spectrum of allowed punctures on the Riemann
surface \cite{Gaiotto:2009we,Gaiotto:2009hg}. This includes so-called irregular punctures, which can be
used to realize 4d SCFTs of Argyres-Douglas type \cite{Bonelli:2011aa, Xie:2012hs, Wang:2015mra}.
The latter exhibit remarkable features:
they possess Coulomb branch operators of fractional dimensions;
they are intrinsically strongly coupled; they can be regarded
as describing the interactions of  massless, mutually non-local BPS
dyons, as in the original paper  \cite{Argyres:1995jj}.

In this work, we investigate 4d $\cN= 2$ SCFTs that originate
from the reduction of the 6d (2,0) SCFT of type $A_{N-1}$, which
 is realized on the worldvolume of a stack of $N$ M5-branes.
Working at large $N$, we may access non-trivial aspects of these 4d $\cN = 2$ SCFTs
by studying the dual $AdS_5$ supersymmetric geometries
in 11d supergravity.
For the cases in which the Riemann surface has no punctures,
or only regular punctures, the dual $AdS_5$ solutions
have been known for quite some time \cite{Maldacena:2000mw, Gaiotto:2009gz}.
More recently,  holographic duals for a family of class $\cS$ constructions
with irregular punctures have been proposed \cite{Bah:2021mzw, Bah:2021hei, Couzens:2022yjl}.

Building on \cite{Bah:2021hei}, in this paper we find new $AdS_5$ 
supersymmetric solutions in M-theory. Our analysis is based
on a direct study of the BPS conditions in eleven dimensions
(as opposed to uplift on $S^4$ from 7d gauged supergravity).
The explicit, closed analytic forms of the new solutions
allow us to compute exactly several holographic observables of interest.
In particular, we identify a class of BPS M2-brane operators
that have fractional conformal dimensions, together
with the correct charges to be mapped to Coulomb branch
operators on the field theory side: these prompt an interpretation in terms of
putative dual SCFTs of Argyres-Douglas
type.

Our geometries exhibit singularities that are interpreted
in terms of internal M5-brane sources, along the lines of \cite{Bah:2021hei}.
Previous results in the literature, see \emph{e.g.}~\cite{Brandhuber:1999np, Apruzzi:2013yva, Gaiotto:2014lca, Apruzzi:2015wna, DHoker:2016ujz, DHoker:2017mds, Bah:2017wxp, Bah:2018lyv},
show that internal sources in holography
can be a powerful ingredient in the construction of non-trivial
holographic pairs.
This work makes further steps in the program
of characterizing internal sources in gauge/gravity duality.

\subsection{Summary of main results}

We begin with an overview of the key steps in our analysis, and a summary of our findings.

\paragraph{The role of an additional $U(1)$ isometry.}

Our starting point is the canonical form of 
$AdS_5$ solutions in 11d supergravity preserving
4d $\cN = 2$ supersymmetry \cite{Lin:2004nb}.
The 11d metric takes the form
\begin{align} \label{LLM_intro}
ds^2_{11} & =  \frac{e^{2 \widetilde \lambda}}{m^2} \, \bigg[
  ds^2_{AdS_5}  + \frac{y^2  \, e^{- 6 \widetilde \lambda}}{4} \, ds^2_{S^2}
+ \frac{ D\chi^2}{1 - y \, \partial_y D}
+ \frac{- \partial_yD}{4\,y} \, \Big( dy^2 + e^D \, (dx_1^2 + dx_2^2) \Big) \bigg] \ .
\end{align}
In particular, the 6d internal space is an $S^2 \times S^1_\chi$ fibration over a 3d base space spanned by the coordinates $x_1$, $x_2$, $y$.
A more complete review of this class of solutions, and the definition of all   quantities entering \eqref{LLM_intro},
is given in section \ref{sec:reminder}.
For the purposes of this introduction, it suffices to recall that
the warp factor $\widetilde \lambda$ and all metric functions are determined in
terms of a single function $D = D(x_1, x_2, y)$, satisfying
 the continual Toda equation
\beq \label{Toda_intro}
\partial_{x_1}^2 D + \partial_{x_2}^2 D + \partial_y^2 e^D =  0 \ .
\eeq

Our primary goal is to construct and analyze new solutions
that may admit an interpretation in terms of
a 4d SCFT of Argyres-Douglas type.
The search for such solutions may be refined from the following considerations.
In the geometric class $\mathcal{S}$ construction of the Argyres-Douglas type SCFTs,
the Riemann surface is a sphere with
an irregular puncture at one pole,
and possibly a regular puncture at the opposite pole.
In this setup, two $U(1)$ symmetries
play an important role. The first is the $U(1)_\phi$
symmetry that rotates the phase of the fiber in
the cotangent bundle to the Riemann surface.
The second is the $U(1)_z$ isometry
of the Riemann surface (rotation along the axis connecting
the two poles of the punctured sphere). In the absence of an irregular puncture,
the superconformal $U(1)_r$ R-symmetry would simply
be identified with $U(1)_\phi$.
In the presence of the irregular puncture, however,
the $U(1)_r$ symmetry is a linear combination 
of $U(1)_\phi$ and $U(1)_z$
\cite{Bonelli:2011aa, Xie:2012hs, Wang:2015mra},  
\beq \label{eq_intro_mixing}
\partial_\chi = \partial_\phi + \alpha \partial_z \ .
\eeq
The mixing coefficient $\alpha$ depends on the specific Argyres-Douglas SCFT under consideration; for the cases relevant to this paper, $\alpha =\frac{N }{N+k}$, where $N$ is the number of M5-branes wrapping the Riemann surface, and $k$ is an integer determined by the choice of irregular puncture.
As indicated on the LHS of \eqref{eq_intro_mixing}, 
the superconformal R-symmetry
is associated to the angle $\chi$ in \eqref{LLM_intro}.

In the search for new M-theory solutions, we would like
the geometry to reflect the $U(1)$ mixing in field theory
recalled above. In order to see this manifestly,
the line element \eqref{LLM_intro} should admit a second
$U(1)$ isometry besides $\partial_\chi$.
We thus input an additional requirement in our construction:
a  $U(1)$ isometry in the 3d base space
spanned by $x_1$, $x_2$, $y$, 
which has to be compatible with the Toda equation
\eqref{Toda_intro}.
This leads to introduce polar coordinates
\beq
x_1 = r \cos \beta \ , \qquad x_2 = r \sin \beta \ ,
\eeq
and restricts the Toda potential $D$ to be 
independent of the angular coordinate~$\beta$.\footnote{Another
possible way to implement the additional $U(1)$ isometry
is to work with the $x_1$, $x_2$ coordinates,
demand that $D$ be independent of $x_2$,
and periodically identify the $x_2$ coordinate.
It may be verified, however, that this procedure
is equivalent to the introduction of the polar coordinates
$(r,\beta)$, up to a conformal scaling 
of $dx_1^2 + dx_2^2$ and a redefinition of $D$.
}
This angular coordinate is associated
to a linear combination of $\partial_\phi$, $\partial_z$
that is independent to the superconformal
R-symmetry generator \eqref{eq_intro_mixing}.

Before proceeding, it is worth clarifying an important physical
point related to the $\partial_\beta$ symmetry.
Na\"ively, the fact that the solution
admits a $\partial_\beta$ isometry seems to imply
an additional global flavor  $U(1)$ symmetry
in the 4d field theory. 
Crucially, however, this conclusion may be invalidated
after a more careful analysis of the 
background 4-flux $\overline G_4$.
In the solutions of \cite{Bah:2021mzw, Bah:2021hei},
 the Kaluza-Klein vector associated to $\partial_\beta$
is massive by virtue of a St\"uckelberg coupling.
This phenomenon originates   from the fact that
the background 4-flux $\overline G_4$ 
is invariant under $\partial_\beta$,
but cannot be completed to an $U(1)_\beta$-equivariant
closed 4-form.
We will observe the same phenomenon in all new
solutions discussed in this work: even though
$\partial_\beta$ is a symmetry of the supergravity
solution, the associated Kaluza-Klein vector
is always massive via a St\"uckelberg mechanism
of the same kind as in \cite{Bah:2021mzw, Bah:2021hei}.

%
%
%
%
%



\paragraph{Separable solutions to the Toda equation.}

%

At this stage, our task is to study solutions
to the Toda equation that are invariant under $\beta$ rotations.
To keep the analysis tractable, at present we restrict our search to exact, analytic solutions. Building on previous experience in
\cite{Bah:2021hei}, we make a change of coordinates
from $r$, $y$ to a new pair of variables $t$, $u$, and impose the following separability condition,
\beq
\label{sep}
y = t u \ , \qquad r =  r_1(t) r_2(u) \ . 
\eeq
\eqref{sep} is a technical assumption that results in a remarkable simplification: we achieve full separation of variables in the Toda equation, thus yielding new analytic solutions.\footnote{
The physics of the ansatz \eqref{sep} is an interesting question  which we reserve for the discussion section, since our comments rely on intuition gained from a closer analysis of the solutions.}

The 11d geometries and flux configurations
given by these new solutions to the Toda equation
are studied in detail in section \ref{sec:catalog}.
Metric regularity and positivity dictate
the allowed region for the coordinates $t$, $u$.
The solutions can be    grouped accordingly into two main classes.
The first class consists of solutions in which 
we have
a rectangular domain in the $(t,u)$ plane.
The second class consists instead of
solutions with a non-rectangular domain.
We perform a classification of solutions in the first class: three possibilities arise, as depicted in Figure
\ref{fig_cases_new}, with Case II reducing to the solutions previously analyzed in \cite{Bah:2021hei}. We refrain from a classification
of solutions in the second class, instead
studying some representative examples,
depicted in Figure \ref{fig_non_rect}.

The solutions of the first class
(with rectangular domains)
are singled out on physical grounds by the structure
of singularities in the 11d metric and warp
factor. Indeed, for all cases in Figure \ref{fig_cases_new} we can furnish an interpretation
of the singularities in the 11d solution in terms
of smeared M5-brane sources of the same kind as in \cite{Bah:2021hei}.
While the physical implications 
of these singularities are difficult to ascertain
purely from a supergravity perspective,
the analysis of \cite{Bah:2021hei} has demonstrated
that they can be consistently utilized
as ingredients in the construction of meaningful,
non-trivial holographic pairs.
This gives us confidence that the new solutions
of this paper can also be interpreted 
as duals to some 4d SCFTs.
We then proceed to compute various
holographic observables: 
central charge; flavor central charges;
dimensions of  some  BPS operators from wrapped M2-branes. A subclass of these BPS operators may be identified in the putative $\CN=2$ field theory duals as Coulomb branch operators of fractional scaling dimension, reinforcing our classification of the dual SCFTs as of Argyres-Douglas type.

\paragraph{Map to an electrostatic problem: more geometries.}

The solutions to the axisymmetric Toda equation can 
be analyzed by means of the
B\"acklund transformation, a functional
transform that maps the Toda equation with
rotational symmetry in the $x_1x_2$ plane to the Laplace equation in
$\mathbb R^3$ with axial symmetry (see \emph{e.g.}~\cite{Gaiotto:2009gz}).
In this electrostatic picture, a solution is specified
by a choice of charge density along the axis of cylindrical
symmetry in $\mathbb R^3$.  
From the perspective of this frame, our task is to 
identify those  charge densities
that correspond to solutions that have an interpretation in terms of 4d 
SCFTs of Argyres-Douglas type.

The next step in our analysis is to determine
explicitly the 
 B\"acklund transformation
 of the analytic Toda solutions found via the separation of
variables \eqref{sep}. This task is addressed in section  \ref{sec_electro},
where in particular, we identify the charge densities
associated to our new Toda solutions.
These charge densities are piecewise linear 
continuous functions, determined by a finite number of
slope and intercept parameters. The latter are related
in a non-trivial way to the flux quanta and geometry
of the solutions in the Toda frame.
Once the charge densities are identified,
they can be generalized systematically, thereby furnishing 
solutions which do not necessarily originate from 
a separable Toda potential.

%
%
%
%
%
%
 
We apply this circle of ideas to the solutions first discussed in \cite{Bah:2021hei}, corresponding to Case II in Figure \ref{fig_cases_new}. Upon identifying and generalizing the
associated charge density in the electrostatic frame,
we find solutions dual to class $\cS$ constructions
with one irregular puncture,
and one puncture labeled by an arbitrary Young
diagram. Our analysis confirms and extends
results first reported in \cite{Couzens:2022yjl}.


\paragraph{Comparison with field theory: central charges and Higgs operators.}
In the final part of this paper,
we perform a systematic comparison
with 4d SCFTs of Argyres-Douglas type.
More precisely, we consider class $\cS$ constructions
in which the irregular puncture
is of type $A_{N-1}^{(N)}[k]$ in the notation of \cite{Xie:2012hs, Wang:2015mra}, while the regular puncture
is specified by a partition of $N$,
or equivalently a Young diagram $Y$.
Building on previous results in the literature \cite{Shapere:2008zf, Xie:2012hs, Xie:2013jc, Cecotti:2013lda, Giacomelli:2020ryy} (see also \cite{Couzens:2022yjl}),
we compute a closed-form expression for the
large-$N$ behavior of the central charge of the
theory with labels $(A_{N-1}^{(N)}[k], Y)$,
for arbitrary Young diagram $Y$.
We similarly derive general expressions
for the central charge of the flavor symmetry associated
to the regular puncture, and for
the Coulomb branch operators. All the solutions constructed here have Coulomb branch operators of fractional dimensions -- one of the hallmarks of $\CN=2$ SCFTs of Argyres-Douglas type.
In the special case in which $N/k$ is an integer,
and the regular puncture is either maximal (full)
or minimal (simple), a 4d $\cN = 1$ Lagrangian description
is available \cite{Agarwal:2017roi, Benvenuti:2017bpg}. In these cases, the Lagrangian description is especially useful
to access Higgs branch operators across the duality.

We identify the Case II solutions of Figure
\ref{fig_cases_new}  with the Argyres-Douglas SCFTs with labels $(A_{N-1}^{(N)}[k], Y)$, for $Y$ a regular puncture labeled by a general partition of $N$ -- these are the same solutions that were identified in \cite{Couzens:2022yjl}, and that generalize the more restrictive class of regular puncture geometries described in \cite{Bah:2021hei}. The Case I solutions are a one parameter generalization of these SCFTs which include an additional smeared M5-brane source, leading to 
additional Higgs branch operators and larger flavor symmetry. 
A hypothesis for the field theory duals of the Case I solutions
is that they correspond to Argyres-Douglas theory 
realized with one regular puncture and one irregular 
 puncture which
is labeled by more refined data compared to $A_{N-1}^{(N)}[k]$. We check that in a certain limit, this extra data is consistent with that of a nested Young tableaux structure of the irregular puncture, as in  \cite{Xie:2012hs, Wang:2015mra}. 
The determination of the precise irregular puncture data
 and more refined tests of this proposal
are left for future work.

\paragraph{Plan of the paper.}
The rest of this paper is organized as follows.
Section \ref{sec:sugra} is devoted to the analysis
of the Toda equation and its solutions obtained via separation of variables. In Section \ref{sec:catalog} we study the
geometry and flux configurations of the M-theory
solutions determined from the Toda solutions of Section \ref{sec:sugra}. In Section \ref{sec_electro} 
we perform the B\"acklund transform, while Section \ref{sec:gen_caseII} is devoted to  
generalizations of the charge density profiles,
and their implications on the M-theory solutions.
In Section \ref{sec:QFT} we perform a detailed comparison
with various large-$N$ quantities for 4d SCFTs of Argyres-Douglas
type. We conclude with a brief discussion.
The appendices collect some derivations and technical material.

\vspace{2mm} 
\noindent 
Reference \cite{Couzens:2022yjl} appeared
 while this work was being completed,
 which has
some overlap with a class of solutions we present.



\section{Supergravity solutions} \label{sec:sugra}

In this section we briefly review the canonical form of 
$AdS_5$ solution of 11d supergravity preserving 4d $\cN = 2$ superconformal
symmetry. These solutions are specified by a choice of Toda potential $D$
satisfying \eqref{Toda_equation} below. We proceed with a construction of analytic
solutions to \eqref{Toda_equation} based on a suitable separation of variables.

\subsection{Canonical form of $AdS_5$ solutions in 11d supergravity} \label{sec:reminder}

The most general $AdS_5$ solution of 11d supergravity preserving 4d $\cN = 2$ superconformal
symmetry was characterized in Lin-Lunin-Maldacena (LLM) \cite{Lin:2004nb}. The 11d metric
and flux are given as  \cite{Gaiotto:2009gz}
\begin{align} \label{LLM}
ds^2_{11} & =  \frac{e^{2 \widetilde \lambda}}{m^2} \, \bigg[
  ds^2(AdS_5)  + \frac{y^2  \, e^{- 6 \widetilde \lambda}}{4} \, ds^2(S^2)
+ \frac{ D\chi^2}{1 - y \, \partial_y D}
+ \frac{- \partial_yD}{4\,y} \, \Big( dy^2 + e^D \, (dx_1^2 + dx_2^2 )\Big) \bigg] \ ,  \nn \\
G_4 & = \frac{1}{4\, m^3}\, {\rm vol}_{S^2} \wedge \bigg[
D\chi \wedge d(y^3 \, e^{- 6 \widetilde \lambda})
+ y \, (1 - y^2  \, e^{-6\widetilde \lambda}) \, dv
- \frac 12 \, \partial_y e^D \, dx_1 \wedge dx_2
\bigg] \ .
\end{align}
The line elements on $AdS_5$ and $S^2$ have unit radius. The quantity $m$ is a mass scale.
The warp factor $\widetilde \lambda$ and the function $D$ depend on $y$,
$x_1$, $x_2$ and are related by
\beq \label{lambda_and_D}
e^{- 6\widetilde \lambda} = \frac{- \partial_y D}{y \, (1 - y \, \partial_y D)} \ .
\eeq
The function $D$ satisfies the Toda equation
\beq \label{Toda_equation}
\partial_{x_1}^2 D + \partial_{x_2}^2 D + \partial_y^2 e^D =  0 \ .
\eeq
The coordinate $\chi$ is an angular coordinate with period $2\pi$. The 1-form
$D\chi$ is defined as
\beq \label{v_def}
D\chi = d\chi + v \ , \qquad v =  - \frac 12 \, \Big(
 \partial_{x_1} D \, dx_2 - \partial_{x_2} D \, dx_1 \Big)   \ .
\eeq
The 2-form  ${\rm vol}_{S^2}$ is the   volume form
on a unit-radius round $S^2$. The Killing vector $\partial_\chi$
is dual to the $U(1)_r$ R-symmetry of the 4d $\cN =2$ SCFT,
while the isometries of $S^2$ are mapped to the $SU(2)_R$ R-symmetry.

It is convenient to 
 introduce polar coordinates $(r,\beta)$ in the $(x_1,x_2)$ plane,
\beq
x_1 + i \, x_2 = r \, e^{i \beta} \  .
\eeq
In particular, the angle $\beta$ has period $2\pi$. 
If the Toda potential $D$ is independent of $\beta$,
 \eqref{Toda_equation} can be   rewritten as
\beq \label{Toda_bis}
\frac 1r \, \partial_r (r \, \partial_r D) + \partial_y^2 e^D = 0 \ .
\eeq

\paragraph{Convenient choice for the mass scale $m$.}
The value of the mass scale $m$ is not physical. It can be
set to any positive value by a   rescaling of the $x_1$, $x_2$, $y$ coordinates
and the Toda potential, of the form $x_1 = a \widehat x_1$,
$x_2 = a \widehat x_2$,  $y = a \widehat y$, $D(x_1,x_2,y) = \widehat D (\widehat x_1, \widehat x_2, \widehat y)$, where $a>0$ is a   constant.
In later sections, we shall find it convenient to set
\beq \label{eq_fix_m}
4\pi m^3 \ell_p^3 = 1 \ ,
\eeq
where $\ell_p$ denotes the 11d Planck length. In our conventions,
$G_4$-flux is 
  quantized as
\beq \label{flux_quant_convention}
\int_{\cC_4} \frac{G_4}{(2\pi \ell_p)^3} \in \mathbb Z \ , 
\eeq
where $\cC_4$ is a 4-cycle in spacetime.




\subsection{Toda equation and separation of variables} \label{sec:separation}

We can analyze the Toda equation \eqref{Toda_bis} by taking the coordinates $y$ and $r$ to be separable functions. That is, we write
\beq \label{var_sep}
y = t  u,\quad r = r_1(t)r_2(u) \ ,
\eeq
in terms of new coordinates $t$, $u$.
Inserting \eqref{var_sep} into the metric given in \eqref{LLM}, we find a cross term of the form
\beq
ds_{11}^2 \supset -\frac{\partial_y D}{2y} \frac{e^{2\tilde{\lambda}}}{m^2} \left(tu + e^D r_1r_1'r_2r_2'\right)dt du \ .
\eeq
Here and in the rest of this section, a prime on a function of one variable denotes differentiation with respect
to that variable.
Imposing that this cross term vanish, we obtain an expression for the
Toda potential in terms of $t$, $u$, $r_{1}(t)$, $r_2(u)$,
\beq \label{eq:eD_ansatz}
e^D = -\frac{tu}{r_1r_1'r_2r_2'} \ .
\eeq
Plugging   this  back into the Toda equation \eqref{Toda_bis}, we find 
a pair of decoupled ODEs for $r_1(t)$ and $r_2(u)$,
\beq  \label{sep_ODEs}
\frac{1}{t}\left(\frac{r_1r_1''}{(r_1')^2}t^2\right)' = \frac{1}{u}\left(\frac{r_2r_2''}{(r_2')^2}u^2\right)'  \qquad \Rightarrow \qquad \frac{r_1' }{r_1} = -\frac{t}{K_1(t)} \ ,  \qquad \frac{r_2'}{r_2} = \frac{u}{K_2(u)} \ .
\eeq
In the previous expression we have introduced the quadratic polynomials
\beq
K_1(t) = -\sigma(t-t_1)(t-t_2) \  ,\qquad K_2(u) = \sigma(u-u_1)(u-u_2) \ ,
\eeq
where $\sigma$, $t_1$, $t_2$, $u_1$, $u_2$ are constant parameters.
In order to have real $K_1$, $K_2$, the parameter $\sigma$ must be real.
The roots $t_1$, $t_2$ of $K_1$ are either both real, or   both 
complex
and complex conjugate of each other. Similar remarks apply to the roots
$u_1$, $u_2$ of $K_2$.

Combining \eqref{eq:eD_ansatz} and \eqref{sep_ODEs} we may write
\beq\label{eq:eD_KK}
e^D = \frac{K_1K_2}{r_1^2r_2^2} \ .
\eeq
If desired, 
the 
first order ODEs in \eqref{sep_ODEs} are readily integrated,
yielding closed form expressions for $r_1(t)$ and $r_2(u)$,
and hence for the Toda potential as a function of $t$, $u$.
For the purposes of computing the 11d metric and flux, however,
the explicit expressions for $r_1(t)$ and $r_2(u)$ are not needed:
when $r_1'$ or $r_2'$ are encountered,
they can be eliminated using the ODEs \eqref{sep_ODEs}.
In conclusion, we can express the metric and flux
in terms of $u$, $t$, $K_1$, and $K_2$,
\begin{align}\label{LLM_yy_metric}
ds_{11}^2 =&\, \frac{e^{2\tilde{\lambda}}}{m^2}\bigg[ds^2(AdS_5) + \frac{t^2u^2 e^{-6\tilde{\lambda}}}{4}ds^2(S^2) + \frac{D\chi^2}{1-tu\partial_y D} \nn  \\
&-\partial_y D \frac{K_1u^2 + K_2 t^2}{4t u}\left(\frac{dt^2}{K_1} + \frac{du^2}{K_2} + \frac{K_1K_2}{K_1u^2 + K_2t^2}d\beta^2\right)\bigg] \ , \nonumber\\
G_4=&\, \frac{1}{4m^3}\text{vol}_{S^2}\wedge d\left[- t^3 u^3 e^{-6\tilde{\lambda}}D\chi
 +  t u v +\frac{1}{2} \mathcal{F}d\beta\right],
\end{align}
where the quantities $v$, $-\partial_y D$, $e^{-6\tilde{\lambda}}$, and $\mathcal{F}$ 
are given by
\begin{align} \label{metric_functions}
v & \equiv v_\beta d\beta =  - \frac 12 r \partial_r D d\beta =
\bigg[ 
1 - \frac \sigma 2 +  \frac \sigma 2 \,  \frac{    u_1 u_2  K_1 + t_1 t_2  K_2  }{  
K_1 u^2 + K_2 t^2
 } 
\bigg] d\beta
\nn \\
-\partial_y D &=
\frac{
(u_1 + u_2) t - (t_1 + t_2) u
}{ K_1 u^2 + K_2 t^2} \sigma
\nonumber\\
e^{-6\tilde{\lambda}} &=\, \frac{(u_1+u_2)t - (t_1+t_2)u}{tu(u_1u_2 t^2 - t_1t_2 u^2)}\nonumber\\
\mathcal{F} &=
2(\sigma-1) ut - \sigma (t_1 + t_2) u
- \sigma (u_1 + u_2) t  \ .
\end{align}
We observe that all quantities written above are real, 
even if we allow for complex roots of $K_1$ and/or $K_2$.


\paragraph{Reflections in $t$ and $u$.}
Let us consider 
a simultaneous flip in the signs of $t$, $t_{1,2}$, and the angular coordinates,
\beq \label{t_flip}
t \mapsto - t \ , \qquad t_{1,2} \mapsto - t_{1,2} \ , \qquad \chi \mapsto - \chi \ , \qquad
\beta \mapsto - \beta \ .
\eeq
The expressions \eqref{LLM_yy_metric} for the 11d metric and flux are invariant under
these redefinitions. By a similar token, one verifies invariance
under the sign flips
\beq \label{u_flip}
u \mapsto - u \ , \qquad u_{1,2} \mapsto - u_{1,2} \ , \qquad \chi \mapsto - \chi \ , \qquad
\beta \mapsto - \beta \ .
\eeq
From \eqref{LLM_yy_metric}, \eqref{metric_functions} we see that the radius squared of the $S^2$,
given by $\frac 14 t^2 u^2 e^{- 6 \widetilde \lambda}$,   changes
sign as we cross $t = 0$ or $u = 0$. It follows that the allowed range for the $(t,u)$
coordinates is necessarily contained in one of the four quadrants of the $(t,u)$ plane.
Performing the sign flips \eqref{t_flip} or \eqref{u_flip} if necessary,
we can assume without loss of generality that the
allowed region in the $(t,u)$ plane lies in the first quadrant,
\beq \label{positive_t_u}
t \ge 0 \ , \qquad u \ge 0 \ .
\eeq

\paragraph{Positivity of metric functions.}
From \eqref{eq:eD_KK} we observe that $K_1$ and $K_2$ must either be both positive, or both negative. As a result, $K_1 u^2 + K_2 t^2$ is positive or negative, respectively.
In either case, we see from \eqref{LLM_yy_metric} that the metric in the directions $t$,
$u$, $\beta$ is non-negative definite if and only if $tu (- \partial_y D) \ge 0$.
This condition also automatically guarantees the non-negativity of the coefficient 
of $D\chi^2$ in the line element.
 
Without loss of generality we can assume $\sigma < 0$. Indeed,
$\sigma = 0$ would give $e^D \equiv 0$, and $\sigma >0$ is equivalent up to
exchanging the roles of $t$ and $u$.
Let us define  
\beq
f_1(t,u) = t_1 t_2 u^2 - u_1 u_2 t^2     \ ,   \qquad
f_2(t,u) = (t_1 + t_2) u - (u_1 + u_2) t    \ .
\eeq
We proceed assuming \eqref{positive_t_u}.
From the expressions of $(- \partial_yD)$ and $e^{- 6 \widetilde \lambda}$,
we infer that we have two possibilities to ensure non-negativity of
the warp factor and metric functions,
\beq \label{inequalities}
\begin{tabular}{c | c}
option (a) & option (b) \\ \hline
$ t \ge 0$, $u \ge 0$ & $ t \ge 0$, $u \ge 0$ \\
$(t -t_1)(t-t_2) \ge 0$ & $(t -t_1)(t-t_2) \le 0$  \\
$(u-u_1)(u-u_2) \le 0$ & 
 $(u-u_1)(u-u_2) \ge 0$ \\
$f_1 \ge 0$, $f_2 \ge 0$ & $f_1 \le 0$, $f_2 \le 0$
\end{tabular}
\eeq
These two sets of inequalities are the starting point
for a systematic discussion of the allowed domains
in the $(t,u)$ plane.

\section{Geometries, fluxes, and observables} \label{sec:catalog}

\subsection{Rectangular domains in the $(t,u)$ plane}

Our next task is to identify 
possible choices for the parameters
$t_{1,2}$, $u_{1,2}$ for which the inequalities \eqref{inequalities}
define a compact region in the $(t,u)$ plane.
In what follows, we focus on option (a) in \eqref{inequalities}, since
option (b)   gives analogous results with the roles of $t$ and $u$
exchanged.

Depending on the values of $u_{1,2}$, $t_{1,2}$, the allowed region in the $(t,u)$
plane, if compact, is a polygon delimited by vertical, horizontal, and oblique lines.
The latter (if present) originate from $f_1 \ge 0$ and/or $f_2 \ge 0$.
In this subsection, we provide a complete classification of the 
choices of $u_{1,2}$, $t_{1,2}$ that yield rectangular domains in the $(t,u)$ plane,
as opposed to polygons admitting oblique sides.
The physical motivation for this restriction on the shape of the allowed domain
originates from the analysis of    singularities
in the supergravity solutions. For rectangular domains,
all singularities that emerge can be interpreted in terms 
of smeared M5-branes sources of the same kind as in \cite{Bah:2021hei}.
For non-rectangular domains, in contrast, novel singularities emerge,
which are interpreted in terms of M5-brane sources that are smeared
in more directions.
While we refrain from a classification of 
non-rectangular domains, we discuss some examples in detail in section \ref{sec_non_rect}.

\begin{figure}
\includegraphics[width= 15 cm]{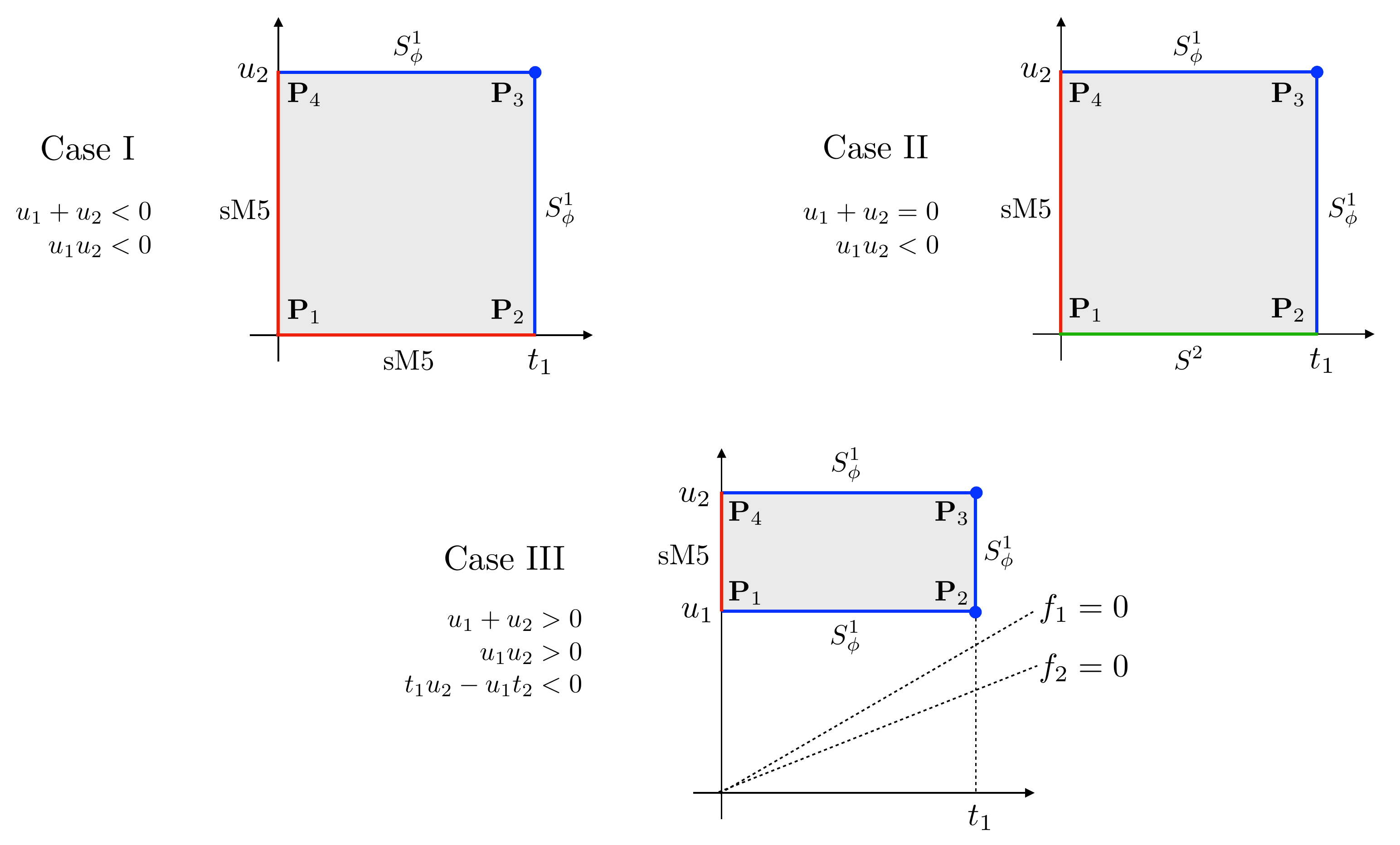}
\caption{\label{fig_cases_new} 
Cases that yield a rectangular domain in the $(t,u)$ plane,
as determined
by  the inequalities \eqref{inequalities} option (a).
In all cases, $0 < t_1 \le t_2$ and $u_1 < u_2$.
The label $S^1_\phi$ indicates that the $\phi$ circle in the base of the $Dz$ fibration
\eqref{with_new_angles} shrinks smoothly along that component of the boundary of the allowed region.
Similarly, the label $S^2$ indicates a smooth shrinking of the $S^2$.
The label sM5    
 stands for smeared M5-brane sources.
Case III is understood to include the  limiting case  $t_1 u_2 - u_1 t_2 = 0$, in which the lines defined by $f_1$ and $f_2$
coincide, and touch the lower right corner of the shaded rectangle. }
\end{figure}

Inspection of the first column of \eqref{inequalities} reveals that
a necessary condition for having a rectangular, compact domain in the $(t,u)$ plane
is that all roots $t_{1,2}$, $u_{1,2}$ be real
and such that $0 < t_1 \le t_2$, $u_1 < u_2$.
As we vary $u_{1,2}$ we obtain different allowed regions.
Restricting to rectangular domains, we find three   cases,
labeled I, II, III, and
summarized in Figure~\ref{fig_cases_new}.

\subsection{Flux quantization and holographic quantities}

\label{sec:CaseI}

While the geometries of Cases I, II, III are different, they   share some
common features. In particular, we observe that the $\beta$ component $v_\beta$
of the 1-form $v$ in \eqref{LLM_yy_metric}, \eqref{metric_functions} is piecewise constant along
vertical and horizontal segments in the $(t,u)$ plane of the form $t = t_i$ or $u =u_i$
($i=1,2$). This implies that, along such segments, a constant linear combination of the 
Killing vectors $\partial_\chi$, $\partial_\beta$ has vanishing norm, with different
linear combinations for each segment.

In order to elucidate the geometry in each case, we find it convenient to introduce
new angular variables $\phi$, $z$. They are related to the Toda angular
variables $\beta$, $\chi$ by a change of coordinates of the form
 \beq \label{new_angles}
\chi = \bigg( 1 + \frac 1 \cC \bigg) \,  \phi -  z \ , \qquad
\beta = - \frac 1 \cC \, \phi + z \ ,
\eeq 
where $\cC$ is a constant, which is given in terms of the value of the
component $v_\beta$ along the horizontal segment $u = u_2$, for each
case in Figure \ref{fig_cases_new}. 
The new angular coordinates are engineered in such a way that,
along the segment $u = u_2$, the Killing vector whose norm vanishes is simply
$\partial_\phi$.

In terms of the new variables $\phi$, $z$, it is convenient to group
the coordinates $t$, $u$, $\phi$ into a 3d base space, with the $S^2$ and the $z$
circle being fibered on top. The 11d line  element and flux
  take  the form
\begin{align} \label{with_new_angles}
ds_{11}^2 &= \frac{e^{2\tilde{\lambda}}}{m^2}\Bigg[ds^2(AdS_5)+ \frac{t^2u^2 e^{-6\tilde{\lambda}}}{4}ds^2(S^2)
+ R_z^2 Dz^2 + R_\phi^2 d\phi^2-\partial_y D \frac{K_1u^2 + K_2 t^2}{4t u}\left(\frac{dt^2}{K_1} + \frac{du^2}{K_2}\right)\bigg] \  , \nn \\
\overline G_4  &:= - \frac{G_4}{(2\pi \ell_p)^3} = 
 \frac{ {\rm vol}_{S^2}  }{4\pi}\wedge d \bigg[ 
Y \, \frac{d\phi}{2\pi}  - W \, \frac{Dz}{2\pi} 
\bigg] \ , \qquad Dz = dz - L d\phi \ .
\end{align}
In the previous expressions,   the fibration of the $z$ circle over the 3d base space
is encoded in the 1-form $Dz$.
We have introduced a rescaled version $\overline G_4$ of $G_4$,
which has the advantage of having integral periods, see \eqref{flux_quant_convention} (the minus
sign is for   convenience). The quantities $L$, $R_z$, $R_\phi$,
$Y$, $W$ are functions of $t$, $u$.
We refer the reader to appendix \ref{sec:fibr_change}
for 
further details on the change of coordinates \eqref{new_angles} and for
the   expressions of $\cC$, $L$, $R_z$, $R_\phi$, $Y$, $W$.

We now consider Cases I, II, III in turn. The results of the remainder of this subsection are summarized in Table \ref{tab:sum}.  
  
 \renewcommand{\arraystretch}{1.55}
 \begin{center}
\begin{table}
\begin{tabular}{|c||c|c|c|}
\hline
		& I &  II & III \\ \hline \hline
flux quanta & 
\renewcommand{\arraystretch}{1}
	$\begin{array}{c}
	\{q,K,M,N_\mathrm W,N_\mathrm S\} \\
	N_{\text{eff}} := q M
	\end{array}$ &  
	\renewcommand{\arraystretch}{1}
	$\begin{array}{c}
	\{q,K,M,N\} \\
	\textcolor{white}{.}\end{array}$ & 
	\renewcommand{\arraystretch}{1}
	$\begin{array}{c}
	\{q_1,q_2,K,M,N_\mathrm W\} \\
	N_{\text{eff}} := q_1 \widetilde{M} + q_2 K ,\ \ \widetilde{M} := M + K
	\end{array}$ \\ \hline 
	\renewcommand{\arraystretch}{1}
	$\begin{array}{c}
\text{relations} \\
\text{among fluxes}
\end{array}$ & 
	$N_{\text{eff}} =  N_\mathrm W - \frac{MN_\mathrm S }{K}$ & 
	$N = q M$ &
	$N_{\text{eff}} = N_\mathrm W -  \frac{ (q_1 + q_2) (\widetilde{M} + K)(q_2 \widetilde{M} + q_1 K)}{ (q_1 - q_2) (\widetilde{M} - K)}$ \\ \hline 
$\partial_\chi = \partial_\phi + (..)\partial_z$ & 
	$\frac{N_{\text{eff}}}{M+K}$ & 
	$\frac{N}{M+K}$ &
	$\frac{N_{\text{eff}}}{\widetilde{M}+K}$ \\  \hline
$c_{\text{hol}}$ & 
	$\frac{1}{12} \frac{ K^2 N_{\text{eff}}^2}{M + K}$ & 
	$\frac{1}{12} \frac{ K^2 N^2}{ M+K}$ &
	$ \frac{1}{12} \frac{K^2\left( N_{\text{eff}}^2 +q_2 (q_1 + q_2)  (\widetilde{M}^2 - K^2)  \right) }{\widetilde{M} + K}  $ \\ \hline
$\left( \Delta(\CO_1),\Delta(\CO_2^i)\right)  $ & 
	$(\frac{ K N_{\text{eff}}}{M+K}, K)$ & 
	$(\frac{ K N}{ M+K}, K)$ & 
	$(\frac{K N_{\text{eff}}}{\widetilde{M} + K}, K)$ \\ \hline 
other operators & 
	$\Delta(\CO_3^j)=M$  & 
	$-$ & 
	\renewcommand{\arraystretch}{1}
	$\begin{array}{c} 
	\Delta(\PP_1) = \frac{K \left( N_{\text{eff}} - (q_1-q_2)(\widetilde{M}-K)\right)}{\widetilde{M} + K} \\
	\Delta(\PP_2^j) = K
	\end{array}$ \\ \hline
\end{tabular}
\caption{The flux quanta, $U(1)_r$ isometry generator, holographic $c$-central charge, and operator dimensions of the Case I, II, III holographic SCFTs, presented in variables most amenable to comparison. Note that Case II is recovered from the limit of Case I for which $N_\mathrm S\to 0$, in which case $N_\mathrm W\to N$. In section \ref{sec:gen_caseII}, the geometries with a monopole of charge $q$ are generalized to monopole profiles corresponding to general Young tableaux. \label{tab:sum} }
\end{table}
\end{center}
\renewcommand{\arraystretch}{1}

\subsubsection{Case I}

\paragraph{Geometry.}  
The allowed region in the $(t,u)$ plane
is $[0,t_1]\times[0,u_2]$ as depicted in Figure \ref{fig_cases_new}. We observe that, in this case,
the inequalities $f_1 \ge 0$, $f_2 \ge 0$ are automatically satisfied
once the other inequalities in \eqref{inequalities} option (a) are satisfied.
The radius $R_\phi$ of the $\phi$ circle in the 3d base space
spanned by $(t,u,\phi)$ goes to zero
along the segments $u = u_2$ and $t = t_1$. Along these segments,
the function $L$ in the 1-form $Dz$ is piecewise constant, 
\beq \label{caseI_monopole}
L(t,u_2) = 0,\qquad L(t_1,u) = -\frac{2(t_2u_2 - t_1u_1)}{\sigma(t_2-t_1)(u_2-u_1)} \equiv q  \ .
\eeq
The jump in $L$ at the corner $(t,u) = (t_1,u_2)$ signals the presence of a monopole
source for the $Dz$ fibration over the 3d base space $(t,u,\phi)$.
The charge of the monopole source
is the quantity $q$ in \eqref{caseI_monopole}, which is automatically positive
for the ranges of the parameters that yield Case I.
In order to have a well-defined geometry, the charge $q$ must be an integer,
\beq
q \in \mathbb N \ .
\eeq
The radius $R_z$ of the $S^1_z$ fiber over $(t,u,\phi)$ 
has an isolated zero at the location of the monopole point. Indeed, near the monopole
 the 4d geometry spanned
by $(t,u,\phi,z)$ is locally $\mathbb R^4/\mathbb Z_q$. Thus, for $q \ge 2$,
the space develops an orbifold singularity.

Along the segment $\mathbf P_1 \mathbf P_2$, the warp factor goes to zero.
In terms of the Toda angular variables $\chi$, $\beta$, 
in the limit $u \rightarrow 0$, the line element takes the form
\begin{align} \label{M5source}
ds_{11}^2 \approx&\; \frac{u^{1/3}}{m^2}\left[\frac{u_1u_2t^{2/3}}{(u_1+u_2)^{1/3}}ds^2(AdS_5) + \frac{u_1u_2t^{2/3}}{(u_1+u_2)^{1/3}}D\chi^2\right]\\
&+ \frac{u^{-2/3}}{m^2}\left[\frac{(u_1+u_2)^{2/3}}{4}\left(du^2 + u^2ds^2(S^2)\right) - \frac{u_1u_2(u_1+u_2)^{2/3}}{4(t-t_1)(t-t_2)}\left(dt^2 + \frac{K_1^2}{t^2}d\beta^2\right)\right].\nonumber
\end{align}
Thus, we have M5-brane sources smeared along the $t$ and $\beta$ directions, with harmonic function $H\propto 1/u$. 
More precisely, the M5-branes are extended along the $AdS_5$ and $\chi$
directions, while the are smeared in two directions, $t$ and $\beta$.
The geometry 
near the segment $\mathbf P_4 \mathbf P_1$ is completely analogous, 
with a harmonic function $H\propto 1/t$.
These smeared M5-branes sources are denoted with the label sM5 in
Figure \ref{fig_cases_new}. They are the same sort of sources   considered in~\cite{Bah:2021hei}.

\paragraph{Flux quantization.}

 The geometry of Case I admits non-trivial 4-cycles,
constructed as follows. The 4-cycle $\cA_{4,\mathrm N}$ is obtained
combining the   segment $\mathbf P_3 \mathbf P_4$ with the $S^2$ and the $Dz$ fiber.
Notice that the $S^2$ shrinks at $\mathbf P_4$, while the $Dz$ fiber shrinks
at the monopole point $\mathbf P_3$. By a similar token,
we define the 4-cycle $\cA_{4,\mathrm E}$ by combining the  
segment   $\mathbf P_2 \mathbf P_3$ with the $S^2$ and the $Dz$ fiber.
The periods of $\overline G_4$ over $\cA_{4,\mathrm N}$, $\cA_{4,\mathrm E}$
are determined by  the values of $W$  in \eqref{with_new_angles}
at the points $\mathbf P_{2,3,4}$. With the expressions recorded in
appendix \ref{sec:fibr_change}, we find
\begin{align}
\int_{\mathcal{A}_{4,\mathrm N}}\overline{G}_4  =  -\frac{\sigma}{2}(u_2-u_1)t_1 \equiv  K \in \mathbb{N}\ ,  
 \qquad
\int_{\mathcal{A}_{4,\mathrm E}}\overline{G}_4  =  -\frac{\sigma}{2}(t_2-t_1)u_2 \equiv  M \in \mathbb{N} \ .
\label{caseI_east}
\end{align}
In these expressions we have fixed the mass scale $m$ as in \eqref{eq_fix_m}.

We also have 4-cycles that measure the charges of the smeared M5-brane loci, 
in the spirit of a ``Gaussian pillbox'' from electrostatics.
We take two 4-cycles\footnote{To be proper, we construct these using intervals away from the edge, then consider the limit as we take either $t$ or $u$ to zero.}, $\mathcal{A}_{4,\mathrm S}$ and $\mathcal{A}_{4,\mathrm W}$, to measure the charges of the brane stacks along the South and West edges respectively. These are comprised of the relevant interval, the $S^2$, and the $\phi$ circle. The associated periods of $\overline G_4$
are determined by the values of $Y$ in \eqref{with_new_angles}
at $\mathbf P_{2,3,4}$. The result reads
\begin{align}
\int_{\mathcal{A}_{4,\mathrm W}}\overline{G}_4  = \frac{(t_1+t_2)u_2^2}{u_2-u_1}\equiv  N_{ \mathrm W} \in \mathbb{N} \ , \qquad
\int_{\mathcal{A}_{4,\mathrm S}}\overline{G}_4   =  \frac{(u_1+u_2)t_1^2}{t_2-t_1}\equiv  N_{\mathrm S}  \in \mathbb{ N} \ .
\end{align}
The expressions of the flux quanta $K$, $M$, $N_{\mathrm W}$, $N_{\mathrm S}$
and the monopole charge $q$ imply
\beq \label{eq:CaseI_flux_relation}
M = \frac{N_{\mathrm W}}{q+\frac{N_{\mathrm S}}{K}}    \ .
\eeq
In particular, integrality of $M$ imposes a constraint on the possible values
for $N_{\rm W}$, $N_{\rm S}$, $K$, $q$:
$Kq + N_{\mathrm S}$ must divide $K N_{\mathrm W}$.
An analogous constraint was found in the solutions of \cite{Bah:2021hei}.

With the flux quanta computed above, the $\chi$ Killing vector can be written as
\beq \label{eq:CaseI_Killing}
\partial_\chi = \partial_\phi + \frac{q M}{K+M}\partial_z  =
\partial_\phi + \frac{N_{\text{eff}}}{K+M}\partial_z \  , \qquad N_{\rm eff} := qM \ .
\eeq
In the second step, we have defined an ``effective brane charge''
$N_{\rm eff}$,  which will prove to be
useful in discussing the dual field theory interpretation of these solutions.

\paragraph{Central charge.}  

For an $AdS_5$ solution in 11d supergravity of the form
\beq
ds_{11}^2 = \frac{e^{2\tilde{\lambda}}}{m^2}\left[ds^2(AdS_5) + ds^2(M_6)\right] \ ,
\eeq
 the holographic central charge is computed via the relation \cite{Gauntlett:2006ai}
\beq \label{eq_c_formula}
c = \frac{1}{2^7\pi^6(m\ell_p)^9}\int_{M_6}e^{9\tilde{\lambda}}\;\mathrm{vol}_{M_6}.
\eeq
Using our metric in \eqref{LLM_yy_metric}, 
for values of the parameters yielding Case I,
we can compute $c$ as 
\begin{align} \label{eq:CaseI_central}
c_{\text{I}} &= \frac{-\sigma (t_1u_2)^2(t_2u_2-t_1u_1)}{2^93\pi^3(m\ell_p)^9}
= \frac{1}{12}\frac{(qKM)^2}{K+M} = \frac{1}{12}\frac{qK^2 N_{\text{eff}}^2}{qK + N_{\text{eff}}} \ . 
\end{align}
In the second step, we have fixed $m$ according to \eqref{eq_fix_m} and we have expressed the result in terms of
the monopole charge $q$ in \eqref{caseI_monopole} and the flux quanta defined in
\eqref{caseI_east}, \eqref{eq:CaseI_Killing}.

\paragraph{M2-brane operators.}  

An M2-brane   wrapping a calibrated 2d submanifold $\cC_2$ in the internal space
yields a BPS operator. The calibration condition reads
\beq \label{eq_calibration}
Y'\big|_{\mathcal{C}_2} = \mathrm{vol}_{M_6}(\mathcal{C}_2) \ ,
\eeq
where the right hand side is the induced volume form from $M_6$.
The calibration 2-form $Y'$ is a   bilinear in the Killing spinors \cite{Gauntlett:2006ai}.
For a solution in canonical LLM form \eqref{LLM}, it reads \cite{Bah:2021hei}
\begin{align} \label{calibration2form}
Y'  =&\;  \frac{1}{4}y^3 e^{-9\tilde{\lambda}}\mathrm{vol}_{S^2} + \frac{1}{2}y e^{-3\tilde{\lambda}}(1-y^2 e^{-6\tilde{\lambda}})d\tau\wedge D\chi
\nn \\
& -\frac{1}{2}\tau e^{-3\tilde{\lambda}}D\chi\wedge dy - \frac{1}{4}\frac{y e^{-9\tilde{\lambda}}\tau e^D}{1-y^2 e^{-6\tilde{\lambda}}} dx^1\wedge dx^2 \ .
\end{align}
In the previous expression, the quantity $\tau$ is a coordinate on the $S^2$,
which is parametrized as
\beq \label{S2_coords}
ds^2(S^2) = \frac{d\tau^2}{1-\tau^2} + (1-\tau^2)d\varphi^2.
\eeq
The explicit expression of $Y'$ for the metric in \eqref{with_new_angles}
is reported in appendix \ref{sec:fibr_change}.

Let us take $\cC_2$ to be the $S^2$ on top
of the monopole point $\mathbf P_3$ (at which both $\partial_\phi$ and $\partial_z$ shrink).
The calibration 2-form restricted on $\cC_2$ reads
\beq
Y'\big|_{\mathcal{C}_2} = \frac{1}{4}(t_1u_2)^3 \left(\frac{(t_1+t_2)u_2 - (u_1+u_2)t_1}{(t_1u_2)(t_1t_2u_2^2 - u_1u_2t_1^2)}\right)^{3/2}\mathrm{vol}_{S^2} = \frac{1}{4}\mathrm{vol}_{S^2} \ .
\eeq
On the other hand, the induced volume form on $\cC_2$ is
\beq
ds^2(\mathcal{C}_2) = \frac{1}{4}\frac{t_1^2u_2^2}{t_1^2u_2^2}ds^2(S^2) = \frac{1}{4}ds^2(S^2) \ .
\eeq
The calibration condition \eqref{eq_calibration} is then satisfied. 
We can compute the conformal dimension of an M2-brane
operator $\cO$
 via \cite{Gauntlett:2006ai}
\beq \label{dim_formula}
\Delta (\mathcal{O}) = \frac{1}{4\pi^2(m\ell_p)^3}\int_{\mathcal{C}_2}e^{3\tilde{\lambda}}\mathrm{vol}_{M_6}(\mathcal{C}_2).
\eeq
If $\cO_1$ denotes the BPS operator associated to $\cC_2$ as above,
we find
\beq
\Delta(\mathcal{O}_1) = t_1u_2 =  \frac{qKM}{M+K} = \frac{qKN_{\text{eff}}}{qK + N_{\text{eff}}}.
\eeq

We can define another submanifold $\mathcal{B}_2$ by considering the interval $\mathbf P_3 \mathbf P_4$ and the linear combination of the $z$ and $\phi$ circles that does not vanish along this interval. That is, we choose the $Dz$ fiber. This 2-manifold sits at a single point on the $S^2$. 
We notice that $\cB_2$ is not a 2-cycle, but rather describes an open M2-brane ending on
a smeared M5-brane source. 
The form $Y'$ can be computed as 
\beq
Y'\big|_{\mathcal{B}_2} = \frac{\tau_*(2+\lambda)(u_1-u_2)}{8}\sqrt{\frac{u_1t + u_2(t-t_1-t_2)}{u_2^2t(u_1t^2 - t_1t_2u_2)}}dt\wedge Dz \ ,
\eeq
while the induced metric is
\beq
ds^2(\mathcal{B}_2) =  R_z^2(t,u_2) Dz^2 + \frac{(t_1+t_2)u_2 - (u_1+u_2)t}{4tu_2(t-t_1)(t-t_2)}dt^2 \ .
\eeq
One can readily check that the calibration condition is satisfied, provided we choose $\tau_* = 1$. 
Let $\cO_2$ denote the operator associated to $\cB_2$. Since 
 $\mathcal{B}_2$ is an open M2-brane,
 we expect $\cO_2$ to admit a degeneracy due to possible choice of boundary
conditions for the M2-brane ending on the  M5-branes.
The dimension of   $\cO_2$   is
\beq
\Delta(\mathcal{O}_2^i ) = -\frac{\sigma}{2}t_1(u_2-u_1) = K \ .
\eeq

We can construct yet another submanifold $\mathcal{D}_2$ using the interval $\mathbf P_3 \mathbf P_2$ and the $Dz$ fiber. We find that $Y'$ takes the form
\beq
Y'\big|_{\mathcal{D}_2} = \frac{\tau_*\sigma(t_2-t_1)}{4}\sqrt{\frac{t_2u + t_1(u-u_1-u_2)}{t_1^2u(t_2u^2 - t_1u_1u_2)}}du\wedge Dz \ ,
\eeq
and the induced metric becomes
\beq
ds^2(\mathcal{D}_2) =  R_z^2(t_1,u) Dz^2 - \frac{(t_1+t_2)u - (u_1+u_2)t_1}{4t_1u(u-u_1)(u-u_2)}dt^2.
\eeq
The calibration condition is satisfied   provided $\tau_* = 1$, and we again have a collection of BPS operators, denoted collectively $\mathcal{O}_3$, with  conformal dimension
\beq
\Delta(\mathcal{O}_3^j) = -\frac{\sigma}{2}u_2(t_2-t_1) = M = \frac{N_{\text{eff}}}{q} \ .
\eeq

\subsubsection{Case II}

 The allowed domain in Case II is $[0,t_1]\times[0,u_2]$,
 the same as in Case I, see Figure \ref{fig_cases_new}. Case II can be regarded
 as a limiting case of Case I, in which $u_1 + u_2 >0$ is sent to zero.
The salient new feature of Case II is the behavior of the metric
near the segment $\mathbf P_1 \mathbf P_2$. One can verify that
in this case the $S^2$ shrinks smoothly, while the warp factor remains finite.
Compared with Case I, we still have a monopole source for $Dz$ at
$\mathbf P_3$, and a smeared M5-branes source along $\mathbf P_4 \mathbf P_1$.

The solutions of Case II are the same as the solutions discussed in \cite{Bah:2021hei}.
We refer the reader to appendix \ref{sec:fibr_change} for the explicit change of coordinates
that makes the correspondence manifest. Since these solutions have already been
studied in detail in \cite{Bah:2021hei}, we will be brief.

Flux quanta $K$, $M$ can be defined for Case II, in complete analogy with
\eqref{caseI_east}. We also have the analog of the flux quantum
$N_{\mathrm W}$, while $N_{\mathrm S}$ is absent, since we no longer have a smeared
M5-brane source along $\mathbf P_1 \mathbf P_2$. The flux quanta
$K$, $M$, $N_{\mathrm W}$ satisfy the same constraint as \eqref{eq:CaseI_flux_relation} with
$N_{\mathrm S}$ set to zero.

The expression of the holographic central charge is
\beq
c_{\text{II}} =  \frac{1}{12}\frac{qK^2 N^2}{qK + N} \  , \qquad N := qM \ .
\eeq 
We have the direct analogs of the $\cO_1$ operators associated to M2-branes
wrapping the $S^2$ on top of the monopole point, as well as analogs of the operators
$\cO_2$ associated to open M2-branes ending on the M5-brane source 
along $\mathbf P_1 \mathbf P_4$. We do not have, however, the analog
of the $\cO_3$ operators, because we only have one set of smeared M5-branes.
The dimensions of $\cO_1$, $\cO_2$ are
\beq
\Delta(\cO_1) = \frac{qKN}{qK + N} \ , \qquad
\Delta(\cO_2^i) = K \ .
\eeq

\subsubsection{Case III}

\paragraph{Geometry.}  

The allowed domain in Case III is the rectangle $[0,t_1] \times [u_1, u_2]$, 
see Figure \ref{fig_cases_new}.   In this case, we find that the $\phi$ circle in the
3d base spanned by $(t,u,\phi)$ shrinks along the three segments
$\mathbf P_1 \mathbf P_2$, $\mathbf P_2 \mathbf P_3$,
and $\mathbf P_3 \mathbf P_4$. The function $L$ is piecewise constant
along these segments,
\beq \label{eq:monopole_charges_yy}
L(t,u_2) = 0,\qquad L(t_1,u) = -\frac{2(t_2u_2 - t_1u_1)}{\sigma(t_2-t_1)(u_2-u_1)},\qquad L(t,u_1) = -\frac{2(u_2+u_1)}{\sigma(u_2-u_1)} \ .
\eeq
We see that $L$ jumps both at $\mathbf P_2$ and at $\mathbf P_3$.
It follows that the $Dz$ fibration has two monopole sources,
located at $\mathbf P_3$, $\mathbf P_2$, with charges $q_1$, $q_2$ respectively,
\beq
\text{at $\mathbf P_3$:} \;\; q_1 = 
 -\frac{2(t_2u_2 - t_1u_1)}{\sigma(t_2-t_1)(u_2-u_1)} \in \mathbb N \  , \qquad
 \text{at $\mathbf P_2$:} \;\; q_2 = 
 -\frac{2(t_2u_1 - t_1u_2)}{\sigma(t_2-t_1)(u_2-u_1)} \in \mathbb N \ .
\eeq
Along the segment $\mathbf P_4 \mathbf P_1$ we have a smeared M5-brane
source of the same kind as explained in Case I.

\paragraph{Flux quantization.}
A first class of 4-cycles is associated to the three components along the boundary
of the allowed region, in which the $\phi$ circle in the base shrinks. Let us define
\begin{align}
\mathcal B_{4, \mathrm N} \; &: \qquad \text{segment $\mathbf P_3 \mathbf P_4 $ combined with the $Dz$ fiber and the $S^2$} \ , \nn \\
\mathcal B_{4, \mathrm E} \; &: \qquad \text{segment $\mathbf P_2 \mathbf P_3$ combined with the $Dz$ fiber and the $S^2$} \  , \\
\mathcal B_{4, \mathrm S} \; &: \qquad \text{segment $\mathbf P_1 \mathbf P_2$ combined with the $Dz$ fiber and the $S^2$} \ . \nn
\end{align}
As before, we can compute the flux through these cycles in terms of the values
of the function $W$ at $\mathbf P_{1,2,3,4}$. The result reads 
\begin{align}
\int_{\mathcal{B}_{4, \mathrm N}}\overline{G}_4   &=
\int_{\mathcal{B}_{4,\mathrm S}}\overline{G}_4 =  -\frac{\sigma}{2}t_1(u_2 -u_1) \equiv K \in \mathbb{N} \  , \nn \\
\int_{\mathcal{B}_{4,\mathrm E}}\overline{G}_4   &= -\frac{\sigma}{2}(t_2-t_1)(u_2 - u_1) \equiv M \in \mathbb{N} \ .
\end{align}
Interestingly we find that the fluxes through $\cB_{4,\mathrm N}$ 
and $\cB_{4,\mathrm S}$ are equal.

Next, we can construct a 4-cycle to measure the flux from the  M5-branes
source along $\mathbf P_4 \mathbf P_1$.
 To this end,  we consider the $S^2$ combined with the $u$ interval and the $\phi$ circle,
 yielding a 4-cycle denoted $\cA_{4, \mathrm W}$. The associated flux quantum is 
\begin{align*}
\int_{\mathcal{A}_{4,\mathrm W}}\overline{G_4}   = \frac{(t_1+t_2)(u_2^2+u_1^2)}{u_2-u_1}\equiv  N_W \in \mathbb{N}.
\end{align*}
We find it convenient to define
\beq
\widetilde{M} = M+K \ ,      \qquad N_{\text{eff}} = q_1\widetilde{M} + q_2 K \ .   
\eeq
We may then write
\beq \label{eq:CaseV_Killing}
\partial_\chi = \partial_\phi + \frac{Mq_1 + K(q_1+q_2)}{M+2K}\partial_z
= \partial_\phi + \frac{N_{\text{eff}}}{\widetilde{M}+K}\partial_z \ .
\eeq 
The quanta found above satisfy the relation  
\beq\label{eq:CaseV_flux_relation}
N_W = \frac{M^2(q_1^2 + q_2^2) + 2K(K+M)(q_1+q_2)^2}{M(q_1 - q_2)},
\eeq
which in terms of $N_{\rm eff}$ and $\widetilde{M}$ can be written
\beq
N_{\text{eff}}  = N_W - \frac{(q_1 + q_2) (\widetilde{M}+K) }{(q_1-q_2) (\widetilde{M}-K)} (q_1 K + q_2 \widetilde{M} ).
\eeq

\paragraph{Central charge.}
Applying formula \eqref{eq_c_formula}, we arrive at
\begin{align} \label{eq:CaseV_central}
c_{\text{III}} &= \frac{-\sigma}{2^93\pi^3(m\ell_p)^9}t_1^2\left(t_1u_1u_2(u_1-u_2) + t_2(u_2^3-u_1^3)\right)\nonumber\\
&= \frac{1}{12}\frac{K^2\left((K+M)^2(q_1 + q_2)^2-M^2q_1q_2\right)}{M+2K}\nonumber\\
&= \frac{1}{12}\frac{K^2(\widetilde{M}^2(q_1+q_2)^2 - (\widetilde{M}-K)^2q_1q_2)}{\widetilde{M}+K}
 \ .
\end{align}

\paragraph{M2-brane operators.}
We still have the direct analogs of the calibrated submanifolds
$\cC_2$ and $\cB_2$ constructed in Case I,
associated to operators $\cO_1$, $\cO_2$, respectively.
In the case at hand, their dimensions are
\beq \label{eq:CaseV_Odim}
\Delta(\mathcal{O}_1) = \frac{K}{M+2K}\big(Mq_1 + K(q_1+q_2)\big)=\frac{KN_{\text{eff}}}{\widetilde{M}+K} \ , \qquad \Delta(\mathcal{O}_2^i) = K \  .
\eeq
We also have a new calibrated 2-cycle, analogous to $\cC_2$,
given by the $S^2$ on top of the 
new monopole point $\mathbf P_2$. If we denote the corresponding 
operator as $\cP_1$, we find
\beq
\Delta(\mathcal{P}_1) = t_1u_1 = \frac{K}{M+2K}\big(Mq_2 + K(q_1 + q_2)\big) = \frac{K N_{\text{eff}}}{\widetilde{M}+K} - \frac{\widetilde{M}-K}{\widetilde{M}+K}(q_1-q_2)K \ .
\eeq
Moreover, we also have a new open calibrated submanifold, analogous to $\cB_2$,
constructed using $\mathbf P_1 \mathbf P_2$ and the $Dz$ fiber. We thus obtain a family of BPS operators, denoted collectively $\cP_2^j$, with dimension
\beq
\Delta(\mathcal{P}_2^j) = -\frac{\sigma}{2}t_1(u_2-u_1) = K.
\eeq
Finally, we can combine the $\mathbf P_2 \mathbf P_3$ segment connecting the two
monopoles and the $Dz$ fiber to obtain a closed calibrated submanifold.
We denote the associated BPS operator as $\cQ$ and we compute its dimension to be
\beq
\Delta(\cQ) = - \frac \sigma 2 (t_2 - t_1)(u_2-u_1) = M  \ . 
\eeq


\subsection{Examples of non-rectangular domains} \label{sec_non_rect}

\begin{figure}
\includegraphics[width= 15.5 cm]{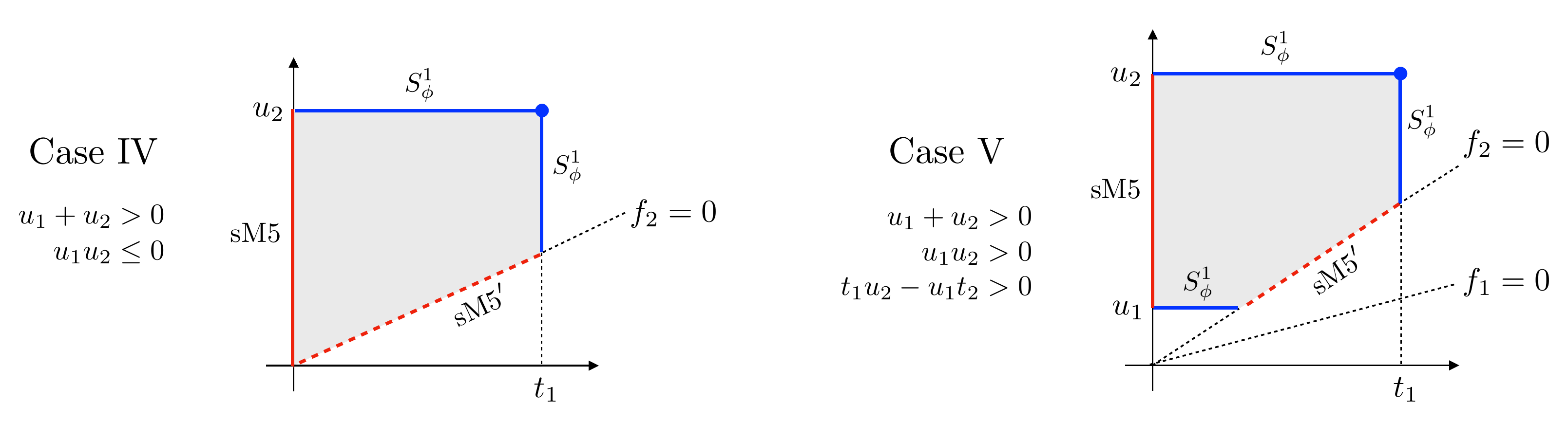}
\caption{\label{fig_non_rect} 
Non-rectangular domains arising from  the inequalities \eqref{inequalities} option (a),
under the assumptions $0 < t_1 \le t_2$ and $u_1 < u_2$.
The label $S^1_\phi$ indicates that the $\phi$ circle in the base of the $Dz$ fibration
\eqref{with_new_angles} shrinks smoothly along that component of the boundary of the allowed region.
The labels  sM5 and $\rm sM5'$    
 stand for two different kinds of smeared M5-brane sources.
 }
\end{figure}

In the previous section, we have assumed $0< t_1 \le t_2$, $u_1 < u_2$,
as these are necessary conditions for having a rectangular domain.
These conditions, however, are not sufficient. Indeed, if we let $u_{1,2}$ vary,
we obtain two more cases, with non-rectangular domains, denoted IV and V
and summarized in Figure \ref{fig_non_rect}.

Let us remark that Cases IV and V do not provide a full classification of all
non-rectangular, compact domains. Further domains (including triangular domains)
can be obtained by relaxing some of the assumptions
$t_{1,2}\in \mathbb R$, $0< t_1 \le t_2$, $u_1 < u_2$. 
In this section, we provide some details on Cases IV and V
as representative examples of non-rectangular domains, but we refrain from a full classification.

Cases IV and V are conveniently described in terms of the same
angular coordinates $\phi$, $z$ introduced  above in \eqref{new_angles} and
entering  \eqref{with_new_angles}.

\subsubsection{Case IV}

\paragraph{Geometry.}

The main novel feature of Case IV is the presence of a boundary
component with positive slope, which is determined by the inequality
$f_2 \ge 0$, see Figure \ref{fig_non_rect}. Near the vertical and horizontal segments
along the boundary of the allowed region, we have similar features
as the Cases I, II, III.

Let us look closer at the geometry along the diagonal line 
defined by $f_2 = 0$. For convenience, and in analogy with \eqref{M5source},
we work with the Toda angular variables $\chi$, $\beta$. 
We rearrange the coordinates $(t,u)$ into linear combinations
$x_\parallel$, $x_\perp$, where $x_\parallel$ runs longitudinally 
to the diagonal line, and  $x_\perp$ runs transversally.
More precisely, 
we can set
\beq
t = \bar t x_\parallel - \bar u x_\perp \ ,  \qquad
u = \bar u x_\parallel + \bar t x_\perp \ , \qquad
\bar t := t_1 + t_2 \ , \qquad \bar u := u_1 + u_2 \ .
\eeq
In the limit $x_\perp \rightarrow 0$, the line element takes the form
\begin{align}\label{eq:CaseI_diag_metric}
ds_{11}^2 \approx&\;  \frac{x_\perp^{-1/3}}{m^2}\Bigg[ x_\parallel^{4/3}\Bigg(\frac{\bar{t}\bar{u}\left(t_1t_2\bar{u}^2-u_1u_2\bar{t}^2\right)}{\left(\bar{t}^2+\bar{u}^2\right)^2}\Bigg)^{1/3}\left(ds^2(AdS_5) + D\chi^2\right)\Bigg] \\
&+ \frac{x_\perp^{2/3}}{m^2}\Bigg[ \frac{1}{4}\Bigg(\frac{\left(t_1t_2\bar{u}^2-u_1u_2\bar{t}^2\right)}{(\bar{t}\bar{u} x_\parallel )^2\left(\bar{t}^2+\bar{u}^2\right)^2}\Bigg)^{1/3}\Bigg(\frac{(\bar{t}\bar{u} x_\parallel )^2}{\left(t_1t_2\bar{u}^2-u_1u_2\bar{t}^2\right)}ds^2(S^2) + \left(\bar{t}^2+\bar{u}^2\right)^2ds^2(M_3)\Bigg)\Bigg]\nonumber \ .
\end{align}
Here
 $ds^2(M_3)$ denotes the metric on the space  described by $(x_\perp ,x_\parallel ,\beta)$,
whose explicit expression is omitted for brevity. 
We interpret \eqref{eq:CaseI_diag_metric} in terms of a smeared M5-brane source
with harmonic function  $H \propto x_\perp$.
The linear behavior of $H$ is indicative of a smearing to effective codimension 1.
Indeed, the M5-branes are now smeared over more directions, compared 
to \eqref{M5source}. More precisely, they are extended along $AdS_5$ and $\chi$,
and smeared in all other directions, except $x_\perp$.
In Figure \ref{fig_non_rect} we use the label $\rm sM5'$ to 
signal this new kind of source.

\paragraph{Flux quantization.} The analysis of $G_4$-flux
quantization is closely analogous to Case I. 
The role of the segment $\mathbf P_1 \mathbf P_2$ of Case I is now played
by the diagonal line determined by $f_2 = 0$.
We can define the analog of the flux quanta 
 $M$, $K$, $N_\mathrm S$, and $N_\mathrm W$ and   verify that they satisfy the same relation
 as in Case I.
 
 \paragraph{Central charge.} Application of \eqref{eq_c_formula} yields the result 
\begin{align} \label{eq:CaseIII_central}
c_{\text{IV}} &=  -\frac{\sigma}{15}\frac{1}{2^9\pi^3(m\ell_p)^9}t_1^2\left(5u_2^2(t_2u_2 - t_1u_1) + \frac{t_1^3(u_1+u_2)^3}{(t_1+t_2)^2}\right)
\nn \\
&= \frac{1}{120}\frac{qKM}{M+K}\left(5qKM + \frac{N_ \mathrm S^3M^3}{N_\mathrm W^2K^2}\right) \ .
\end{align}

 

\subsubsection{Case V}

The salient features of the geometry of Case V are a combination
of the ingredients already introduced above,
as can be seen from Figure \ref{fig_non_rect}.
The charge of the monopole   at $(t_1, u_2)$ is
\beq
q = \frac{2 (t_1 u_1-t_2 u_2)}{\sigma  (t_1-t_2)
   (u_1-u_2)}  \ .
\eeq
We can define the analog of the flux quanta $K$ and $M$,
given by differences of the values of $W$ at the points
$(t_1, u_2)$ and $(0,u_2)$, $(t_1 , (u_1 + u_2) t_1/(t_1+t_2))$,
\beq
K =   - \frac \sigma 2 t_1(u_2-u_1) \  , \qquad
M = - \frac \sigma 2 u_2 (t_2-t_1) \ .
\eeq
We observe that the analogous flux constructed with the horizontal
segment at $u = u_1$ is vanishing.
We can also define the analog of the flux $N_\mathrm W$, which measures
the charge of the $\rm sM5$ smeared source, and a new flux
$N_{\rm d}$ which measures the $\rm sM5'$ charge along the diagonal
component of the boundary of the allowed region.
These fluxes are determined by the values of $Y$, and are given by
\beq
N_{\rm W} = -\frac{(t_1+t_2) \left(u_1^2+u_2^2\right)}{u_1-u_2} \ , \qquad
N_{\rm d} = \frac{u_1^2 (t_1+t_2)}{u_1-u_2}-\frac{t_1^2
   (u_1+u_2)}{t_1-t_2} \ .
\eeq
The flux quanta and the monopole charge satisfy the following identity,
\begin{align} \label{caseV_id}
0 =&\; 2 q (N_{\rm d}+N_{\rm W}) \left(K^2 N_{\rm W}-K M (N_{\rm d}+N_{\rm W})+M^2
   (2 N_{\rm d}+N_{\rm W})\right)
   \nn \\
   &-2 K M q^3 \left(K^2+M^2\right)+q^2
   (K-M)^2 (K N_{\rm W}-M (2 N_{\rm d}+N_{\rm W}))
   \nn \\
   &+(N_{\rm d}+N_{\rm W})^2
   (K N_{\rm W}-M (2 N_{\rm d}+N_{\rm W})) \ .
\end{align}
Let us define
\beq
\widetilde N = N_{\rm d} + N_{\rm W} \ .
\eeq
We can use \eqref{caseV_id} to express $N_{\rm W}$ in terms
of $q$, $K$, $M$, $\widetilde N$,
\begin{align}
N_{\rm W} & = \frac{2 M (K q+{\widetilde N}) \left(q^2 \left(K^2+M^2\right)-2 M {\widetilde N}
   q+{\widetilde N}^2\right)}{(K+M) (q (K-M)+{\widetilde N})^2} \ .
\end{align}
The expression of the central charge
in terms of the parameters $\sigma$, $t_{1,2}$, $u_{1,2}$ is
\begin{align} \label{eq:CaseIV_central}
c_{\text{V}} &=  -\frac{\sigma}{15}\frac{1}{2^9\pi^3(m\ell_p)^9}\left(\frac{t_1^5(u_1+u_2)^3}{(t_1+t_2)^2} - \frac{u_1^5(t_1+t_2)^3}{(u_1+u_2)^2} + 5t_1^2 u_2^2 (t_2u_2 - t_1u_1)\right) \ . 
\end{align}
We may also express this quantity in terms of the flux quanta and monopole charge,
\begin{align}
c_{\rm V} &=  \frac{K^2 M^2 q^2}{60 (K+M)^2 (K q+{\widetilde N})^2 ({\widetilde N}-M q)^2 (q (K-M)+{\widetilde N})^5} \times 
\nn \\
&\qquad  \times \bigg[
5 K^7 q^6 \left(19 M^2 {\widetilde N} q^2-2 M^3 q^3-22 M {\widetilde N}^2 q+7
   {\widetilde N}^3\right)
   \nn \\
   &\qquad +5 K^6 q^5 \left(68 M^2 {\widetilde N}^2 q^2-27 M^3 {\widetilde N} q^3+6
   M^4 q^4-69 M {\widetilde N}^3 q+23 {\widetilde N}^4\right)
   \nn \\
   &\qquad +K^5 {\widetilde N} q^4 \left(765 M^2
   {\widetilde N}^2 q^2-470 M^3 {\widetilde N} q^3+125 M^4 q^4-595 M {\widetilde N}^3 q+176
   {\widetilde N}^4\right)
   \nn \\
   &\qquad +5 K^4 q^3 ({\widetilde N}-M q) \left(104 M^2 {\widetilde N}^3 q^2-31 M^3
   {\widetilde N}^2 q^3-11 M^4 {\widetilde N} q^4+4 M^5 q^5-101 M {\widetilde N}^4 q+34
   {\widetilde N}^5\right)
   \nn \\
   &\qquad +5 K^3 q^2 ({\widetilde N}-M q)^2 \left(24 M^2 {\widetilde N}^3 q^2+22
   M^3 {\widetilde N}^2 q^3-15 M^4 {\widetilde N} q^4+2 M^5 q^5-54 M {\widetilde N}^4 q+23
   {\widetilde N}^5\right)
   \nn \\
   &\qquad +5 K^2 q (M q-{\widetilde N})^3 (M q+{\widetilde N}) \left(-14 M^2
   {\widetilde N}^2 q^2+M^3 {\widetilde N} q^3+M^4 q^4+23 M {\widetilde N}^3 q-10 {\widetilde N}^4\right)
   \nn \\
   &\qquad +5
   K^8 M q^8 (3 M q-{\widetilde N})+5 K^9 M q^9+K^{10} q^9
   \nn \\
   &\qquad +5 K
   ({\widetilde N}-M q)^4 \left(-3 M^2 {\widetilde N}^3 q^2+3 M^4 {\widetilde N} q^4-M^5 q^5+2
   {\widetilde N}^5\right)
   \nn \\
   &\qquad +M^3 q^2 (M q-{\widetilde N})^5 \left(M^2 q^2-5 M {\widetilde N} q+5
   {\widetilde N}^2\right)
\bigg] \ .
\end{align}

%
%
%
%




\section{Electrostatic picture}  \label{sec_electro}

In this section we review the map from the axisymmetric Toda 
system to an electrostatic problem, and we describe
the electrostatic interpretation of the solutions discussed in section~\ref{sec:catalog}.

\subsection{Review of the B\"acklund transform}

In this work, we study solutions for which the Toda function $D$
is independent of the angular coordinate $\beta$ in the $x^1$, $x^2$ plane.
For such solutions, it is possible to perform a B\"acklund transform,
which furnishes an electrostatic interpretation for the BPS conditions \cite{Gaiotto:2009gz}.

The B\"acklund transform
takes the coordinates $(r,y)$ and the function $D(r,y)$ of the canonical LLM
form to new coordinates $(\rho,\eta)$ and a new function $V(\rho,\eta)$.
The B\"acklund transform is 
defined implicitly by the relations  
\beq \label{Backlund_def}
\rho^2 = r^2 \, e^D \ , \qquad y = \rho \, \partial_\rho V \ , \qquad \log r = \partial_\eta V   \ .
\eeq
The 11d metric and flux can be written as
\begin{align} \label{Backlund_form}
ds^2_{11} & = \frac{1}{m^2} \, \bigg[ \frac{\dot V \, \widetilde \Delta}{2 \, V''} \bigg]^{1/3}
\bigg[ ds^2_{AdS_5}
+ \frac{ V'' \, \dot V}{2\, \widetilde \Delta} \, ds^2_{S^2}
+ \frac{ V''}{2\, \dot V} \bigg(
d\rho^2 + d\eta^2 + \frac{2 \, \dot V}{2 \, \dot V - \ddot V} \, \rho^2 \, d\chi_{\rm B}^2 \bigg)
\nn \\
& \qquad \qquad \qquad \qquad   + \frac{ \, \dot V  - \ddot V}{2\, \dot V \, \widetilde \Delta} \, \bigg(
d\beta - \frac{2 \, \dot V \, \dot V'}{2 \, \dot V  - \ddot V} \, d\chi_{\rm B} \bigg)^2 \bigg] \ , \nn \\
G_4 & = \frac{1}{4\,m^3} \, {\rm vol}_{S^2} \wedge d \bigg[
- \frac{2 \, \dot V^2 \, V'' }{\widetilde \Delta } \, d\chi _{\rm B} + \bigg(
\eta - \frac{\dot V  \, \dot V'}{ \widetilde \Delta } \bigg) \, d\beta 
\bigg] \ ,
\end{align}
where we used the notation $\dot V = \rho \, \partial_\rho V$, $V' = \partial_\eta V$, and so on, and we
introduced
\beq
\widetilde \Delta = (2 \, \dot V - \ddot V ) \, V'' + (\dot V')^2 \ .
\eeq 
The angle $\chi_{\rm B}$ after the B\"acklund transform
is related to the angle $\chi$ in the canonical LLM form  as  \cite{Bah:2019jts}
\beq \label{chi_B}
\chi_{\rm B}  = \chi + \beta\ .  
\eeq
The function $V$ obeys the 3d Laplace equation in cylindrical
coordinates: away from sources,
\beq \label{Laplace_eq}
\partial_\eta^2 V + \frac 1 \rho \, \partial_\rho (\rho \, \partial_\rho V) = 0  \ .
\eeq
This equation motivates the interpretation of $V$ as an electrostatic potential
in three dimensions.
We allow for electric charges localized along the $\eta$ axis,
with a charge density $\lambda(\eta)$. The charge density can be extracted 
from the potential $V$ via
\beq \label{get_the_charge}
\lambda(\eta) =\lim_{\rho \rightarrow 0^+} \, \rho \, \partial_\rho \, V  \ .
\eeq

\paragraph{Some redundancies in the parametrization.}

We have already observed that the mass scale $m$ can
be fixed according to \eqref{eq_fix_m} without loss of generality.
We also notice that 
the 11d metric and flux depend on $V''$, $\dot V$, $\ddot V$, $\dot V'$,
but not on $V'$.
It follows that a replacement of the form
\beq \label{eq_eta_shift}
V(\rho,\eta) \mapsto V(\rho,\eta) + k \, \eta \ ,
\eeq
where $k$ is an arbitrary constant, 
 has no effect on the 11d metric and flux.


%
 
\paragraph{Line element and flux in terms of $V$, $z$, $\phi$.}

In what follows, it will be convenient to
rewrite the metric and flux in \eqref{Backlund_form}
by trading the angular variables $\chi_{\rm B}$, $\beta$
with the angular variables $z$, $\phi$ first introduced in section \ref{sec:CaseI}
and discussed in greater detail in appendix \ref{sec:fibr_change}.
The change of coordinates that related the Toda angular
variables $\chi$, $\beta$ to the new variables  
 $\chi_{\rm B}$, $\beta$ is of the form
 \beq \label{angular_vars}
\chi = \bigg( 1 + \frac 1 \cC \bigg) \,  \phi -  z \ , \qquad
\beta = - \frac 1 \cC \, \phi + z \ ,
\eeq 
where $\cC$ is a positive constant. 
Below, we demonstrate that $\cC$ can be identified with a ratio
of $G_4$-flux quanta.
Combining \eqref{chi_B} and \eqref{angular_vars}, we can recast 
\eqref{Backlund_form} in the following form
\begin{align} \label{Backlund_form_BIS}
ds^2_{11} & =  (4\pi)^{2/3} \, \ell_p^2 \, \bigg[ \frac{\dot V \, \widetilde \Delta}{2 \, V''} \bigg]^{1/3}
\bigg[ds^2_{AdS_5}
+ \frac{ V'' \, \dot V}{2\, \widetilde \Delta} \, ds^2_{S^2}
+ \frac{ V''}{2\, \dot V} (d\rho^2 + d\eta^2 )
+ R_\phi^2 \, d\phi^2 + R_z^2 \, Dz^2
\bigg] \ , \nn \\
\overline G_4  &: = - \frac{G_4}{(2\pi \ell_p)^3} =   \frac{ {\rm vol}_{S^2}  }{4\pi}\wedge d \bigg[ 
Y \, \frac{d\phi}{2\pi}  - W \, \frac{Dz}{2\pi} 
\bigg] \ ,
\end{align}
where we have introduced
\begin{align} \label{stuff_from_V}
R_\phi^2 & =   \frac{V''}{2 \, \dot V - \ddot V } \, \rho^2  \  , &
R_z^2 & =
\frac{2 \, \dot V - \ddot V}{ 2 \, \dot V \, \widetilde \Delta}
\ , \nn \\
Dz & = dz - L \, d \phi \ , &
L & =  
\frac 1 \cC + \frac{2 \, \dot V \, \dot V '}{2 \, \dot V - \ddot V}
\ , \nn \\
W & =   
\eta  - \frac{\dot V \, \dot V'}{\widetilde \Delta}   
\ , &
Y & =  \frac{2 \, \dot V \, (\dot V - \dot V' \, \eta)}{ 2 \, \dot V - \ddot V} \ .
\end{align} 
Notice that we have fixed the mass scale $m$ according to
\eqref{eq_fix_m}.






\subsection{Electrostatic potential from Toda solutions}  \label{sec:new_solus}


Let us study the B\"acklund transform reviewed above in the cases discussed in Section
\ref{sec:catalog}. The expressions in this subsection apply to all cases.

From \eqref{eq:eD_ansatz} and \eqref{Backlund_def}, we can immediately find the expression for $\rho(t,u)$,
\beq\label{eq:rho_yy}
\rho = \sqrt{K_1K_2} = \sqrt{-\sigma^2(t-t_1)(t-t_2)(u-u_1)(u-u_2)} \ .
\eeq
The function $\eta(t,u)$ can be found in the following way. Let us treat $V_{\text{T}}$ as     function of $t$ and $u$, where we have included the   subscript `T' to remind ourselves that this is the electrostatic potential as inferred from the Toda coordinates via the Backl\"und transformation. We can use the chain rule, equation \eqref{Backlund_def}, and the unspecified $\eta(t,u)$ to write
\begin{align}\label{eq:V_derivs_yy}
\partial_t V_{\text{T}} &= \partial_t \eta \left(\log r_1 + \log r_2\right) + \frac{tu}{2K_1}\partial_t K_1  \ , \\
\partial_u V_{\text{T}} &= \partial_u \eta \left(\log r_1 + \log r_2\right) + \frac{tu}{2K_2}\partial_u K_2 \ . \nonumber
\end{align}
Imposing the integrability condition $\partial_t \partial_u V_{\text{T}} = \partial_u\partial_t V_{\text{T}}$ yields 
\begin{align} \label{eq:eta1_yy}
K_2t\partial_t K_1 - K_1u\partial_u K_2 &= 2\left(K_2t\partial_u\eta + K_1u\partial_t\eta\right) \ ,
\end{align}
while the condition that $V_{\text{T}}$ is a solution to Laplace's equation gives
\begin{align} \label{eq:eta2_yy}
\frac{1}{\rho}(\rho\partial_\rho V_{\text{T}}) + \partial_\eta^2 V_{\text{T}}=0\quad\Rightarrow\quad 2\left(u\partial_u\eta - t\partial_t\eta\right) - \left(t\partial_uK_2 + u\partial_tK_1\right)=0 \ .
\end{align}
By combining \eqref{eq:eta1_yy} and \eqref{eq:eta2_yy}, we can solve for $\partial_t\eta$,
$\partial_u \eta$ and find the simple relations
\beq \label{eq:eta_yy}
\partial_t\eta = \frac{1}{2}\frac{\partial K_2}{\partial u} \qquad\text{and}\qquad \partial_u\eta = -\frac{1}{2}\frac{\partial K_1}{\partial t}.
\eeq
We can use these to solve for the total $\eta(t,u)$ as 
\begin{align} \label{eta_expr_gen}
\eta(t,u) 
&= \sigma\left(tu - \frac{(u_1+u_2)t + (t_1+t_2)u}{2}\right) +A \ . 
\end{align}
where $A$ is an integration constant.


Now that we have expressions for both $\rho$ and $\eta$ in the $(t,u)$ coordinates, we can use \eqref{eq:V_derivs_yy} to find an expression for the potential in $(t,u)$ coordinates. After rearranging we find 
\begin{align} \label{eq:VT_yy}
V_{\text{T}}(t,u) =&\; (\eta-A)\log(r_1r_2) +  t\left(\frac{u_1+u_2}{2}\right) + u\left(\frac{t_1+t_2}{2}\right)\\
&-\frac{1}{2}\Bigg(\frac{u_1+u_2}{t_2-t_1}\left(t_1^2\log ( t_1-t) - t_2^2\log (t_2-t) \right)\nonumber\\
& + \frac{t_1+t_2}{u_2-u_1}\left(u_1^2\log (u-u_1) - u_2^2\log( u_2 -u) \right)\Bigg) + V_0 \ , \nonumber
\end{align}
where $V_0$ is another integration constant. 
We have used the fact that, for all Cases I through V, in the interior
of the allowed region in the $(t,u)$ plane we have
we have $0 < t< t_1 < t_2$ and $u_1 < u < u_2$.

Having determined the explicit change of coordinates from $(\rho,\eta)$ to $(t,u)$,
for each Case I through V we can map the allowed region
in the $(t,u)$ plane to the allowed region in the $(\rho,\eta)$ plane,
up to a constant shift in $\eta$ related to the integration constant $A$,
see Figure \ref{fig:Cases134_re}. In particular, we observe that, in each case,
the components of the boundary in the $(t,u)$ plane where the $\phi$ circle in the base
shrinks are mapped to segments along the $\eta$ axis.

\subsection{Electrostatic interpretation of Cases I and II}

We now analyze the electrostatic interpretation of Case I. The interpretation
for Case II follows by taking the limit $u_2 \rightarrow - u_1$, or equivalently
$N_{\rm S} \rightarrow 0$.


\paragraph{Charge density profile.}

\begin{figure}
	\centering
	\includegraphics[width= \textwidth]{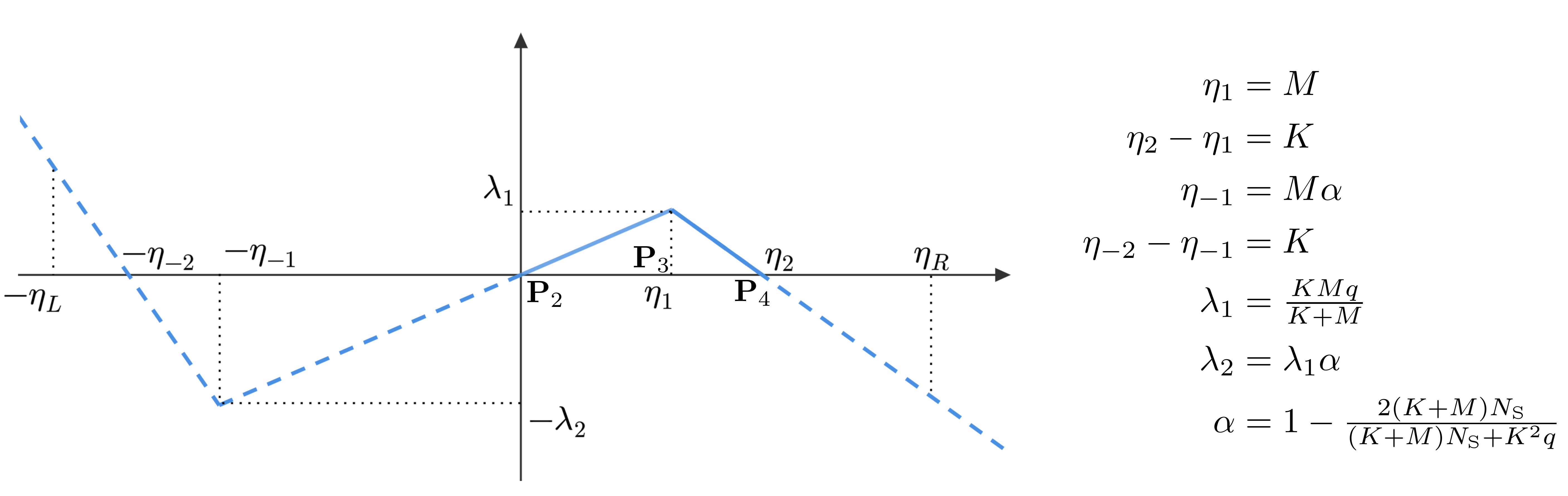}
	\caption{The  extended charge density profile $\lambda(\eta)$ for Case I,
	featuring a ``mirror monopole'' located at $-\eta_{-1}$ paired with the known monopole located at $\eta_1$. The labels $\mathbf P_2$, \dots, $\mathbf P_4$ refer to Figure \ref{fig_cases_new}.
	The extended charge density for Case II can be
	obtained by considering the limit $N_{\rm S} \rightarrow 0$,
	which implies $\eta_{-1,-2} =\eta_{1,2}$, $\lambda_2 = \lambda_1$
	and gives an odd function of $\eta$.}
	\label{fig:2Sided_Lambda}
\end{figure}

The electrostatic potential $V_{\rm T}$ satisfies the Laplace equation
for any $\rho >0$, but there are localized electric sources on the $\eta$ axis.
 We can find their charge density using the formula
\beq
\lambda_{\text{T}} = \lim_{\rho\to 0}\rho\partial_\rho V_{\text{T}}.
\eeq
Firstly, we have the segments on the $\eta$ axis that correspond to the 
 edges $\mathbf P_3 \mathbf P_4 $ and $\mathbf P_2\mathbf  P_3$ in Figure \ref{fig_cases_new}. For these segments we find 
\beq
\lambda_{\text{T}}(t) = u_2 t \ ,  \qquad \lambda_{\text{T}}(u) = t_1 u \ ,
\eeq
respectively.
The charge density is   piecewise linear. At the monopole the charge density takes the form
\beq\label{eq:lambda_1}
\lambda_1 := \lambda_{\rm T}(\mathbf{P_4}) = t_1u_2 = \frac{qKM}{K+M} \ .
\eeq
Using this and the expression \eqref{eta_expr_gen} for $\eta$, we can write the charge density as a piecewise linear function in $\eta$,
\beq
\lambda_{\text{T}}(\eta) = \begin{dcases}
\frac{t_1}{t_2-t_1}\left(-\frac{2}{\sigma}(\eta-A) - t_1(u_1+u_2)\right), & \tilde{\eta}_{\text{min}}\leq (\eta-A) <\tilde{\eta}_1 \ , \\
-\frac{u_2}{u_2-u_1}\left(-\frac{2}{\sigma}(\eta-A) - u_2(t_1+t_2)\right), & \tilde{\eta}_1 \leq (\eta-A) < \tilde{\eta}_2 \ .
\end{dcases}
\eeq
We have defined the special $\eta$ positions as
\beq
\tilde{\eta}_{\text{min}} = -\frac{\sigma}{2}t_1(u_1+u_2),\qquad \tilde{\eta}_1 = -\frac{\sigma}{2}(t_2u_2 + t_1u_1),\qquad \tilde{\eta}_2 = -\frac{\sigma}{2}u_2(t_1+t_2) \ .
\eeq
We find it convenient to fix the integration constant $A$
to the value $A = - \tilde \eta_{\rm min}$,
that is
\beq \label{A_choice}
A = \frac{\sigma}{2}t_1(u_1+u_2) \ .
\eeq
With this choice, the linear charge density takes the form
\beq \label{charge_CaseI}
\lambda_{\text{T}}(\eta) = \begin{dcases}
\frac{\lambda_1}{\eta_1}\eta, & 0 \leq \eta < {\eta}_1 \ , \\
-\frac{\lambda_1}{\eta_2-\eta_1}\left(\eta - \eta_2\right), & {\eta}_1 \leq \eta < {\eta}_2 \ ,
\end{dcases}
\eeq
where we have defined
\beq\label{eq:eta_12}
{\eta}_1 = -\frac{\sigma}{2}u_2(t_2-t_1) =  M \ ,\qquad {\eta}_2 = -\frac{\sigma}{2}(t_2u_2-t_1u_1) = (K+M) \ .
\eeq
This charge density $\lambda(\eta)$    is depicted 
as a solid line in Figure \ref{fig:2Sided_Lambda}.



%

\paragraph{Improved Form for $V$.} The expression for $V_{\rm T}$ in \eqref{eq:VT_yy} suffers from some drawbacks: it is not a closed form expression in the $(\rho,\eta)$ coordinates, and it is only determined within the region in the $(\rho,\eta)$ plane
that corresponds to the allowed region in the $(t,u)$ plane.
 The charge density profile $\lambda_{\text{T}}(\eta)$ is thus similarly confined. 
 
To fix these issues, we replace $V_{\rm T}$ with a new electrostatic potential $V$,
defined throughout the entire $(\rho,\eta)$ plane.
We observe that $V$ and $V_{\rm T}$ need not be identical:
if they differ by a transformation of the form \eqref{eq_eta_shift}, they yield the same
11d metric and flux.

We write the new electrostatic potential $V$ using the standard Green's
function for the Laplace operator in $\mathbb R^3$,
\beq \label{nice_V}
V (\rho,\eta)= - \frac 12 \int_{-\infty}^{+\infty} \frac{\lambda(\eta')}{\sqrt{\rho^2 + (\eta - \eta' )^2}} d\eta' \ .
\eeq
Notice that the charge density profile is now extended to be a function
on the whole $\eta$ axis. The new $\lambda(\eta)$ must necessarily
agree with $\lambda_{\rm T}(\eta)$ computed above for values of $\eta$
within the allowed region. Outside this region, we need an educated guess
for the form of $\lambda(\eta)$.

With the benefit of hindsight, we choose $\lambda(t)$ to be the piecewise linear
function
\beq \label{nice_lambda_caseI}
\lambda(\eta) = \begin{dcases}
-\frac{\lambda_{2}}{\eta_{-2}-\eta_{-1}}\left(\eta+\eta_{-2}\right), & -\eta_{-2} \leq \eta < -\eta_{-1} \ ,\\
\frac{\lambda_1+\lambda_{2}}{\eta_1+\eta_{-1}}\left(\eta - \eta_1 + \frac{\eta_1+\eta_{-1}}{\lambda_1+\lambda_{2}}\lambda_1\right), & -\eta_{-1} \leq \eta < \eta_1 \ ,\\
-\frac{\lambda_1}{\eta_2-\eta_1}\left(\eta-\eta_2\right), & {\eta}_1 \leq \eta < {\eta}_2 \ .
\end{dcases}
\eeq
This charge density profile is depicted in Figure \ref{fig:2Sided_Lambda}.
It agrees with $\lambda_{\rm T}$ in \eqref{charge_CaseI} for $0 \le \eta \le \eta_2$.
It extends $\lambda_{\rm T}$ including 
a second monopole, \emph{i.e.}~a location on the $\eta$ axis where the slope changes.
The extension is governed by the   parameters $\eta_{-2}$, $\eta_{-1}$, $\lambda_2$,
which will be fixed below.

We may now insert $\lambda(\eta)$ from \eqref{nice_lambda_caseI} into \eqref{nice_V} to compute
the improved electrostatic potential $V$. Due to the linear growth
of $\lambda$ as $\eta \rightarrow +\infty$ and $\eta \rightarrow -\infty$,
the $\eta'$ integral in \eqref{nice_V} is divergent.
We treat it by regularization and ``minimal subtraction''.
More explicilty, we write 
\beq \label{eq:Vreg_gen}
V = \lim_{\substack{\eta_L \to \infty\\ \eta_R\to\infty}}
\left[\widehat V(\eta_L,\eta_R)   + \frac{\lambda_2 [\eta_L - (\eta + \eta_{-2}) \log (2\eta_L) ] }{2(\eta_{-2} - \eta_{-1})}
- \frac{\lambda_1 [ \eta_R + (\eta - \eta_2) \log (2\eta_R) ]}{2 (\eta_2 - \eta_1)}
\right] \ .
\eeq
In the previous expression, $\widehat V(\eta_L,\eta_R)$ is the same as
$V$ in \eqref{nice_V}, but with  region of integration $[- \eta_L , \eta_R]$.
We add two ``counterterms'' to $\widehat V(\eta_L,\eta_R)$
to remove all divergences. Notice that the counterterms are
independent of $\rho$ and at most linear in $\eta$, of the same form as the transformation
\eqref{eq_eta_shift}.

\begin{figure}
	\centering
	\includegraphics[width=.8\textwidth]{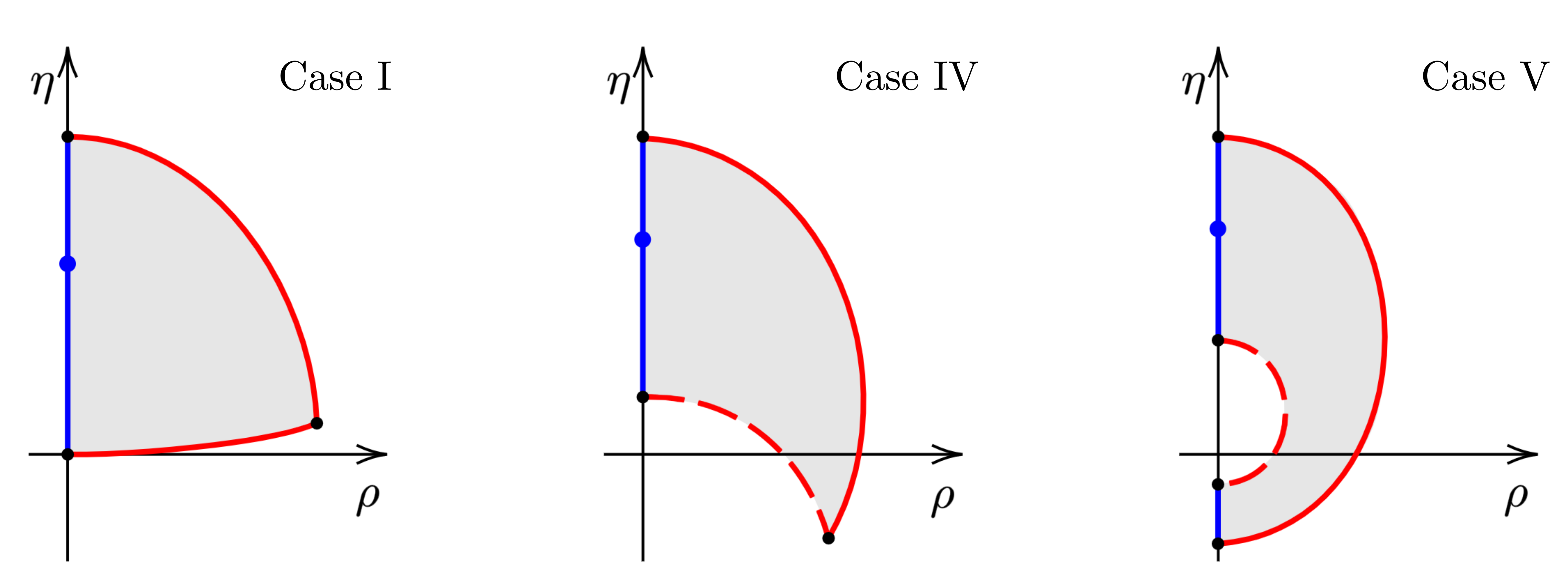}
	\caption{The allowed regions in the $(\rho,\eta)$ plane for Cases I, IV, and V respectively. The solid red arcs signify the loci where $\partial_\rho V = 0$, while the dashed red lines represent the loci where $\partial_\eta^2 V = 0$. 
The similarity with Figures \ref{fig_cases_new}, \ref{fig_non_rect} is intentional, as the solid and dashed lines in each figure correspond to one another. 
As the combination
$t_1 u_2 - u_1 t_2$ goes from positive to negative, we transition 
from Case V to Case III. The red dashed arc of Case V shrinks
and  is replaced by the second monopole present in Case III.
}
	\label{fig:Cases134_re}
\end{figure}

\paragraph{Allowed regions.} Now that we have an expression for $V(\rho,\eta)$ coming from our ansatz for the linear charge density $\lambda(\eta)$, we can look at the regularity conditions coming from the metric and find our allowed regions in the $(\rho,\eta)$ plane. Looking at the 11d line element, we see that we must satisfy
\beq
\rho \geq 0 \ ,\qquad \partial_\rho V \geq 0 \ ,\qquad \partial_\eta^2 V \geq 0.
\eeq
These inequalities determine a region in the $(\rho,\eta)$ plane, which depends on the
unfixed parameters $\eta_{-1}$, $\eta_{-2}$, $\lambda_2$ in the 
extended charge density profile \eqref{nice_lambda_caseI}. On the other hand,
we know what the allowed region in the $(\rho,\eta)$ plane must be,
since it can be deduced from the allowed region in the $(t,u)$ plane
using \eqref{eq:rho_yy}, \eqref{eta_expr}. By comparison, we determine the unfixed parameters to be
\beq\label{eq:eta_ab}
\eta_{-1} =  \frac{\sigma}{2}u_1(t_2-t_1)  \ ,\qquad \eta_{-2} = \frac{\sigma}{2}(t_2u_1-t_1u_2) \ ,\qquad \lambda_2 = -t_1u_1 \ ,
\eeq
The allowed region can be seen in Figure \ref{fig:Cases134_re}.

\subsection{Electrostatic interpretation of Case III and beyond}

\paragraph{Case III.} 

\begin{figure}
	\centering
	\includegraphics[width= 0.9\textwidth]{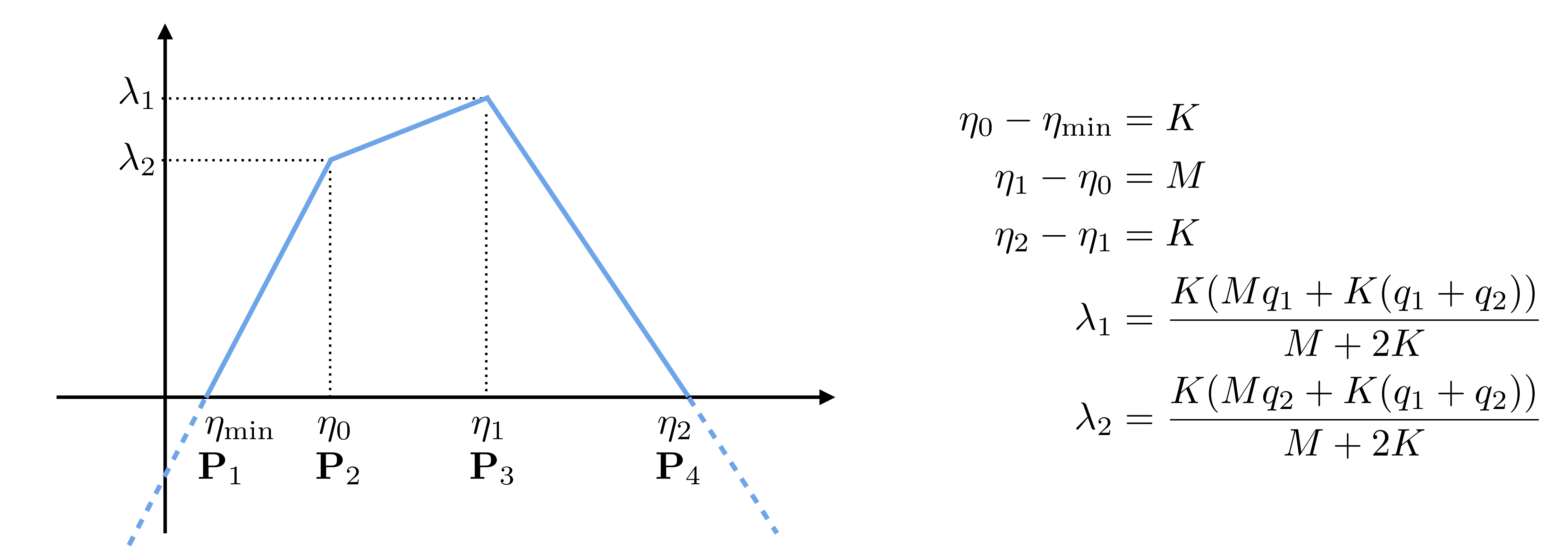}
	\caption{The  extended charge density profile $\lambda(\eta)$ for Case III.
	We notice the absence of ``mirror monopoles'': the 
	charge density outside the interval $[\eta_{\rm min}, \eta_2]$
	is obtained by prolonging the outermost segments  in a trivial way.
	The labels $\mathbf P_1$, \dots, $\mathbf P_4$ refer to Figure \ref{fig_cases_new}.}
	\label{fig:lambdaIII}
\end{figure}

Case III can be studied in an analogous way.
Compared to Case I, 
we also have the edge $\mathbf P_1 \mathbf P_2 $ with charge density
\beq
\lambda_{\text{T}}(t) = u_1 t,
\eeq
with another monopole at $\mathbf P_2$ such that
\beq
 \lambda_2:= \lambda_{\rm T}(\mathbf P_2 ) = t_1u_1 = \frac{K(Mq_2 + K(q_1+q_2))}{M+2K}.
\eeq
The charge density, after fixing $A$ as in \eqref{A_choice}, reads
\beq
\lambda_{\text{T}}(\eta) = \begin{dcases}
\frac{\lambda_2}{\eta_0-\eta_{\text{min}}}\left(\eta -\eta_{\text{min}}\right), & {\eta}_{\text{min}} \leq \eta < {\eta}_0 \ , \\
\frac{\lambda_1}{\eta_1}\eta, & {\eta}_0 \leq \eta < {\eta}_1 \ , \\
-\frac{\lambda_1}{\eta_2-\eta_1}\left(\eta - \eta_2\right), & {\eta}_1 \leq \eta < {\eta}_2 \ ,
\end{dcases}
\eeq
where the $\eta_i$ are given as
\begin{align}
{\eta}_{\text{min}} &= \frac{q_2(M+2K)}{q_1 - q_2} \ , &
 {\eta}_0 &=  \frac{ Mq_2 + K(q_1+q_2)}{q_1 - q_2} \ ,\\
\eta_1 &= \frac{ Mq_1 + K(q_1+q_2))}{q_1 - q_2},& \eta_2 &= \frac{q_1(M+2K)}{q_1 - q_2} \ .
\end{align}
The profile $\lambda_{\rm T}$ is depicted in Figure \ref{fig:lambdaIII}
as a solid line.

As before, we seek an extension of $\lambda_{\rm T}$ to the entire $\eta$ axis,
in such a way as to reproduce the allowed region in the $(\rho,\eta)$ plane
as determined via the B\"acklund transform.
The outcome of this analysis is that, in contrast with Case I, 
the extended charge density 
profile in Case III does not feature any ``mirror monopole'': the profile $\lambda_{\rm T}$
 is extended to the whole $\eta$ axis by simply
extending the outermost linear pieces with constant slopes.

\paragraph{Cases IV and V.} 

These cases can be studied in a similar fashion. Overall, 
we observe that for all Cases I through V the extended charge
density profile has exactly two monopoles.
We refrain from giving explicit expressions for Cases IV and V,
but we depict the allowed regions in the $(\rho,\eta)$ plane
in Figure \ref{fig:Cases134_re}.

%
%
%
%
%
%





\section{More monopoles} \label{sec:gen_caseII}

In the previous section, we have identified the extended charge
density profiles associated to Cases I through V. This allows us to
consider natural generalizations of these solutions, obtained by decorating the
extended charge density profiles with additional monopoles.
The explicit expression for the electrostatic potentials
sourced by these multi-monopole charge densities are
reported in appendix \ref{app_Vs}.

In this section we study in detail generalizations of Case II,
for which the charge density profile is an odd function of $\eta$.
The solutions discussed here will be
identified in section \ref{sec:QFT} as gravity duals of class $\cS$ constructions
with one irregular puncture, and one regular puncture associated to a Young
diagram of arbitrary shape.

%
%

\subsection{Charge density profile}

The sought-for generalizations of Case II solutions 
are all of the form \eqref{Backlund_form_BIS}.
They are specified by a choice of the positive constant $\cC$,
and a choice of electrostatic potential $V$.
The latter can in turn be written in terms 
of the standard Green's function on $\mathbb R^3$
and a charge density profile $\lambda(\eta)$ along the $\eta$ axis, see \eqref{nice_V},
repeated here for convenience,
\beq \label{from_charge_to_V}
V (\rho,\eta)= - \frac 12 \int_{-\infty}^{+\infty} \frac{\lambda(\eta')}{\sqrt{\rho^2 + (\eta - \eta' )^2}} d\eta' \ .
\eeq
Notice that the charge density $\lambda(\eta)$ 
is defined along the entire $\eta$ axis. The actual physical range of 
the coordinate $\eta$ is determined by  
the regularity and positivity of the metric functions.

\begin{figure}
\centering
\includegraphics[width = 10 cm]{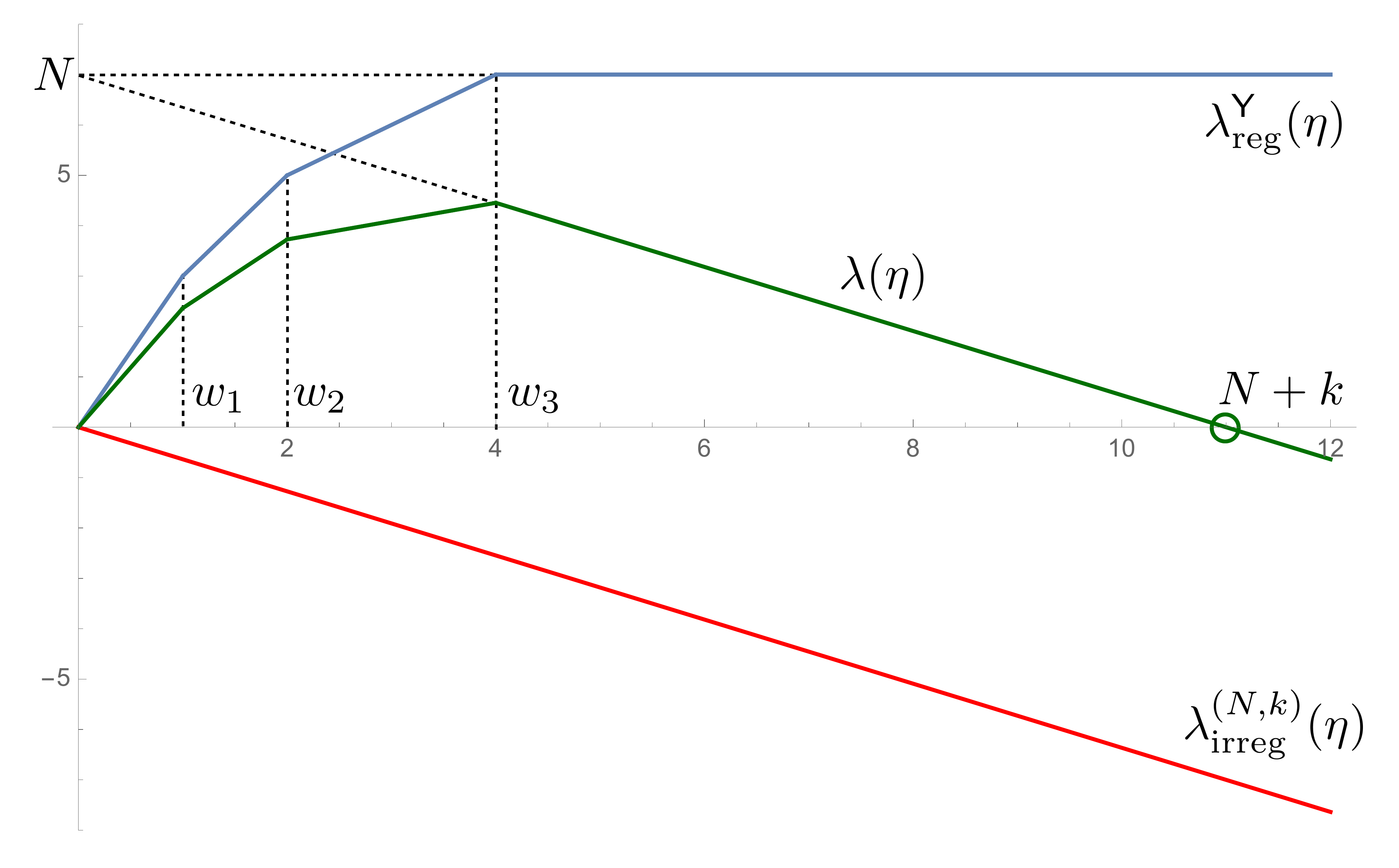}
\caption{
The total charge density $\lambda(\eta)$ can be written
as the sum of the contributions $\lambda_{\rm reg}^{ Y}(\eta)$
and $\lambda_{\rm irreg}^{(N,k)} (\eta)$. 
Both $\lambda_{\rm reg}^{ Y}(\eta)$ and $\lambda_{\rm irreg}^{(N,k)} (\eta)$
are odd functions of $\eta$.
The plot depicts $\lambda(\eta)$,
 $\lambda_{\rm reg}^{ Y}(\eta)$, and $\lambda_{\rm irreg}^{(N,k)} (\eta)$ on the semiaxis $\eta \ge 0$ in the case $N = 7$, $p=3$, $w_1 = 1$, $w_2 = 2$, $w_3 = 4$,
$k_1 = k_2 = k_3 = 1$, $N+k = 11$.
}
\label{fig_sum_plot}
\end{figure}

The total charge density profile $\lambda$ that enters  
 in \eqref{from_charge_to_V} is conveniently 
written as the sum of two contributions, see Figure \ref{fig_sum_plot},
\beq \label{eq_decomposition}
\lambda(\eta) = \lambda_{\rm reg}^{ Y}(\eta) + \lambda_{\rm irreg}^{(N,k)}(\eta)  \ .
\eeq
In the first term, $ Y$ denotes the Young diagram associated to the partition
\beq \label{my_partition}
N = \sum_{a=1}^p k_a w_a \ ,
\eeq
in which $p \ge 1$ is an integer,   $\{ w_a\}_{a=1}^p$ is an increasing sequence
of positive integers, and $k_a$ are positive integers.
The function $\lambda_{\rm reg}^{ Y}(\eta)$
is continuous and piecewise linear.  
It is also assumed to be odd in $\eta$,
\beq
\lambda_{\rm reg}^{ Y}( - \eta) = - \lambda_{\rm reg}^{ Y}(\eta) \ .
\eeq
It is sufficient to specify $ \lambda_{\rm reg}^{ Y}(\eta)$ for $\eta \ge 0$:
it is given as
\beq \label{eq_lambda_reg}
\lambda_{\rm reg}^{ Y}(\eta) = \left\{
\begin{array}{ll}
\displaystyle \eta\,  \sum_{b=1}^p k_b   & \quad \text{for $  0  \le  \eta <   w_1$} \ ,
\\[4mm] 
\displaystyle \eta \, \sum_{b=a+1}^p k_b   + \sum_{b=1}^a w_b k_b  & \quad \text{for $  w_{a}  \le \eta <  w_{a+1}$, \quad $a=1,2,\dots,p-1$} \ ,
\\[4mm] 
\displaystyle N & \quad \text{for $\eta \ge  w_p$} \ .
\end{array}
\right.
\eeq
The profile $\lambda_{\rm reg}^{ Y}(\eta)$
is exactly the same charge density profile that enters the 
$AdS_5$ solutions that describe the local
geometry near a regular puncture
\cite{Gaiotto:2009gz}, hence the label ``reg''.

The second term $\lambda_{\rm irreg}^{(N,k)}(\eta)$ in \eqref{eq_decomposition}
is determined by two integer parameters $N$ and $k$,
where $N$ is the same as in the partition \eqref{my_partition}, and $k$ satisfies
\beq
k > w_p - N \ .
\eeq
The quantity $\lambda_{\rm irreg}^{(N,k)}(\eta)$
 is a simple linear function of $\eta$,
\beq
\lambda_{\rm irreg}^{(N,k)}(\eta) = - \frac{N}{k+N} \, \eta \ .
\eeq
The label ``irreg'' is motivated by the analysis of section \ref{sec:QFT},
which shows that $\lambda_{\rm irreg}^{(N,k)}(\eta)$ is naturally associated
to the irregular puncture in the dual class $\cS$ field theory construction.

Notice that, due to the linear growth of $\lambda_{\rm irreg}^{(N,k)}(\eta)$
as $|\eta| \rightarrow \infty$, the formula \eqref{from_charge_to_V}
for the electrostatic potential $V$ is formally divergent. These divergences 
are treated by regularizing the $\eta'$ integral with a cutoff $\lambda_{\rm max}$, 
and performing a ``minimal subtraction'' of divergent terms.
We refer the reader to appendix \ref{app_Vs} for further details,
and for the explicit expression for $V$.

Finally, the positive constant $\cC$ is also determined by the parameters $N$, $k$,
\beq \label{the_value_of_C}
\cC = \frac{N+k}{N} \ .
\eeq
Having prescribed $\cC$ and $\lambda(\eta)$, we have fully specified the solution.
For $p=1$ we recover precisely the electrostatic description
of the solutions in Case II.

Notice that we have not justified the form \eqref{eq_decomposition} of the charge density
or the value of $\cC$. We refer the reader to appendix \ref{app_caseII} for a detailed analysis
of the problem, which demonstrates that \eqref{eq_decomposition} and \eqref{the_value_of_C} can be inferred
from
 metric regularity and flux quantization.

\subsection{Geometry of the solutions}

\paragraph{Allowed region in the $(\rho,\eta)$ plane.}

Since we have chosen an odd profile for $\lambda(\eta)$, the electrostatic potential
satisfies
\beq
V(\rho, - \eta) = - V(\rho,\eta) \ .
\eeq
In particular $V(\rho,0) = 0$, signaling the presence of a conducting plane
at $\eta = 0$. The radius of the $S^2$ shrinks there. As a result,
we restrict to $\eta \ge 0$.

\begin{figure}
\centering
\includegraphics[width = 7.5 cm]{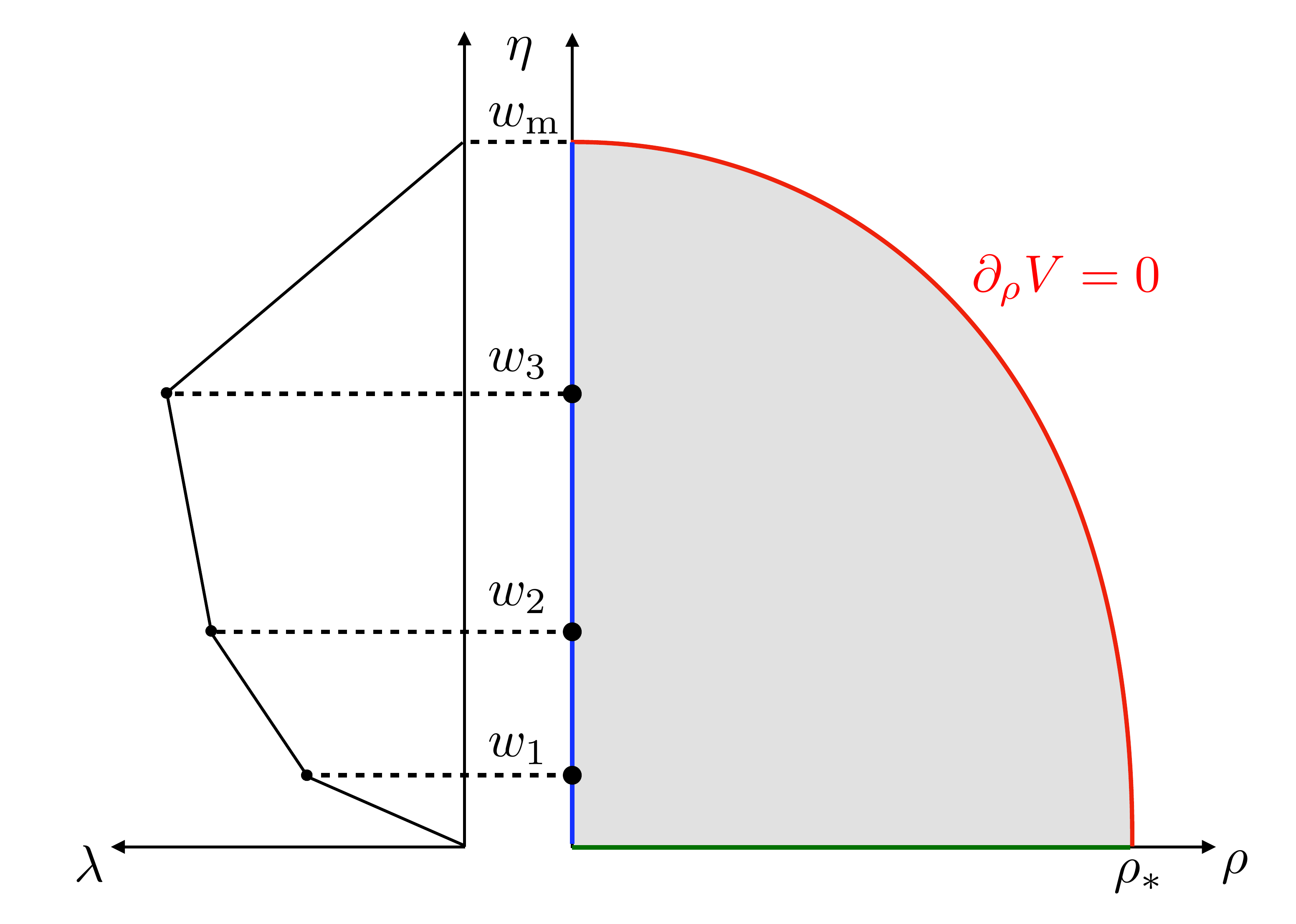}
\caption{
A schematic depiction of the allowed region
in the $(\rho,\eta)$ plane, for a case with $p=3$
monopoles along the $\eta$ axis.
The arc connecting $w_{\rm m}$ and $\rho_*$
corresponds to $\partial_\rho V = 0$ and is defined by the equation \eqref{eq_boundary_arc}.
On the right, we also include the plot of the charge density $\lambda(\eta)$
on the interval $[0,w_{\rm m}]$.
}
\label{fig_more_monopoles}
\end{figure}

Using the explicit expressions of the metric functions that can be obtained
using \eqref{stuff_from_V}, we verify that all positivity requirements are satisfied,
provided that we consider the region in the $(\rho, \eta)$ plane determined by the conditions
\beq \label{eq_allowed_region}
\rho \ge 0 \ , \qquad \eta \ge 0 \ , \qquad \partial_\rho V \ge 0 \ .
\eeq
This domain is depicted in Figure \ref{fig_more_monopoles}.
The  curve
$\partial_\rho V = 0$
 is given more explicitly as
\beq \label{eq_boundary_arc}
- \frac{ 2 \,  \eta}{\cC} + \sum_{a=1}^p k_a \, \Big( 
\sqrt{(\eta + w_a)^2 + \rho^2}
- \sqrt{(\eta - w_a)^2 + \rho^2}
\Big)=  0 \ .
\eeq
One can verify that this curve intersects the $\eta$ axis  at
\beq
w_{\rm m} = N+k \ ,
\eeq
which is precisely the location of the positive zero of the function
$\lambda(\eta)$, see Figure \ref{fig_sum_plot}.
The curve \eqref{eq_boundary_arc} intersects the $\rho$ axis
at a point $\rho_*$, where $\rho_*$ is
the positive solution to the equation
\beq
\sum_{a=1}^p \frac{k_a \, w_a}{\sqrt{\rho_*^2 + w_a^2}} = \frac{1}{\cC} \ .
\eeq
We also observe that, 
in the case $p=1$, the equation \eqref{eq_boundary_arc}
can be equivalently written as
\beq
\frac{\rho^2}{w_1^2 \, (k_1^2 \cC^2 -1)}
+ \frac{\eta^2}{k_1^2 w_1^2 \cC^2} = 1 \ ,
\eeq
which describes an ellipse in the $(\rho,\eta)$ plane.
For $p \ge 2$, this is no longer the case, but
the locus $\partial_\rho V = 0$
has the same qualitative shape as for $p=1$.

\paragraph{Geometry of the internal space.}
The internal space $M_6$ can be regarded as an $S^2 \times S^1_z$
fibration over the 3d space spanned by $\rho$, $\eta$, $\phi$.
The radius of the $\phi$ circle in the 3d base goes to zero smoothly
along the $\eta$ axis.

The size of the $S^2$ depends on $\rho$, $\eta$,
but it is not twisted over the 3d base.
In contrast, the $z$ circle is twisted over the $\phi$ circle,
as prescribed by the metric function $L$ in \eqref{stuff_from_V}.
In particular, $L$ is piecewise constant along the $\eta$ axis,
with jumps at the locations $\eta = w_a$ where the slope
of the charge density $\lambda$ changes.
It follows that the points $(\rho,w_a)$ are monopole sources
for the $Dz$ fibration over $(\rho, \eta, \phi)$.
Notice that the radius of $Dz$ goes to zero at the monopole points.

\paragraph{Behavior near the boundary of the allowed region.}

The boundary of the allowed region
depicted in Figure \ref{fig_more_monopoles}
 consists of several components, which we discuss in turn.

As mentioned above, along the segment $[0,\rho_*]$ on $\rho$ axis the $S^2$ shrinks smoothly, and caps off the
internal space. This is the green horizontal line in Figure \ref{fig_more_monopoles}.

Let us now consider the segment $[w_a, w_{a+1}]$ ($a = 0,1\dots, p-1$) along the $\eta$ axis.
Here
a combination of the $S^1_z$ and $S^1_\phi$ circles shrinks smoothly. 
More precisely, 
the following linear combination
of $\partial_z$, $\partial_\phi$,
\beq
\partial_\phi + \sum_{b=a+1}^p k_b \, \partial_z \ , 
\eeq
has vanishing norm
as we approach the $[w_a, w_{a+1}]$ segment.
By a similar token, 
it is the Killing vector $\partial_\phi$ that has vanishing norm
along the $[w_p, w_{\rm m}]$ segment.
We have thus accounted for the whole vertical blue line
in Figure \ref{fig_more_monopoles}.

Finally, we have to discuss the arc $\partial_\rho V = 0$, depicted in red
in Figure \ref{fig_more_monopoles}. As shown in appendix \ref{app_caseII},
this locus corresponds to an M5-brane source, of total charge $N$,
which is extended along $AdS_5$ and $z$, smeared along $\phi$
and the $\partial_\rho V = 0$ arc in the $(\rho,\eta)$ plane,
and localized in the remaining directions.

\subsection{Holographic central charge}

The general formula \eqref{eq_c_formula} is easily specialized
to solutions of the form \eqref{Backlund_form_BIS}. It reads
\beq
c = \frac{1}{128 \, (\pi m^3 \ell^3_p)^3} \, \int_{\cB_2} \rho \, \dot V \, V'' \, d\rho \, d\eta  \ ,
\eeq
where
we have reinstated the mass scale $m$ and 
 $\cB_2$ denotes the allowed region for the $\rho$, $\eta$ coordinates,
determined by the conditions \eqref{eq_allowed_region}.
To proceed, 
 we   observe that
\beq
\rho \, \dot V \, V'' =  \partial_\rho \bigg[
- \frac 12 \, \rho^2 \, (\partial_\rho V)^2
\bigg] + \rho \, (\rho \, \partial_\rho V) \, \bigg[ 
\partial_\eta^2 V + \frac 1 \rho \, \partial_\rho (\rho \, \partial_\rho V)
\bigg] \ .
\eeq
We argue that the second term can be dropped.
We know that 
the combination $\partial_\eta^2 V + \frac 1 \rho \, \partial_\rho (\rho \, \partial_\rho V)$
is zero, except for terms localized on the $\eta$ axis.
The quantity $\rho \, \partial_\rho V$ is finite
as we approach the $\eta$ axis: 
it is given by the charge density $\lambda(\eta)$.
Because of the extra $\rho$ factor in front,
we conclude that this term drops away in the limit $\rho$ going to zero.

%

The quantity of interest can then be cast as
\beq
\int_{\cB_2} \, \partial_\rho \bigg[
- \frac 12 \, \rho^2 \, (\partial_\rho V)^2
\bigg]  \, d\rho \, d\eta
= \int_{\cB_2} \, d\bigg[  - \frac 12 \, \rho^2 \, (\partial_\rho V)^2 \, d\eta\bigg]
= \int_{\partial \cB_2} \, \bigg[  - \frac 12  \, (\rho \, \partial_\rho V)^2 \, d\eta\bigg] \ .
\eeq
Recall that $\partial \cB_2$ consists of three components:
the segment $[0,\rho_*]$ along the $\rho$ axis,
the arc defined by the equation $\partial_\rho V =0$,
and the segment $[0, w_{\rm m}]$ along the $\eta$ axis.
The 1-form $- \frac 12  \, (\rho \, \partial_\rho V)^2 \, d\eta$
vanishes along the $\rho$ axis (because it only has a leg along $\eta$),
and it also vanishes along the arc where $\partial_\rho V =0$.
It follows that the only non-trivial contribution
originates from the integral over the segment $[0, w_{\rm m}]$
along the $\eta$ axis. Along this segment,
we can make the replacement $\rho \, \partial_\rho V \rightarrow \lambda$.
In conclusion, collecting all factors, the holographic
central charge can be written as
\beq \label{holo_c_from_charge}
c = \frac{1}{256 \, (\pi m^3 \ell_p^3)^3}  \, \int_0^{w_{\rm m}}  \, \lambda(\eta)^2 \, d\eta
=  \frac 14 \, \int_0^{w_{\rm m}}   \lambda(\eta)^2 \, d\eta \ . 
\eeq
In the second step 
we have fixed the value of the mass scale $m$ according to \eqref{eq_fix_m}.

The integral in \eqref{holo_c_from_charge} is readily
evaluated making use of the expression \eqref{eq_decomposition} for the charge density.
Let us introduce the notation
\beq \label{eq_y_and_m}
y_a = \sum_{b=1}^a w_b k_b \ , \qquad m_a = - \frac {N}{N+k}
+ \sum_{b=a+1}^p k_b \ , \qquad a = 0,1,\dots,p \ .
\eeq
For $a=p$,   the sum in the second expression is understood to be zero.
We also use the convention $w_0 :=0$.
We may then write
\beq
c  = \frac 14 \, \sum_{a=0}^p \bigg[ 
\frac 13 \, m_a^2 \, (w_{a+1}^3 - w_a^3)
+ m_a \, y_a \, (w_{a+1}^2 - w_{a}^2)
+ y_a^2 \, (w_{a+1} - w_a)
\bigg] \ .
\eeq
We can also write $c$ 
directly
in terms of 
 $k$ and the partition $N  = \sum_{a=1}^p k_a w_a$,
\begin{align} \label{conjecture}
c  &= \frac{1}{12}  (N+k) \, N^2  + \frac{1}{12}  \frac {N}{N+k} \, \sum_{a=1}^p k_a \, w_a^3
 -   \frac 16  \,  \sum_{a=1}^p  k_a^2 \, w_a^3
  - \frac{1}{12} \sum_{a=1}^p \, \sum_{b=a+1}^p  (w_a^3 + 3 \, w_a \, w_b^2) \, k_a \, k_b \ .
\end{align}

\subsection{'t Hooft anomalies from inflow}

The inflow methods of \cite{Bah:2019rgq} (based on \cite{Freed:1998tg,Harvey:1998bx})
provide a systematic way of extracting 't Hooft anomalies 
from a holographic solution in 11d supergravity.
The key ingredient in this approach is the construction of the 4-form $E_4$,
which is the closed equivariant completion of the background
flux $\overline G_4$ in \eqref{Backlund_form_BIS}. 
We refer the reader to appendix \ref{app_caseII} for a discussion of $E_4$ and the
derivation of the inflow anomaly polynomial.
The results of our analysis are as follows.

\paragraph{Symmetries from isometries.}
Even though
the geometry possesses two $U(1)$ isometries,
generated by $\partial_z$ and $\partial_\phi$,
only one linear combination of these Killing vectors 
yields a massless $U(1)$ gauge field in the 5d low-energy effective theory
that describes 11d supergravity reduced on $M_6$.
The other linear combination of putative massless $U(1)$ gauge fields
gets massive by a St\"uckelberg mechanism:
it ``eats'' an axionic scalar, which comes from the expansion of the M-theory
3-form onto a cohomologically non-trivial closed 3-form in $M_6$.
In terms of the angular variables $\chi$, $\beta$ in the canonical LLM form,
the linear combination of Killing vectors that yields
a massless $U(1)$ gauge field is simply $\partial_\chi$.
This is to be expected, since this isometry corresponds to the superconformal
$U(1)_r$ symmetry.  Using \eqref{angular_vars}, we can also write
\beq
\label{rs}
\partial_\chi = \partial_\phi + \frac 1 \cC \, \partial_z= 
\partial_\phi +\frac {N}{ N+k}  \, \partial_z   \ .
\eeq
To summarize,
the relevant isometries on $M_6$ that correspond
to massless 5d gauge fields are $U(1)_\chi$,
and the $SO(3)$ isometry of the $S^2$.
Let  $c_1(U(1)_\chi)$, $p_1(SO(3))$ denote the first Chern class
 and the first Pontryagin class constructed in terms of these massless gauge fields.
These quantities are identified on the field theory side with 
the Chern classes $c_1^r := c_1(U(1)_r)$ and $c_2^R := c_2(SU(2)_R)$
constructed with the background fields for the $U(1)_r \times SU(2)_R$
R-symmetry. More precisely, we have the relations
\beq \label{background_identifications}
c_1(U(1)_\chi) =  -2 \, c_1^r \ , \qquad p_1(SO(3)) = - 4 \, c_2^R \  . 
\eeq

\paragraph{Orbifold points and flavor symmetries.}
From the expressions recorded in \eqref{Backlund_form_BIS}, \eqref{stuff_from_V},
we verify that the monopole locations $(\rho,\eta) = (0,w_a)$, $a=1,\dots,p$
are orbifold points for the internal space geometry $M_6$,
More precisely,  near the $a$th monopole location 
$(\rho,\eta) = (0,w_a)$, the internal space $M_6$
is locally given as $S^2 \times (\mathbb R^4/\mathbb Z_{k_a})$,
where $k_a \in \mathbb Z_{>0}$ is the charge of the $a$th monopole. 
Following the same logic as in \cite{Gaiotto:2009gz},
we conclude that these orbifold points correspond holographically 
to non-Abelian summands $\mathfrak{su}(k_a)$ in the
global 0-form symmetry algebra.
Accordingly, in our inflow analysis we introduce background gauge fields
for these symmetries. We use $c_2(SU(k_a))$ for the second Chern class
constructed with these background fields.

\paragraph{Inflow anomaly polynomial.}

The anomaly inflow analysis in appendix \ref{app_caseII}
yields the following expression for the leading terms in the inflow anomaly polynomial
at large $N$,
\begin{align} \label{eq_inflow_summary}
- I_6^{\rm inflow} & = \cA_{r,R} \, c_1^r \, c_2^R 
+  \sum_{a=1}^p \,  k_{SU(k_a)} \, c_1^r \, c_2(SU(k_a))  \ .
\end{align}
The 't Hooft anomaly coefficient $\cA_{r,R}$ is given as
\begin{align} \label{A_anom_coeff}
\cA_{r,R} & =  \sum_{a=0}^{p}
\bigg[   \frac 23 \, m_a^2 \, (w_{a+1}^3 -w_a^3 ) + m_a \, y_a \, (w_{a +1} ^2- w_a^2)  \bigg]  \ ,
\end{align}
where $m_a$, $y_a$ are as in \eqref{eq_y_and_m}.
One can verify that 
\beq
\cA_{r,R} = - 4 \, c \ ,
\eeq
where $c$ is given by \eqref{conjecture}. This identification is the expected 
4d $\cN = 2$ SCFT 
relation that holds at large $N$ between the central charges $a=c$ and the mixed $U(1)_r\, SU(2)_R$ anomaly. It provides a consistency check of the
result \eqref{conjecture} for the holographic central charge.

More interestingly, the inflow analysis provides the following
values for the flavor central charges of the $SU(k_a)$ symmetries,
\beq \label{flavor_from_inflow}
k_{SU(k_a)} = 2 \, (y_a + m_a \, w_a) \ ,
\eeq
with $m_a$, $y_a$ defined in  \eqref{eq_y_and_m}.

\subsection{Operators from wrapped M2-branes}

The expression of the calibration 2-form $Y'$
in terms of the electrostatic potential $V$
is  
\begin{align}
&Y'  = 2 \, \bigg[ \frac{V'' \, \dot V}{2 \,  \widetilde \Delta}  \bigg]^{3/2} \,   {\rm vol}_{S^2}
+ \sqrt{ \frac{V''}{2 \, \dot V \, \widetilde \Delta}  } \, \frac{ (\dot V')^2 + \rho^2 \, (V'')^2 }{(\dot V')^2 - \ddot V \, V''} \, \tau \, Dz \, d\eta    \\
& + \sqrt{ \frac{V''}{2 \, \dot V \, \widetilde \Delta}  } \, \frac{ (\Delta V)  \, \dot V' \, \rho }{(\dot V')^2 - \ddot V \, V''} \, \tau \, 
Dz \, d\rho
+ \sqrt{ \frac{V''}{2 \, \dot V \, \widetilde \Delta}  } \, \frac{\dot V \, \dot V '}{\widetilde \Delta }  \, Dz \, d\tau
 -  \sqrt{ \frac{V''}{2 \, \dot V \, \widetilde \Delta}  } \, \frac{\dot V \, \ddot V }{
2 \, \dot V - \ddot V 
} \, d\tau \, d\phi 
\nn \\
& -   \sqrt{ \frac{V''}{2 \, \dot V \, \widetilde \Delta}  } \, 
\frac{ \dot V '  \, (\ddot V \, \widetilde \Delta  + 2 \, \rho^2  \, \dot V \, (V'')^2) }{
(2 \, \dot V - \ddot V) \, ((\dot V')^2 - \ddot V \, V'')
} \, \tau \, d\eta \, d\phi
 -   \sqrt{ \frac{V''}{2 \, \dot V \, \widetilde \Delta}  } \, 
\frac{  
\ddot V ^2 \, \widetilde \Delta + 2 \, \dot V \, (\dot V ')^2 \, V'' \, \rho^2
}{ 
(2 \, \dot V - \ddot V) \, ((\dot V')^2 - \ddot V \, V'') \, \rho
} \, \tau \, d\rho \, d\phi
\ . \nn
\end{align}
In the previous expression, we have used the shorthand notation
\beq
\Delta V = V'' + \frac{1}{\rho^2} \ddot V \ ,
\eeq
and we have used the coordinates $\tau$, $\varphi$ on $S^2$
as in \eqref{S2_coords}.   

\paragraph{M2-branes located at monopoles.}

Let us consider an M2-brane wrapping the $S^2$ and located at one of the monopole
points, $\rho =0$, $\eta = w_a$.
The relevant terms in the calibration 2-form $Y'$ and the 6d line element are
\beq
Y' \supset  2 \, \bigg[ \frac{V'' \, \dot V}{2 \,  \widetilde \Delta}  \bigg]^{3/2} \,   {\rm vol}_{S^2} \ , \qquad
ds^2(M_6) \supset\frac{ V'' \, \dot V}{2\, \widetilde \Delta} \, ds^2_{S^2} \ .
\eeq
In order to verify the calibration condition, we have to evaluate the limit
of the quantity $\frac{V'' \, \dot V}{2 \,  \widetilde \Delta}$ as we approach
the monopole location. This can be done by setting
\beq
\rho = R \, \sqrt{1-t^2}  \ , \qquad
\eta = w_a + R \, t  \ ,
\eeq
and considering the limit $R \rightarrow 0$ at fixed $t$.
With this prescription, we find that
\begin{align} \label{near_monopoles}
\dot V & =  
  \sum_{b=1}^{a-1} k_b \, w_b
  + w_a \, \sum_{b=a}^p k_b 
  - \frac {N}{N+k} \, w_a
   + \cO(R) \  , \nn \\
 \ddot V & = 0 + \cO(R) \  , \quad V''   = \cO(1/R) \ , \nn \\
 \dot V' & = \text{ finite but $t$-dependent quantity} + \cO(R) \ .
\end{align}
Because of the $1/R$ pole in $V''$, near the $a$th monopole we can write
\beq
\frac{V'' \, \dot V}{2 \, \widetilde \Delta}
= \frac{V'' \, \dot V}{2 \, (2 \, \dot V - \ddot V) \, V'' + 2 \, (\dot V ')^2}
=  \frac 14 + \cO(R) \ .
\eeq
Notice how the finite value of $\dot V$ drops from the result.
The calibration condition is then satisfied by virtue of 
$2 \, \left( \tfrac 14 \right)^{3/2} = \tfrac 14$.
Let us denote the BPS operator associated to an  M2-brane located
at $\eta = w_a$ as $\cO_a$.

The dimension of wrapped M2-brane operators  is computed with the   formula 
\eqref{dim_formula}, which in the present context takes the form (temporarily reinstating
the mass scale $m$)
\beq \label{eq_dim_formula}
\Delta 
=    
 \frac{1}{ \pi \, (4 \pi m^3 \ell_p^3)} \, \int_{\cC_2} \bigg[  \frac{\dot V \, \widetilde \Delta}{ 2 \, V''}  \bigg]^{1/2} \, {\rm vol}_{\cC_2} \ .
\eeq
Let us specialize to the operators $\cO_a$ defined above.
We have
\beq
{\rm vol}_{\cC_2} =  \frac{V'' \, \dot V}{2 \, \widetilde \Delta}  \bigg|_{(\rho,\eta) = (0,\eta_a)}\, {\rm vol}_{S^2} \ .
\eeq
The dimension of the operator $\cO_a$ is then given by ($ {\rm vol}_{S^2}$ gives a factor $4\pi$)
\beq
\Delta (\cO_a) =  \frac{1}{(4 \pi m^3 \ell_p^3)} \cdot   4\, \bigg[  \frac{\dot V \, \widetilde \Delta}{ 2 \, V''}  \bigg]^{1/2} \,  \frac{V'' \, \dot V}{2 \, \widetilde \Delta} \bigg|_{(\rho,\eta)=(0,\eta_a)} \ .
\eeq 
We have already observed that 
$\frac{V'' \, \dot V}{2 \, \widetilde \Delta}$ approaches 1/4, 
for all monopoles. Near the $a$th monopole, we also have
\beq
 \frac{\dot V \, \widetilde \Delta}{ 2 \, V''} = 
  \frac{\dot V \,(2 \, \dot V - \ddot V) \, V'' + \dot V \, (\dot V')^2 }{ 2 \, V''} 
  = \dot V^2 + \cO(R) \ .
\eeq
 In conclusion, making use of \eqref{near_monopoles} for the value of $\dot V$, we arrive at
\beq
\Delta(\cO_a) =   \sum_{b=1}^{a-1} k_b \, w_b
  + w_a \, \sum_{b=a}^p k_b
 - \frac {N}{N+k} \, w_a 
   \ .
\eeq

\paragraph{M2-branes stretching along the interval $[w_p, w_{\rm m}]$.}

These M2-branes wrap the 2d submanifold  obtained by considering the final segment  $[w_p, w_{\rm m}]$
combined with the $Dz$ circle. 
Notice that this 2d submanifold is not closed: we are considering open M2-branes,
which end on the smeared M5-brane source.
The relevant terms in $Y'$ are  
\begin{align}
Y' & \supset
 \sqrt{ \frac{V''}{2 \, \dot V \, \widetilde \Delta}  } \, \frac{ (\dot V')^2 + \rho^2 \, (V'')^2 }{(\dot V')^2 - \ddot V \, V''} \, \tau \, Dz \, d\eta  \ .
\end{align}
This quantity has to be evaluated in the limit $\rho \rightarrow 0$,
for a generic $\eta$ with $w_p < \eta < w_{\rm m}$. 
In this limit, $\dot V$ is finite (it is given by the charge density),
 $\dot V' \rightarrow - N/(N+k)$  (independent of  $w_p < \eta < w_{\rm m}$),
$\ddot V$ goes to zero, $V''$ is finite.
It follows that
\beq
\frac{ (\dot V')^2 + \rho^2 \, (V'')^2 }{(\dot V')^2 - \ddot V \, V''}  \rightarrow 1 \ .
\eeq
The relevant parts of the $M_6$ line element are
\beq
ds^2(M_6) \supset \frac{V''}{2 \, \dot V} \,  d\eta^2 + \frac{2 \dot V - \ddot V}{ 2 \, \dot V \, \widetilde \Delta} \, Dz^2 \ ,
\eeq
which implies
\beq
{\rm vol}_{\cB_2} =  \frac{1}{2 \, \dot V}  \, \sqrt{   \frac{V'' \, (2 \dot V - \ddot V)}{\widetilde \Delta}   }  \, d\eta \, Dz 
=     \sqrt{  \frac{  V''    }{2 \, \dot V \, \widetilde \Delta}   }  \, d\eta \, Dz\ ,
\eeq
where in the second step we used the fact that $\ddot V$ vanishes
as $\rho$ goes to zero.
Comparing with $Y'$, we see that the calibration condition is satisfied
for $\tau = -1$.
We denote the BPS operators associated to these wrapped M2-branes
as $\cO_{[w_p, w_{\rm m}]}$.

The dimension of these operators is computed from \eqref{eq_dim_formula},
The result reads
\beq
\label{higgs1}
\Delta (\cO_{[w_p, w_{\rm m}]})= \frac{1}{ \pi \, (4 \pi m^3 \ell_p^3)} \, \int_{\cB_2} \frac 12 \, d\eta \, Dz  
=  \frac{1}{(4 \pi m^3 \ell_p^3)} \, \int_{[w_p, w_{\rm m}]}  d\eta 
=  N+k - w_p  \ .
\eeq

%

\paragraph{M2-branes stretching along the intervals $[w_a, w_{a+1}]$.}

The discussion above generalizes directly to M2-branes
that wrap an interval $[w_a, w_{a+1}]$ combined with the $Dz$ circle on that interval.
In this case, this 2d submanifold is closed.
The calibration condition holds for the same reason as in the $[w_p, w_{\rm m}]$ case:
$\ddot V$ goes to zero, with finite $\dot V$, $\dot V'$, $V''$.
The dimensions of these operators are again computed from \eqref{eq_dim_formula}, with the result
\beq
\label{higgs2}
\Delta (\cO_{[w_a , w_{a+1}]}) = w_{a+1} - w_a \ .
\eeq







\section{Comparison with field theory} \label{sec:QFT}

In this section, we present evidence that the supergravity solutions presented in section \ref{sec:gen_caseII} are dual to the four-dimensional $\CN=2$ Argyres-Douglas SCFTs that arise from wrapping $N$ M5-branes on a sphere with one irregular puncture labeled by $k$, and one regular puncture labeled by a partition of $N$. 
In particular, we identify Case II with the generalized monopole profiles constructed in section \ref{sec:gen_caseII} with the Argyres-Douglas SCFTs labeled by an irregular puncture $A_{N-1}^{(N)}[k]$, for integer $k > -N$, and one regular puncture labeled by a partition of $N$. This proposal is checked in detail in the remainder of this section. It also matches the proposal of \cite{Couzens:2022yjl}. 

It is natural to propose that Cases I and III also correspond to SCFTs of this type, with an irregular puncture with possibly more finely-grained structure, and one regular puncture labeled by a partition of $N$. Indeed, the three solutions  share key features, as we briefly summarize below. In order to simplify this discussion, let us temporarily restrict the regular puncture to be labeled by a single monopole of charge $q$. By inspection of Table \ref{tab:sum}, we then observe the following:

\begin{itemize}

\item In all cases, an integer $N_{\text{eff}}$ may be identified which plays the role of an effective number of M5-branes; and an integer $\widetilde{M}$ may be identified which plays a similar role to the integer $M$ in the Case II solution. 

\item In terms of $N_{\text{eff}}$, $\widetilde{M}$, and the other fluxes, we observe that the R-symmetry twist, dimension of the Coulomb branch operator $\CO_1$, and dimensions of the Higgs branch operators $\CO_2^i$ are identical between the three cases. 

\item Case I and Case II share more features. Case I represents a one-parameter generalization of the Case II solution labeled by the flux quantum $N_\mathrm S$, and which reproduces the Case II geometry when $N_\mathrm S\to 0$. In terms of the effective number of branes $N_{\text{eff}}$, the holographic central charge is identical in these two cases. The difference lies in the additional stack of smeared M5-branes labeled by $N_\mathrm S$, and additional associated Higgs branch operators $\CO_3^j$. At the end of this section, we speculate on a possible field theory interpretation of these features. 

\item By contrast, Case III is evidently {\it not} continuously connected to the other two by tuning flux quanta. At present we refrain from further speculation as to the specific field theory dual of Case III, but point out some intriguing features.  
The central charge differs from that of Cases I and II by a term proportional to the new monopole charge $q_2$, and there are new operators in the spectrum. Furthermore, the two monopole charges $q_{1,2}$ are mixed by the relation $N_{\text{eff}}  = q_1 \widetilde{M} + q_2 K$. It would be interesting to understand the mapping of these features to the field theory dual.

\end{itemize}

We now return to the proposed Case II duality, beginning with a summary of the partition of $N$ labeling the regular puncture is mapped to the electrostatic charge profile.

\subsection{Map to the Young tableaux} 
\label{sec:map}

As discussed in section \ref{sec:gen_caseII}, the geometry of the blue sides of the square depicted in Figure \ref{fig_cases_new} is specified by a charge density profile $\lambda(\eta)$. We now summarize the relationship between the profile $\lambda(\eta)$ and the data of a Young tableaux specifying a regular puncture in this region of the geometry. Our notation for this mapping closely follows the discussion in \cite{Bah:2019jts}.    

There are $p$ monopoles located on the $\eta$ axis at locations $\eta_{a=1,\dots,p}$, with positive integer charges $k_a$ given by 
\ba{
k_a = \ell_a-\ell_{a+1}\ ,\qquad \ell_a = \sum_{b=a}^p k_b\ ,\qquad \ell_{p+1}  = 0\ .\label{kl}
}  
The configuration is labeled by a partition of $N$,
\ba{
N = \sum_{a=1}^p k_a w_a\ ,
} 
where $w_a$ are an increasing series of positive integers. (We remind the reader that $k$ without a subscript labels the irregular puncture, and is not to be confused with the monopole charges $k_a$!) 
This data is related to the data of a Young tableaux as follows. Rewriting the partition of $N$ as
\ba{
N = \sum_{a=1}^p (w_a-w_{a-1}) \ell_a\ ,\qquad w_0=0\ ,
}
we identify a corresponding Young tableaux with distinct row lengths $\ell_a$ with multiplicities $(w_a-w_{a-1})$. Changing variables, the lengths of all rows {\it including} repetitions are given by the quantities 
\ba{
\tilde{\ell}_i=\ell_a\quad  \text{for all}\  i=w_{a-1}+1,\dots,w_a\ ,
}
with $\tilde{p}=w_p$ the total number of rows. This reformulates the partition as $N=\sum_{i=1}^{\tilde{p}} \tilde{\ell}_i$. 
We furthermore define the quantities $\tilde{k}_i = \tilde{\ell}_i - \tilde{\ell}_{i+1}$ (equivalently, $\tilde{\ell}_i  = \sum_{j=i}^{\tilde{p}} \tilde{k}_j$) such that
\ba{
\tilde{k}_i = k_a\quad \text{if} \ i=w_a\ ,\quad \text{otherwise}\quad \tilde{k}_{i\neq w_a}=0\ .
}
The flavor symmetry of the associated regular puncture is given in terms of the monopole charges as
\ba{
G_F = S\left[ \prod_{a=1}^p U(k_a) \right] = S \Bigg[   \prod_{{i =1 } }^{\tilde{p}} U(\tilde{k}_i) \Bigg] \ ,
}
where the product over $i$ is understood to not include the cases $\tilde{k}_i=0$. 

Other useful quantities related to the Young tableaux are as follows. We introduce the notation 
\ba{
N_a = \sum_{b=1}^a (w_b-w_{b-1} )\ell_b = \sum_{b=1}^{a-1} w_b k_b + w_a \ell_a\ ,\quad N_p=N\ . \label{na}
}
The corresponding variables $\tilde{N}_i$ in terms of the Young tableaux data are
\ba{
\tilde{N}_i = \sum_{j=1}^i \tilde{\ell}_j = \sum_{j=1}^{i-1} j \tilde{k}_j + i \sum_{j=i}^{\tilde{p}} \tilde{k}_j\ ,\quad \tilde{N}_{\tilde{p}}=\tilde{N}_{\tilde{p}+1} = N\ ,
}
which satisfy $\tilde{\ell}_i = \tilde{N}_i - \tilde{N}_{i-1}$. We also will utilize the pole structure, a set of $N$ integers $p_i=i-(\text{height of }i\text{'th box})$ labeling the $i$'th box in the diagram, starting with $i=1$ and $p_1=0$ in the bottom left corner and increasing from left to right across a row (in a convention in which row lengths decrease from bottom to top). The $p_i$ are related to the $\tilde{N}_i$ by
	\ba{
	\sum_{i=1}^N (2i-1) p_i = \frac{1}{6} \left( 4 N^3 - 3 N^2 - N\right) - \sum_{i=1}^{\tilde{p}} (N^2-\tilde{N}_i^2)\ . \label{idd}
	}
For reference, these data for some special cases are as follows:

\begin{description}

\item[Rectangular box puncture,] with $N/\ell$ rows of length $\ell$, and flavor symmetry $G_F=SU(\ell)$. The case $\ell=N$ corresponds to the maximal puncture, while $\ell=1$ corresponds to the ``non-puncture''.
\ba{\bs{
\text{tableaux data:}&\quad \tilde{p}= N/\ell\ ;\quad \tilde{\ell}_{1,\dots,\tilde{p}} = \ell\ ; \quad \tilde{k}_{\tilde{p}}=\ell,\ \tilde{k}_{1,\dots,\tilde{p}-1}=0\ ;\quad \tilde{N}_i = i \ell  \\
\text{geometric data:}&\quad p=1\ ;\quad \ell_1=\ell\ ;\quad k_1=\ell;\quad N_1=N\ ; \quad w_1=N/\ell \\
\text{pole structure:}&\quad p_{i=1+(m-1) \ell,\dots, m\ell} = i- m\  ,\quad m=1 ,\dots,N/\ell 
}\label{box}}

\item[Minimal puncture,] with one row of length 2 and $N-2$ rows of length 1, and flavor symmetry $G_F=U(1)$. 
\ba{\bs{
\text{tableaux data:}&\quad \tilde{p}=N-1\ ;\quad \tilde{\ell}_1=2 , \ \tilde{\ell}_{2,\dots,N-1} = 1\ ;\quad \tilde{k}_1=\tilde{k}_{N-1}=1,\ \\
&\quad \tilde{k}_{2,\dots,N-2}=0\ ;\quad \tilde{N}_i=i+1\\
\text{geometric data:}&\quad p=2\ ;\quad \ell_1=2 ,\ \ell_2=1\ ;\quad k_1=k_2=1\ ;\\
&\quad N_1=2 ,\  N_2=N\ ;\quad w_1=1,\ w_2=N-1 \\
\text{pole structure:}&\quad p_1=0,\ p_{2,\dots,N}=1
}\label{min}}
\end{description}
These are drawn for the case $N=4$ in Figure \ref{fig:reg}.

\afterpage{\clearpage}

\tikzset{
    square/.style={%
        draw=none,
        circle,
        append after command={%
            \pgfextra \draw[black] (\tikzlastnode.north-|\tikzlastnode.west) rectangle 
                (\tikzlastnode.south-|\tikzlastnode.east);\endpgfextra}
    }
}

\begin{figure}[h]
\centering
 \begin{subfigure}[b]{0.95\textwidth}
 \centering
 \resizebox{0.68\textwidth}{!}{
 \begin{tikzpicture}
 \begin{scope}[shift={(-3,-0.5)}]
 \node [square,scale=1.2] at (0,0) (s1) { };
  \node [square,scale=1.2] at (0.45,0) (s2) { };
  \node [square,scale=1.2] at (0.9,0) (s3) { };
  \node [square,scale=1.2] at (1.35,0) (s4) { };
 \end{scope}
    \begin{scope}[shift={(0,0)}]
    \draw[very thin,color=gray,step=0.5] (-0.5,-2) grid (2.5,0.5);
   \draw[->] (0,-2) -- (0,0.5);
   \draw[->] (-0.5,0) -- (2.5,0);
   
   \filldraw[color=blue] (0,0) circle (0.05);
   \filldraw[color=blue] (0.5,0) circle (0.05);
   \filldraw[color=blue] (1,0) circle (0.05);
   \filldraw[color=black] (1.5,0) circle (0.05);
   \filldraw[color=black] (2,0) circle (0.05);
   
   \filldraw[color=blue] (0,-0.5) circle (0.05);
   \filldraw[color=blue] (0.5,-0.5) circle (0.05);
   \filldraw[color=blue] (1,-0.5) circle (0.05);
   \filldraw[color=blue] (0,-1) circle (0.05);
   \filldraw[color=blue] (0.5,-1) circle (0.05);
   \filldraw[color=blue] (0,-1.5) circle (0.05);
   
         \draw[color=red] (0,-1.5) circle (0.15);
   
   \draw[densely dotted] (0,-1.5) -- (2,0);
   \end{scope}
   \begin{scope}[shift={(4,-2)}]

   \draw[->] (0,0) -- (0,2.5);
   \draw[->] (0,0) -- (3,0);
   
   \node at (3.2,0) {$\eta$};
   \node at (-0.2,2.5) {$\lambda$};
   
   \draw[black] (0.5,-0.1) -- (0.5,0.1);
   \node at (0.5,-0.3) {$1$};
   
   \draw[black] (-0.1,2) -- (0.1,2);
   \node at (-0.3,2) {$N$};
   
   \draw[black] (0,0) -- (0.5,2);
   \draw[black] (0.5,2) -- (3,2);
   \end{scope}
\end{tikzpicture}
}
 \caption{The maximal puncture with $G_F=SU(N)$, corresponding to the box diagram $Y_{\ell}$ with $\ell=N$. Here, $p_\ell = \ell-1$, such that $p_\ell = \{0,1,2,3\}$ in the above diagram (allowing $\ell = 1,\dots, N$).}
 \end{subfigure}
 %
 %
\begin{subfigure}[b]{0.95\textwidth}
\centering
 \resizebox{0.68\textwidth}{!}{
  \begin{tikzpicture}
   \begin{scope}[shift={(-3,-1)}]
 \node [square,scale=1.2] at (0,0) (s1) { };
  \node [square,scale=1.2] at (0.45,0) (s2) { };
    \node [square,scale=1.2] at (0,0.45) (s3) { };
  \node [square,scale=1.2] at (0.45,0.45) (s4) { };
 \end{scope}
     \begin{scope}[shift={(0,0)}]
    \draw[very thin,color=gray,step=0.5] (-0.5,-2) grid (2.5,0.5);
   \draw[->] (0,-2) -- (0,0.5);
   \draw[->] (-0.5,0) -- (2.5,0);
   
   \filldraw[color=blue] (0,0) circle (0.05);
   \filldraw[color=blue] (0.5,0) circle (0.05);
   \filldraw[color=blue] (1,0) circle (0.05);
   \filldraw[color=black] (1.5,0) circle (0.05);
   \filldraw[color=black] (2,0) circle (0.05);
   
      \filldraw[color=blue] (0,-0.5) circle (0.05);
   \filldraw[color=blue] (0.5,-0.5) circle (0.05);
   \filldraw[color=blue] (1,-0.5) circle (0.05);
   \filldraw[color=blue] (0,-1) circle (0.05);
   
         \draw[color=red] (0,-1) circle (0.15);
   
      \draw[densely dotted] (0,-1) -- (2,0);
     \end{scope}
        \begin{scope}[shift={(4,-2)}]
  
   \draw[->] (0,0) -- (0,2.5);
   \draw[->] (0,0) -- (3,0);
   
   \node at (3.2,0) {$\eta$};
   \node at (-0.2,2.5) {$\lambda$};
   
   \draw[black] (1,-0.1) -- (1,0.1);
   \node at (1,-0.3) {$2$};
   
   \draw[black] (-0.1,2) -- (0.1,2);
   \node at (-0.3,2) {$N$};
   
   \draw[black] (0,0) -- (1,2);
   \draw[black] (1,2) -- (3,2);
   \end{scope}
\end{tikzpicture}
}
  \caption{The puncture with $G_F=SU(N/2)$, corresponding to the box diagram $Y_{\ell}$ with $\ell = N/2$. Here, $p_{\ell = 1,\dots, N/2} = \ell -1$ and $p_{\ell = N/2+1,\dots,N} = \ell -2$, such that in the diagram shown $p_\ell = \{0,1,1,2\}$.}
  \end{subfigure}
\begin{subfigure}[b]{0.95\textwidth}
\centering
 \resizebox{0.68\textwidth}{!}{
  \begin{tikzpicture}
    \begin{scope}[shift={(-3,-1)}]
 \node [square,scale=1.2] at (0,0) (s1) { };
  \node [square,scale=1.2] at (0.45,0) (s2) { };
   \node [square,scale=1.2] at (0.9,0.0) (s3) { };
    \node [square,scale=1.2] at (0,0.45) (s4) { };
 \end{scope}
     \begin{scope}[shift={(0,0)}]
    \draw[very thin,color=gray,step=0.5] (-0.5,-2) grid (2.5,0.5);
   \draw[->] (0,-2) -- (0,0.5);
   \draw[->] (-0.5,0) -- (2.5,0);
   
   \filldraw[color=blue] (0,0) circle (0.05);
   \filldraw[color=blue] (0.5,0) circle (0.05);
   \filldraw[color=blue] (1,0) circle (0.05);
   \filldraw[color=black] (1.5,0) circle (0.05);
   \filldraw[color=black] (2,0) circle (0.05);
   
   \filldraw[color=blue] (0,-0.5) circle (0.05);
   \filldraw[color=blue] (0.5,-0.5) circle (0.05);
   \filldraw[color=blue] (1,-0.5) circle (0.05);
   \filldraw[color=blue] (0,-1) circle (0.05);
   \filldraw[color=blue] (0.5,-1) circle (0.05);
   
      \draw[color=red] (0,-1) circle (0.15);
     \draw[color=red] (0.5,-1) circle (0.15);
   
     \draw[densely dotted] (0.5,-1) -- (2,0);
     \draw[densely dotted] (0.5,-1) -- (0,-1);
	\end{scope}
   \begin{scope}[shift={(4,-2)}]
   
   \draw[->] (0,0) -- (0,2.5);
   \draw[->] (0,0) -- (3,0);
   
   \node at (3.2,0) {$\eta$};
   \node at (-0.2,2.5) {$\lambda$};
   
   \draw[black] (1,-0.1) -- (1,0.1);
   \node at (1,-0.3) {$2$};
   
     \draw[black] (0.5,-0.1) -- (0.5,0.1);
   \node at (0.5,-0.3) {$1$};
   
   \draw[black] (-0.1,2) -- (0.1,2);
   \node at (-0.3,2) {$4$};
   
    \draw[black] (-0.1,1.5) -- (0.1,1.5);
   \node at (-0.3,1.5) {$3$};
   
   \draw[black] (0,0) -- (0.5,1.5);
    \draw[black] (0.5,1.5) -- (1,2);
   \draw[black] (1,2) -- (3,2);
   \end{scope}
\end{tikzpicture}
}
   \caption{The puncture with $G_F=S(U(2)\times U(1))$. Here, $p_\ell = \{0,1,2,2\}$.}
   \end{subfigure}
 %
 %
\begin{subfigure}[b]{0.95\textwidth}
\centering
 \resizebox{0.68\textwidth}{!}{
  \begin{tikzpicture}
      \begin{scope}[shift={(-3,-1.5)}]
 \node [square,scale=1.2] at (0,0) (s1) { };
  \node [square,scale=1.2] at (0.45,0) (s2) { };
   \node [square,scale=1.2] at (0,0.45) (s3) { };
    \node [square,scale=1.2] at (0,0.9) (s4) { };
 \end{scope}
    \begin{scope}[shift={(0,0)}]
    \draw[very thin,color=gray,step=0.5] (-0.5,-2) grid (2.5,0.5);
   \draw[->] (0,-2) -- (0,0.5);
   \draw[->] (-0.5,0) -- (2.5,0);
   
   \filldraw[color=blue] (0,0) circle (0.05);
   \filldraw[color=blue] (0.5,0) circle (0.05);
   \filldraw[color=blue] (1,0) circle (0.05);
   \filldraw[color=black] (1.5,0) circle (0.05);
   \filldraw[color=black] (2,0) circle (0.05);
   
         \filldraw[color=blue] (0,-0.5) circle (0.05);
   \filldraw[color=blue] (0.5,-0.5) circle (0.05);
   \filldraw[color=blue] (1,-0.5) circle (0.05);
   
   \draw[color=red] (1,-0.5) circle (0.15);
   \draw[color=red] (0,-0.5) circle (0.15);
   
        \draw[densely dotted] (0,-0.5) -- (1,-0.5);
         \draw[densely dotted] (1,-0.5) -- (2,0);
\end{scope}
   \begin{scope}[shift={(4,-2)}]
   \draw[->] (0,0) -- (0,2.5);
   \draw[->] (0,0) -- (3,0);
   
   \node at (3.2,0) {$\eta$};
   \node at (-0.2,2.5) {$\lambda$};
   
   \draw[black] (1.5,-0.1) -- (1.5,0.1);
   \node at (1.5,-0.3) {$N-1$};
   
     \draw[black] (0.5,-0.1) -- (0.5,0.1);
   \node at (0.5,-0.3) {$1$};
   
   \draw[black] (-0.1,2) -- (0.1,2);
   \node at (-0.3,2) {$N$};
   
    \draw[black] (-0.1,1) -- (0.1,1);
   \node at (-0.3,1) {$2$};
   
   \draw[black] (0,0) -- (0.5,1);
    \draw[black] (0.5,1) -- (1.5,2);
   \draw[black] (1.5,2) -- (3,2);
   \end{scope}
\end{tikzpicture}
}
    \caption{The minimal puncture with $G_F=U(1)$, correpsonding to the tableaux with one row of length 2 and $N-2$ rows of length 1. Here, $p_\ell = \{0,1,\dots,1\}$.}
    \end{subfigure}
 %
\begin{subfigure}[b]{0.95\textwidth}
\centering
\resizebox{0.68\textwidth}{!}{
  \begin{tikzpicture}
       \begin{scope}[shift={(-3,-1.7)}]
 \node [square,scale=1.2] at (0,0) (s1) { };
  \node [square,scale=1.2] at (0,0.45) (s2) { };
   \node [square,scale=1.2] at (0,0.9) (s3) { };
    \node [square,scale=1.2] at (0,1.35) (s4) { };
 \end{scope}
    \begin{scope}[shift={(0,0)}]
    \draw[very thin,color=gray,step=0.5] (-0.5,-2) grid (2.5,0.5);
   \draw[->] (0,-2) -- (0,0.5);
   \draw[->] (-0.5,0) -- (2.5,0);
   
   \filldraw[color=blue] (0,0) circle (0.05);
   \filldraw[color=blue] (0.5,0) circle (0.05);
   \filldraw[color=blue] (1,0) circle (0.05);
   \filldraw[color=black] (1.5,0) circle (0.05);
   \filldraw[color=black] (2,0) circle (0.05);
   \end{scope}
  \begin{scope}[shift={(4,-2)}]

   \draw[->] (0,0) -- (0,2.5);
   \draw[->] (0,0) -- (3,0);
   
   \node at (3.2,0) {$\eta$};
   \node at (-0.2,2.5) {$\lambda$};
   
   \draw[black] (2,-0.1) -- (2,0.1);
   \node at (2,-0.3) {$N$};
   
   \draw[black] (-0.1,2) -- (0.1,2);
   \node at (-0.3,2) {$N$};
   
   \draw[black] (0,0) -- (2,2);
   \draw[black] (2,2) -- (3,2);
   \end{scope}
   
\end{tikzpicture}
}
     \caption{The non-puncture, corresponding to the box diagram $Y_{\ell}$ with $\ell=1$. Here, $p_\ell = 0$ for all $\ell$.}
     \end{subfigure}
     \caption{The Young tableaux, Newton polygons, and charge density profiles $\lambda_{\text{reg}}$ of the possible regular punctures for $N=4$. The blue dots represent deformation parameters of the Seiberg-Witten curve. The red circled dots are the Coulomb branch operators $\CO_a$ identified in \eqref{dfinal}. \label{fig:reg}}
 \end{figure}
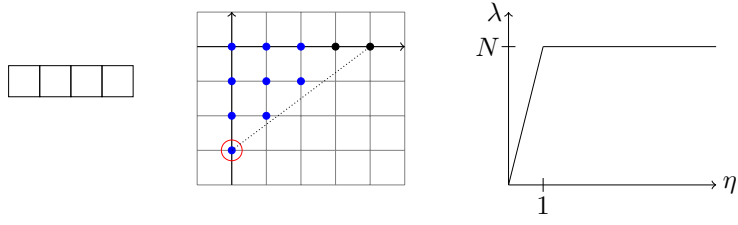
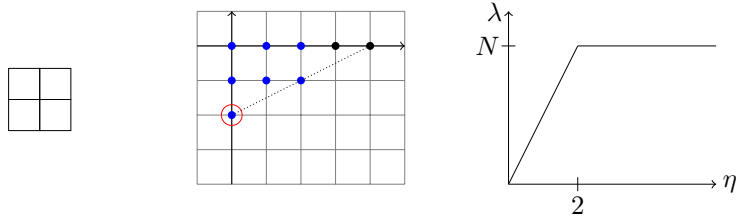
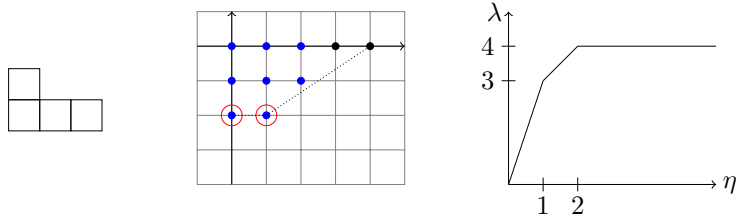
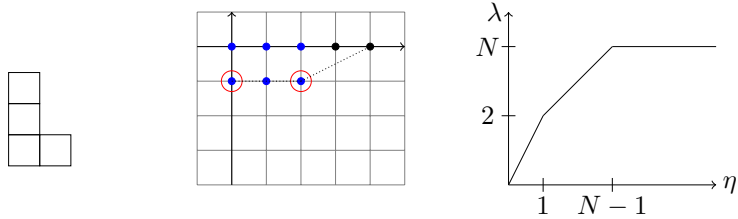
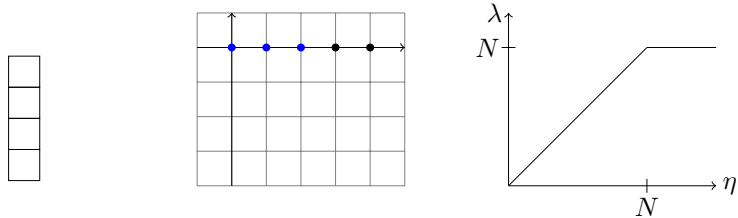

\subsection{Checks of the holographic duality} 

We now review some properties of the field theories $(A_{N-1}^{(N)}[k],Y)$ which participate in the proposed duality, and match these properties with those of the supergravity solutions presented above. The $(A_{N-1}^{(N)}[k],Y)$ field theories are 4d $\CN=2$ SCFTs of Argyres-Douglas type, engineered by wrapping $N$ M5-branes on a sphere with one irregular puncture labeled by the integer $k$ denoted $A_{N-1}^{(N)}[k]$ \cite{Xie:2012hs,Wang:2015mra}, and one regular puncture labeled by the Young tableaux $Y$ which is a partition of $N$. Since \cite{Bah:2021hei} includes a review of the classification of irregular singularities, as well as a detailed review of the properties of the $(A_{N-1}^{(N)}[k],Y)$ SCFTs when $Y$ is a Young tableaux of rectangular box type (including the trivial case with no regular puncture on the sphere), in this section we focus primarily on how the data of the general Young tableaux enters the checks between properties of the SCFTs and the holographic solutions\footnote{We note that most of the checks performed in this section are not sensitive to differences between the irregular singularities $A_{N-1}^{(b)}[k]$ with $b=N$ (Type I) versus $b=N-1$ (Type II), although we explicitly perform checks for the former case. In particular, the central charges, R-symmetry twist, maximum Coulomb-branch operator dimension, and rank of the flavor symmetry are all unchanged at leading order in $N$ between the Type I and Type II theories, differing only at subleading order. The Case II identification of the holographic parameter $K$ with the field theory parameter $k$ would have an additional order 1 term in matching to the Type II singularity, but this difference does not affect the leading order duality checks. One difference is that we do not have access to a check of Higgs branch operator dimensions for the Type II singularities, since these cases do not have a known quiver Lagrangian description.}.

\subsubsection{R-symmetry}

The $U(1)_r$ superconformal R-symmetry of these $\CN=2$ SCFTs is independent of the details of the regular puncture. It is given by a linear combination of the global $U(1)_z$ isometry of the sphere and the would-be $U(1)_\phi$ R-symmetry that would be preserved in the absence of an irregular defect, as
\ba{
r = R_{\phi} + \frac{N}{k+N} R_z\ .
}
This matches the R-symmetry identified in \eqref{rs}.

\subsubsection{Central charge}

The central charge of the $(A_{N-1}^{(N)}[k],Y)$ theories may be decomposed as the sum of the central charge $c_{(A_{N-1},A_{k-1})}$ for the SCFTs without the regular puncture, plus the additional contribution $\Delta c$,\footnote{Since we are interested in checks at large $N$ where $a=c$, here we focus only on the $c$ central charge.}
\ba{
c = c_{(A_{N-1},A_{k-1})} + \Delta c\ . \label{ccft}
}
The quantity $c_{(A_{N-1},A_{k-1})}$ is given by \cite{Xie:2012hs}
\ba{
 c_{(A_{N-1},A_{k-1})}  &= \frac{k^2 (N^2-1) - (k+N) (N - 2 + \text{GCD}(k,N) )}{12 (k+N)} \\
 &\overset{N\to \infty}{=} \frac{N^2 (M-N)^2}{12 M}\ ,\quad M = k+N \label{clargen}
}
where for later reference we also evaluated the large $N$ limit of large $N, k$ at fixed $k/N$, and are using $M=k+N$. 
$\Delta c$ depends on the details of the Young tableaux $Y$, and can be derived by partial closure of the maximal puncture to an arbitrary regular puncture \cite{Giacomelli:2020ryy},  
\ba{
\Delta c &=   \frac{1}{12} \left(n_h(Y) + 2 n_v(Y) +2 N^3 - N - 1+ \frac{N}{(k+N)}  \left(  6 I_{\rho_Y} -N^3+N \right)  \right) \ .
\label{dc}
}
This is given in terms of the quantities $n_v(Y)$ and $n_h(Y)$ which represent the effective number of vector multiplets and hypermultiplets respectively contributed by the regular puncture labeled by $Y$, 
\ba{
n_v(Y) &= - \sum_{i=1}^{\tilde{p}} (N^2 - \tilde{N}_i^2 ) - \frac{1}{2} N^2  + \frac{1}{2} \ ,\\
n_h(Y) &= n_v(Y) + \frac{1}{2} \sum_{i=1}^{\tilde{p}} \tilde{N}_i \tilde{k}_i - \frac{1}{2}  \ ,
}
as well as the embedding index $I_{\rho_Y}$ of $\mathfrak{su}(2)$ into $\mathfrak{su}(N)$ labeling the partial closure of the full puncture, 
\ba{
I_{\rho_Y} &= \frac{1}{6} \sum_{i=1}^{\tilde{p}} i (i^2-1) \tilde{k}_i \ . 
}
These are given in terms of the Young tableaux data $(\tilde{N}_i,\tilde{k}_i,\tilde{p})$ defined in section \ref{sec:map}. For example, using \eqref{box} one can verify that for the box diagram labeled by $\ell$, these evaluate to $I_{\rho_{Y_\ell}} = \frac{1}{6} N (\frac{N^2}{\ell^2} - 1)$, $n_{h}(Y_\ell) = \frac{2N}{3\ell}(\ell^2 - N^2)$, and $n_v(Y_{\ell}) = \frac{1}{6} (3 + \ell N - \frac{4 N^3}{\ell})$, which upon substituting into \eqref{dc} and \eqref{ccft} yield the $c$ central charge evaluated in \cite{Bah:2021mzw,Bah:2021hei}. One can similarly verify that using \eqref{min} for the case of the minimal puncture, the resulting central charge matches the result obtained in \cite{Xie:2013jc}.

The field theory central charge \eqref{ccft} can be compared with the holographic central charge, computed above in \eqref{conjecture} as
\ba{
c_{\text{hol}} = \frac{1}{12} \left( M N^2 + \sum_{a=1}^p \left( \frac{N}{M} k_a w_a^3 - 2 k_a^2 w_a^3 - \sum_{b=a+1}^p (w_a^3 + 3 w_a w_b^2)k_ak_b \right)  \right)\ . \label{chol}
}
In order to check that these two quantities match at large $N$, we need to evaluate the large-$N$ limit of \eqref{dc}. First we change from Young tableaux variables 
$(\tilde{N}_i,\tilde{k}_i,\dots,i=1,..,\tilde{p})$ to variables $({N}_a,{k}_a,\dots,a=1,..,p)$. Since $\tilde{k}_i$ is only nonzero at the location of a monopole $i=w_a$, at which point $\tilde{k}_i=k_a$ and $\tilde{N}_i=N_a$, we can replace $\sum_i \tilde{N}_k \tilde{k}_i$ with $\sum_a N_a k_a$ in $n_h(Y)$, as well as $\sum_i i(i^2-1)\tilde{k}_i$ with  $\sum_a w_a(w_a^2-1){k}_a$ in $I_{\rho_Y}$. We also make use of the following identity ({\it e.g.}  \cite{Bah:2019jts}),
\ba{
 \sum_{i=1}^{\tilde{p}} (N^2 - \tilde{N}_i^2 )& =\sum_{a=1}^p \left( \frac{2\ell_a^2}{3}  (w_a^3 - w_{a-1}^3 ) + \ell_a (N_a-w_a \ell_a) (w_a^2 - w_{a-1}^2) - \frac{N_a k_a}{6}    \right) - \frac{N^2}{2}  + \frac{1}{2} \ .
}
Terms of $\CO(N^2)$ and less are subleading in the large $N$ limit and can be dropped. These include the sum $\sum_a N_a k_a$, as well as the $-1$ part of the sum $\sum_a w_a(w_a^2-1)k_a$. Putting all this together, and adding $\Delta c$ to the large $N$ limit of $c_{(A_{N-1},A_{k-1})}$ given in \eqref{clargen}, we evaluate
\ba{
c\overset{N\to \infty}=\frac{1}{12} \left( {M N^2}+\sum_{a=1}^{{p}} \left(  \frac{N w_a^3 {k}_a}{M}      -{2\ell_a^2}  (w_a^3 - w_{a-1}^3 ) -3 \ell_a (N_a-w_a \ell_a) (w_a^2 - w_{a-1}^2)    \right)  \right)  \ .}
Substituting for $N_a$ and $\ell_a$ from \eqref{kl} and \eqref{na},  we evaluate
\ba{ 
\bs{
c\overset{N\to \infty}=&\frac{1}{12} \Bigg[{M N^2}+\sum_{a=1}^{{p}}\frac{N}{M}  w_a^3 {k}_a    \\
&-\sum_{a=1}^{{p}}\left({2\left( \sum_{b=a}^p k_b \right)^2}  (w_a^3 - w_{a-1}^3 ) +3 \left( \sum_{b=a}^p k_b \right)\left( \sum_{b=1}^{a-1} w_b k_b \right) (w_a^2 - w_{a-1}^2)    \right) \Bigg]
}\\ 
\bs{
=&\frac{1}{12} \left[{M N^2}+\sum_{a=1}^{{p}} \left( \frac{N}{M} {k}_aw_a^3  -  2 k_a^2 w_a^3   -   \sum_{b=a+1}^p  w_a^3  k_a k_b - 3 \sum_{b=1}^{a-1} w_a^2  w_b k_a k_b  \right) \right]  \ , \label{ccc}
}
}
where in the second equality we converted the sums over $w_{a-1}$ to sums over $w_a$, and then expanded the remaining sums in order to cancel some terms. The final sum in \eqref{ccc} can be recast as 
\ba{
\sum_{a=1}^p \sum_{b=1}^{a-1} w_a^2  w_b k_a k_b = \sum_{a=1}^p \sum_{b=a+1}^p w_a w_b^2 k_a k_b\ .
}
Substituting into \eqref{ccc} and comparing with the holographic central charge \eqref{chol}, we find agreement. 

\subsubsection{Coulomb branch operators}

The Seiberg-Witten curve of the SCFT takes the form
\ba{
y^2 = x^N + z^k + \dots + u_{ab} x^a z^b+ \dots
} 
where $u_{ab}$ are deformations of the curve with scaling dimension 
\ba{
\Delta(u_{ab}) = \frac{k N - ak - b N}{k+N} \ .
}
The regular puncture with associated Young tableaux $Y$ contributes terms 
\ba{
y^2  
\supset \sum_{l=2}^N \sum_{n=1}^{p_l} v_{l,n}z^{-n} x^{N-l}\ ,\qquad \Delta(v_{l,n}) = \frac{l (M-N) + n N}{M} \ , 
\label{ypunc}
} 
where $p_l$ is the pole structure defined above \eqref{idd}.

Coulomb branch operators are scalar primaries of protected $\CN=2$ chiral multiplets with superconformal $SU(2)_R\times U(1)_r$ R-charges satisfying $r=2\Delta, R=0$. These correspond to deformation parameters $u_{ab}$ with $\Delta(u_{ab})>1$. These deformations are nicely encoded in a Newton polygon, by plotting the $(a,b)$ coordinates associated to operators $u_{ab}$ on a grid. In particular, the operators $v_{l,n}$ associated contributed by the addition of the regular puncture on the sphere will correspond to points below the horizontal axis of this grid, since the associated powers of $x$ and $z$ in \eqref{ypunc} are negative. Examples of the quadrants of the Newton polygon associated to the regular puncture deformations are shown in Figure \ref{fig:reg}.

In \eqref{near_monopoles}, we identified Coulomb branch type operators associated to wrapping an M2-brane on the $S^2$ at each of the $a=1,\dots,p$ monopole locations, and computed their scaling dimensions
\ba{
\Delta(\CO_a) &= \sum_{b=1}^{a-1} k_b w_b + w_a \sum_{b=a}^p k_b - \frac{Nw_a}{M} = N_a - \frac{N w_a}{M} \label{dco} .
}
In the second equality we used the definition of $N_a$ from \eqref{na}.
We can match these onto operators in the SCFT as follows. Firstly we note that these operators correspond to some value of $l$ and $n=p_l$ in \eqref{ypunc}, since at a given value of $l$ we are interested in the largest dimension Coulomb branch operator (which  has $n=p_l$). In the Newton polygon, these are points bounding the lower edge of the triangle.

Since the $\CO_a$ are associated with fluxes through the $S^2$ surrounding the monopole locations, we restrict attention to these points. The locations of the monopoles $\eta_{a}$ coincide with changes in the slope of the density profile $\lambda(\eta)$, and correspondingly with changes in the lengths between subsequent rows of the associated Young tableaux (or equivalently, changes in the slope of the Newton polygon). Since the distinct lengths of rows of the Young tableaux are given by the $\ell_a$, with multiplicities $(w_a-w_{a-1})$, evidently these changes in slope occur at box numbers $l_a = \sum_{b=1}^a (w_b-w_{b-1})\ell_b$, which is none other than $N_a$. Thus, we identify the operators $\CO_a$ with the $v_{l,n}$ for $l=N_a$ and $n=p_{N_a}$, with dimensions
\ba{
\Delta(v_{l=N_a,n=p_{N_a}}) =  N_a -  \frac{N}{M} \left( N_a - p_{N_a} \right)\ .
} 
Finally, we evaluate $p_{N_a} = N_a - $(height of $N_a$'th box) by again appealing to the partition of $N$. Since there are $(w_a-w_{a-1})$ rows of length $\ell_a$ in the tableaux, the height of the $N_a$'th box is equal to the number of rows that have already been surpassed, namely $\sum_{b=1}^a (w_b-w_{b-1}) = w_a$. Therefore,
\ba{
\Delta(v_{l=N_a,n=p_{N_a}}) = N_a - \frac{Nw_a}{M} \ , \label{dfinal}
}
which exactly matches the dimensions of the holographic operators \eqref{dco}. These operators are circled in red in the example Newton polygons drawn in Figure \ref{fig:reg}.

\subsubsection{Flavor central charge}

The flavor central charges associated to the $SU(k_a)$ flavor symmetries at the monopole locations were computed in \eqref{flavor_from_inflow} as
\ba{
k_{SU(k_a)} = 2 (y_a + m_a w_a)
}
Using  $m_a = \ell_{a+1} - \frac{N}{M}$, $\ell_a = \sum_{b=a}^p k_b$ (with $\ell_{p+1}=0$), $y_a=\sum_{b=1}^a w_b k_b$, this can be rewritten
\ba{
k_{SU(k_a)} &= 2 \left(\sum_{b=1}^a  k_bw_b +w_a \sum_{b=a+1}^p k_b - \frac{N}{M}w_a\right) = 2 \Delta(\CO_a) \ , \label{fcent}
}
where $\Delta(\CO_a)$ are the scaling dimensions given in  \eqref{dco}, which we showed in that section match the maximal-dimension Coulomb branch operators associated to the $a$'th monopole. This confirms the conjecture that the flavor symmetry central charge of the associated Argyres-Douglas SCFT is equal to twice the scaling dimension of the Coulomb branch operator of maximal dimension \cite{Xie:2013jc}.

\subsubsection{Higgs branch operators for the minimal puncture}

In the special case that $Y$ corresponds to either a maximal puncture or a minimal puncture, a Lagrangian quiver description of the Argyres-Douglas SCFT is known. In \cite{Bah:2021hei}, a class of Higgs branch operators are matched in the case of the maximal puncture between baryonic operators in the quiver, and a class of M2-brane probes in the holographic solution. With the generalized holographic solutions for any regular puncture, we can now perform a similar check for the minimal puncture case.

The Lagrangian description for the theories with a simple puncture in addition to the irregular puncture is known for the case that $k=mN$ is an integer multiple of $N$. 
The UV quiver and IR quiver are depicted in Figure \ref{fig:quiver2}. They consists of $N-1$ gauge nodes $SU(m\ell + 1)$, $\ell = 1,\dots, N-1$. Bifundamental hypermultiplets $(Q_\ell,\tilde{Q}_\ell)$ connect the $\ell$'th node to the $\ell+1$'th node. There is one fundamental hypermultiplet  $(q_1,\tilde{q}_1)$ at the first node, and one $(q_{N-1},\tilde{q}_{N-1})$ at the last node. There are adjoints $\phi_\ell$ for each of the gauge nodes. There are also singlets $M_{{j}}$, ${j}= 1,\dots, m(N-1)$ (${j} = \hat{j}-m$ for $\hat{j}$ the index in \cite{Agarwal:2017roi}). The $\CN=1$ R-charges of the quarks and adjoints are
	\ba{\bs{
	R_{\CN=1}(Q_\ell) = R_{\CN=1}(q_1) = \frac{3m+2}{3(m+1)}\ ,\qquad R_{\CN=1}(q_{N-1} ) = \frac{ 3m+2-mN}{3(m+1)}\ , \\
	R_{\CN=1}(\phi_\ell) =  \frac{2}{3(m+1)}\ ,
	}}
	from which the dimensions can be computed $\Delta = \frac{3}{2} R_{\CN=1}$. 

	From the quivers, we can evidently construct sets of baryonic operators of the form 
		\ba{
		\bs{
	B_1& = \epsilon_{i_1\dots i_{(N-1)m+1}} (q_{N-1})^{i_1} (q_{N-1}\phi)^{i_2} \dots (q_{N-1} \phi^{(N-1)m})^{i_{(N-1)m + 1} }\ ,\\
	B_2 &= q_1 Q_1\dots Q_{N-2} q_{N-1}\ , 
	}}
	with dimensions
	\ba{
	 \Delta(B_1) = k-\frac{k}{N} +1\ ,\qquad \Delta(B_2) = N\ . \label{opsh}
	}

	\usetikzlibrary{shapes.geometric}
	\usetikzlibrary{arrows, decorations.markings}
	
		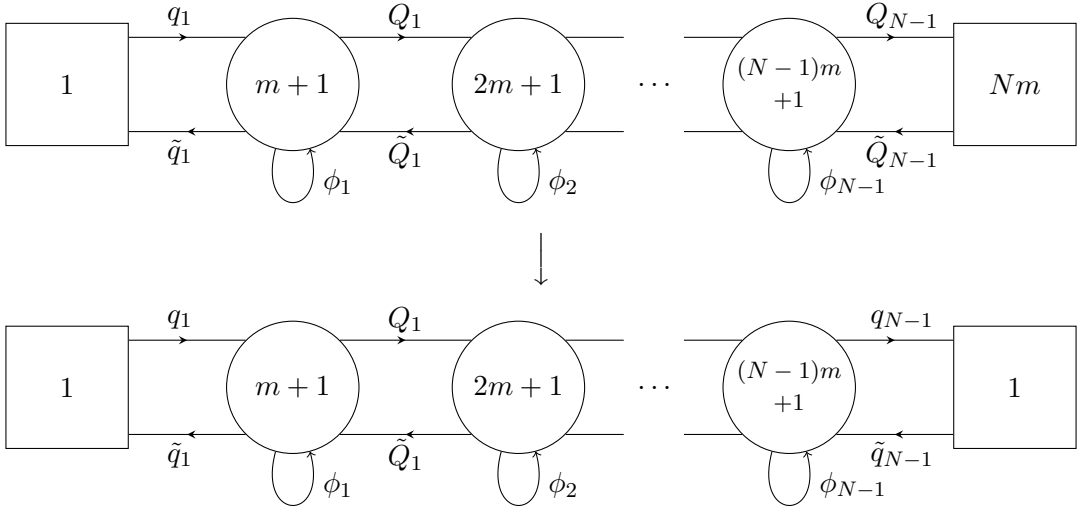
\begin{figure}[t!]
	\centering
	
 		\begin{tikzpicture}[square/.style={regular polygon,regular polygon sides=4}]

\tikzset{
  ->-/.style={decoration={markings, mark=at position 0.5 with {\arrow{stealth}}},
              postaction={decorate}},
}

	\node[square,draw=black,minimum size=2.3cm,fill=none]  (5) at (-3,0) { };
	\node at (-3,0) {$1$};
		
	\node[circle,draw=black,minimum size=1.76cm,fill=none]  (1) at (0,0) { };
	\node at (0,0) {$m+1$};
        \draw[->] (1) to [out=-105,in=-75,looseness=5.5] (1);
        \node at (0.6,-1.3) {$\phi_1$};
	
	\node[circle,draw=black,minimum size=1.76cm,fill=none]  (2) at (3,0) { };
	\node at (3,0) {$2m+1$};
	 \draw[->] (2) to [out=-105,in=-75,looseness=5.5] (2);
	\node at (3.6,-1.3) {$\phi_2$};
	
	\node[circle,draw=black,minimum size=1.76cm,fill=none]  (3) at (6.6,0) { };
	\node at (6.6,0) {\footnotesize$\begin{array}{c} (N-1)m \\ +1 \end{array}$};
	 \draw[->] (3) to [out=-105,in=-75,looseness=5.5] (3);
	 \node at (7.45,-1.3) {$\phi_{N-1}$};
	
	\node[square,draw=black,minimum size=2.3cm,fill=none]  (4) at (9.6,0) { };
	\node at (9.6,0) {$Nm$};
	
	\draw[->-] (-2.18,0.627) -- (1.north west);
	\draw[->-] (1.south west) -- (-2.18,-0.627);
	\node at (-1.5,0.9) {${q}_1$};
	\node at (-1.5,-0.9) {$\tilde{q}_1$};
	
	\draw[->-] (1.north east) -- (2.north west);
	\draw[->-] (2.south west) -- (1.south east);
	\node at (1.5,0.9) {${Q}_1$};
	\node at (1.5,-0.9) {$\tilde{Q}_1$};
	
	\draw (2.north east) -- (4.4,0.627);
	\draw (2.south east) -- (4.4,-0.627);
	
	\node at (4.84,0) {$\dots$};
	
	\draw (5.2,0.627) -- (3.north west);
	\draw (5.2,-0.627) -- (3.south west);
	
	\draw[->-] (3.north east) -- (8.78,0.627);
	\draw[->-] (8.78,-0.627) -- (3.south east);
	\node at (8.1,0.9) {${Q}_{N-1}$};
	\node at (8.1,-0.9) {$\tilde{Q}_{N-1}$};
  
		\end{tikzpicture}
		
		\begin{tikzpicture}
		
		\node at (4.8,0) {$\Big\downarrow$};
		\end{tikzpicture}
		
	 	\begin{tikzpicture}[square/.style={regular polygon,regular polygon sides=4}]

\tikzset{
  ->-/.style={decoration={markings, mark=at position 0.5 with {\arrow{stealth}}},
              postaction={decorate}},
}
	\node[square,draw=black,minimum size=2.3cm,fill=none]  (5) at (-3,0) { };
	\node at (-3,0) {$1$};
		
	\node[circle,draw=black,minimum size=1.76cm,fill=none]  (1) at (0,0) { };
	\node at (0,0) {$m+1$};
        \draw[->] (1) to [out=-105,in=-75,looseness=5.5] (1);
        \node at (0.6,-1.3) {$\phi_1$};
	
	\node[circle,draw=black,minimum size=1.76cm,fill=none]  (2) at (3,0) { };
	\node at (3,0) {$2m+1$};
	 \draw[->] (2) to [out=-105,in=-75,looseness=5.5] (2);
	   \node at (3.6,-1.3) {$\phi_2$};
	
	\node[circle,draw=black,minimum size=1.76cm,fill=none]  (3) at (6.6,0) { };
	\node at (6.6,0) {\footnotesize$\begin{array}{c} (N-1)m \\ +1 \end{array}$};
	 \draw[->] (3) to [out=-105,in=-75,looseness=5.5] (3);
	  \node at (7.45,-1.3) {$\phi_{N-1}$};
	
	\node[square,draw=black,minimum size=2.3cm,fill=none]  (4) at (9.6,0) { };
	\node at (9.6,0) {$1$};
	
	\draw[->-] (-2.18,0.627) -- (1.north west);
	\draw[->-] (1.south west) -- (-2.18,-0.627);
	\node at (-1.5,0.9) {${q}_1$};
	\node at (-1.5,-0.9) {$\tilde{q}_1$};
	
	\draw[->-] (1.north east) -- (2.north west);
	\draw[->-] (2.south west) -- (1.south east);
	\node at (1.5,0.9) {${Q}_1$};
	\node at (1.5,-0.9) {$\tilde{Q}_1$};
	
	\draw (2.north east) -- (4.4,0.627);
	\draw (2.south east) -- (4.4,-0.627);
	
	\node at (4.84,0) {$\dots$};
	
	\draw (5.2,0.627) -- (3.north west);
	\draw (5.2,-0.627) -- (3.south west);
	
	\draw[->-] (3.north east) -- (8.78,0.627);
	\draw[->-] (8.78,-0.627) -- (3.south east);
	\node at (8.1,0.9) {${q}_{N-1}$};
	\node at (8.1,-0.9) {$\tilde{q}_{N-1}$};
	
		\end{tikzpicture}		
	\caption{The upper figure is the UV quiver and the lower figure the IR quiver for the flow to the $(A_{N-1}^{(N)}[k=mN],S)$ SCFTs. \label{fig:quiver2}}
	\end{figure}

	On the gravity side, we computed the dimensions of M2-brane probes corresponding to Higgs branch operators in \eqref{higgs1} and \eqref{higgs2}. For the minimal puncture with data summarized in \eqref{min}, these correspond to operators with dimensions
	\ba{
	\Delta(\CO_{[w_p,w_{\rm m}]}) = k + \CO(1) \ ,\qquad \Delta(\CO_{[w_1,w_2]}) = N + \CO(1)\,. 
	}
	At large $N$, these operator dimensions match \eqref{opsh}. We thus identify the M2-brane probes $\CO_{[w_p,w_{\rm m}]}$ and $\CO_{[w_1,w_2]}$  in the minimal puncture geometry with the baryons $B_1$ and $B_2$ in the Lagrangian quiver of the proposed dual SCFT.

\subsection{Speculations on nested Young tableaux}

We end this section with some comments on another Argyres-Douglas SCFT, which we speculate might be dual to the Case I geometries constructed in this work. 

One generalization of the $(A_{N-1}^{(N)}[k], Y)$ Argyres-Douglas SCFTs takes the Higgs field in the Hitchin system to be specified by a series  $T_1\subseteq T_2\subseteq \dots \subseteq T_{2 + \frac{k}{N}}$ of semisimple elements of $\mathfrak{su}(N)$ that are not necessarily regular ({\it e.g.}~see \cite{Xie:2012hs,Wang:2015mra}). Then, the data of the irregular puncture is refined to depend on a sequence of Young tableaux  $Y_n \subseteq Y_{n-1} \dots \subseteq Y_1$, $n = \frac{k}{N} + 2$. The Young tableaux encode the degeneracies of the eigenvalues of the matrices $T_i$. The ``maximal'' irregular singularity $A_{N-1}^{(N)}[k]$ is recovered  in the limit that there is no degeneration of the eigenvalues, so that all matrices are of maximal type.

In this case, the Seiberg-Witten curve has additional parameters,
\ba{
\label{sw2}
x^N + \sum_{i=2}^N \left(  \dots + \sum_{j=1}^{m_i} u_j ^{(i)}z^{m_i-j}  \right) x^{N-i}= 0\ ,\quad m_i = \sum_{l} p_i^{(l)} - 2i + 1 \ .
}
Here $p_i^{(l)}$ is the pole structure of the $i$'th box in the $l$'th Young tableaux of the sequence. The independent Coulomb branch operators thus have dimensions 
\ba{
\Delta(u_{j}^{(i)}) =  \frac{i k  - (m_i-j) N}{k+N}\ .
}

\paragraph{Suggestive example: $\{ Y_Q,\dots,Y_Q\}$. }
Let us illustrate such a scenario with a suggestive example. 
Suppose there is just one type of tableaux labeled by $Q$, which occurs with multiplicity $n= \frac{k}{N} + 2$, 
\ba{\rho = \{Y_Q,\dots,Y_Q\}\ ,
}
where we assume that $k$ is an integer multiple of $N$. The limit $Q=N$ reduces to the ``maximal'' irregular puncture without the additional nested structure. For simplicity let us take the regular puncture on the other pole of the sphere to be trivial, {\it i.e.} $q=1$. Then, the pole structure parameters are given by,
\ba{
p_i^{(l)} = i - m\ ,\quad i = 1 + (m-1)Q,\dots,m Q\ ,\quad m = 1,\dots, N/Q\ ,\\
 m_i = \left( \frac{k}{N}+2\right)p_i -2i + 1\ .
}
For $Q \sim \CO(1)$, the quantities $m_i$ are negative, and we cannot use this description \eqref{sw2} to enumerate the Coulomb branch operators. However for $Q\sim \CO(N)$, one may verify that the $m_i$ are positive. In this case, we check that the Coulomb branch operator $u_{m_N}^{(N)}$ has dimension of order $N$, given by
\ba{
\Delta(u_{m_N}^{(N)}) = \frac{N k}{k+N}\ ,
}
and that the rank of the Coulomb branch is given by\footnote{One check on this formula is that it satisfies the expected relation $\text{rank}(\text{CB}) = \text{dim}_\rho(\text{CB}) - \text{dim}(SU(N))$, where $\text{dim}_\rho(\text{CB})$ is equal to one half times the sum over the dimensions of the nilpotent orbits of $\mathfrak{su}(N)$ that appear in the sequence $\rho$. }
\ba{
\text{rank}(\text{CB}) = \frac{k N}{2} - \frac{N}{2Q}(k+2N) + 1 \ .
}
We may compute the central charge at large $N$ by summing over the Coulomb branch operator dimensions \cite{Shapere:2008zf}, yielding
\ba{
2a-c &= \frac{1}{4} \sum_i (2 \Delta(u_i) - 1)=   \frac{1}{2} \sum_{i=2}^N\sum_{j=1}^{m_i} \frac{i k - (m_i - j) N}{k+N} - \frac{1}{4} \text{rank}(\text{CB})  \nn \\
&= \frac{k^2 N^2}{12 (k+N)} + \CO(N^2) \ .
}
In particular, we note that at leading order for large $N$, these data -- the large-$N$ central charge, maximal dimension Coulomb branch operator, and rank of the Coulomb branch -- are indistinguishable from the ``maximal'' irregular puncture theory, $(A_{N-1},A_{k-1})$. This is not surprising, since this singularity labeled by $Q\sim \CO(N)$ is quite similar to the maximal irregular puncture case.  
However, even for $Q\sim \CO(N)$ there is additional data contributed by the nested structure: the flavor symmetry of the field theory is enhanced. The number of mass parameters  is equal to the number of distinguished eigenvalues of $T_1$ \cite{Wang:2015mra}, which in this example contributes $\text{rank}_{\rho}(F)$ to the rank of the flavor symmetry as,
\ba{
\text{rank}_{\rho}(F) = \frac{N}{Q}- 1\ ,\qquad \text{rank}(F) = \text{rank}_{\rho}(F)  + \text{rank}_{\rho}(SU(N))\ .
}

Now, return to the Case I holographic solutions, again for simplicity taking $q=1$ such that there is no regular puncture on the sphere. Their data is summarized in Table \ref{tab:sum}, where since $q=1$ we may replace $M=N_{\text{eff}}$. If we identify the number of M5-branes $N$ and the irregular puncture parameter $k$ with the holographic fluxes $N_{\text{eff}}$ and $K$ in a way identical to the Case II dictionary, 
\ba{
N = N_{\text{eff}}\ ,\qquad k = K + M - N = K\ , 
}
 then evidently the R-symmetry, central charges, and dimensions of the operators $\CO_1$ match both data of the Argyres-Douglas SCFT labeled by ``maximal'' regular puncture $A_{N-1}^{(N)}[k]$, and the Argyres-Douglas SCFT labeled by the nested Young tableaux $\rho = \{Y_Q,\dots,Y_Q\}$ with $Q\sim \CO(N)$.  

In addition, both Case I and the $\rho = \{Y_Q,\dots,Y_Q\}$ irregular puncture theory possess enhanced flavor symmetry. Naively, we expect a flavor symmetry coming from the two stacks of $N_\mathrm W$ and $N_\mathrm S$ smeared M5-branes in the geometry, where (again, naively) we might expect that the maximal rank of this flavor symmetry is identified with the maximal possible rank of the sources, equaling $N_W + N_S - 1$. This identification would provide a proposed map between the flux quantum $N_\mathrm S$ on the gravity side, and the field theory parameters $N,k,$ and $q$. 

While a proposed duality along these lines seems promising, more precise checks would be required to put it on firmer footing. In particular, a more precise understanding of the flavor symmetry of the smeared M5-brane geometry would be necessary. We leave this direction to future work. At present, we view this discussion as a hint that the dual to the Case I geometry is consistent with Argyres-Douglas SCFT whose irregular puncture has refined structure.




\section{Discussion} \label{sec:discussion}

The results of this work suggest several interesting directions
for future research, some of which are discussed below.

\paragraph{Exploring the geometry of irregular singularities.}

The separable Toda solutions of Case I and II studied in this paper
exhibit an interesting qualitative feature that sets them apart
from the Maldacena-Nu\~nez \cite{Maldacena:2000mw} and
Gaiotto-Maldacena  \cite{Gaiotto:2009gz} solutions. Recall that the latter
describe the holographic duals of class $\cS$ setups 
without punctures, and with regular punctures, respectively.

To elucidate the novel features of the solutions of Case I and II,
let us examine in closer detail the locus $y=0$ in the 
standard form of the line element \eqref{LLM}.
Recall that the internal space $M_6$ can be regarded as an
$S^2 \times S^1_\chi$ fibration over a 3d base space $B_3$, with coordinates
$x_1$, $x_2$, $y$. The line element \eqref{LLM} reveals that
the locus $y=0$ in $B_3$ has an intrinsic geometric meaning,
because it corresponds to the region where the $S^2$ shrinks.
Both in the Maldacena-Nu\~nez and in the Gaiotto-Maldacena solutions, the locus $y=0$ consists of a single component,
at which the $S^2$ shrinks smoothly, while the warp factor remains
non-singular.
This should be contrasted with the solutions of Case I and II, 
in which the locus $y=0$ splits into two components,
meeting transversally at a point. This structure is made manifest
by the introduction of the $t$, $u$ coordinates, see \eqref{var_sep}.
In particular, the relation $y = tu$ can be regarded as a  way
of parametrizing the split of the locus $y=0$ into two components,
$t=0$ and $u=0$, intersecting transversally at $u=t=0$.
In Case II, the splitting of the $y=0$ locus
is particularly important for the structure of the internal geometry,
because the behavior of the warp factor and
line element along the two components $u=0$ and $t=0$
exhibits clear qualitative differences:
along the component $u=0$, the $S^2$ shrinks but the warp
factor remains non-singular, while along the component
$t=0$ we find a smeared M5-brane source,
see Figure~\ref{fig_cases_new}.

The considerations of the previous paragraphs
suggest a more physical way of thinking about
the change of coordinates $y = tu$: it describes
 setups in which the $y=0$ locus has a non-trivial
sub-structure. From this point of view, it is natural to wonder
whether the locus $y=0$ might admit richer structures than those
studied in this paper. For example, one might ask whether
$y=0$ could split into three components, $\cC_1$, $\cC_2$, $\cC_3$
say, with $\cC_1$ and $\cC_2$ meeting at a point,
and similarly for $\cC_2$ and $\cC_3$.
It is not clear whether such more complicated setups
would still allow for a separation of variables in the Toda equation;
this might pose a technical challenge to exploring 
such possibilities analytically in the Toda frame.

A possible strategy to study a multi-component $y=0$ locus
might be inspired by the methods of \cite{Bah:2015fwa}.
This reference studies the most general $AdS_5$ M-theory solution preserving
4d $\cN = 1$ supersymmetry in which the internal space
is a fibration of a compact 4-manifold over a (punctured) Riemann
surface. The internal geometry is assumed to preserve at least a $U(1)^2$ isometry. The 4-dimensional fiber over the Riemann
surface can be described as a $U(1)^2$ fibration over a 2d base space. The latter is a region in $\mathbb R^2$ with a boundary
consisting of several segments, where on each segment, a different linear combination of the two $U(1)$ Killing vectors shrinks to zero size.
These setups bear some formal analogies to the
multi-component $y=0$ locus we would like to explore. 

Alternatively, one might work in the electrostatic picture
after the B\"acklund transform. In this frame, the task is to identify
those charge densities that give rise to a multi-component $y=0$ locus, and to study them systematically.

\paragraph{Connections with Painlev\'e equations and integrable systems.}

In the Introduction, we have motivated our assumption
that the internal geometry admits an additional Killing vector
$\partial_\beta$, in addition to the Killing vector $\partial_\chi$
dual to the superconformal $U(1)_r$ symmetry.
We have also noticed, however, how this $\partial_\beta$ isometry
does not give rise to a continuous $U(1)$ flavor symmetry on the field
theory side.
It is natural to wonder whether we could relax the assumption
of having the additional isometry $\partial_\beta$, and search
for solutions that can be interpreted in terms of irregular punctures. In particular, one may wonder whether, 
upon relaxing the $\partial_\beta$ isometry, non-singular
(or less singular) solutions could be found in which the smeared
M5-brane sources encountered in this work
are resolved.

The study of the Toda equation \eqref{Toda_intro} in non-axisymmetric
setups is particularly challenging.
A possible inroad into this problem is furnished by
the analysis in \cite{Petropoulos:2014rva}.
The main idea is to consider a coordinate change
from $x_1$, $x_2$, $y$ to a new set of coordinates
$\tau$, $\vartheta$, $\varphi$, of the form
\beq
x_1 = e^{a\tau} \Omega_1(\tau) \sin \vartheta \cos \varphi \ , \qquad
x_2 = e^{a\tau} \Omega_2(\tau) \sin \vartheta \sin \varphi \ , \qquad
y = \Omega_2(\tau) \cos \vartheta \ ,
\eeq
where $a$ is a constant parameter,   $\Omega_{1,2}$
are functions of $\tau$, $0 \le \vartheta \le \pi$,
and $\varphi$ is an angle of period $2 \pi$.
In the generic case in which $\Omega_1(\tau)$
is not identically equal to $\Omega_2(\tau)$, this ansatz
describes a non-axisymmetric configuration.
Nonetheless, one retains analytic control as follows.
If we introduce a new coordinate $s$ via
$as  = e^{-a\tau}$, 
and we set the Toda potential $D$ to be $e^D = a^2 s^2$,
we can verify that the Toda equation
reduces to an ODE for a single function $w = w(s)$.
The ODE is of Painlev\'e III type,
\beq
 w'' = \frac{(w')^2}{w} - \frac{w'}{ s } + \gamma  w^3 + \frac{ {\delta}}{w}  \ ,
\eeq 
where $\gamma$, $\delta$ are constant parameters.   

Relating the search of M-theory solutions dual to irregular punctures
to a Painlev\'e equation is particularly tantalizing,
given that Painlev\'e equations have natural links to
the Hitchin integrable system on  punctured Riemann surfaces,
see \emph{e.g.}~\cite{Bonelli:2016qwg}.
A functional transform from the Toda equation to the 
Painlev\'e III equation might thus be a way to
establish a precise correspondence between the holographic description of the class $\cS$ model,
and its description in terms of the Higgs field entering the
Hitchin system. We plan to investigate this direction further in the future.

\paragraph{Holographic realizations of renormalization group flows.}
Another interesting  direction is to understand the holographic analogues of known renormalization group flows between Argyres-Douglas SCFTs. For example, there is an RG flow between the $(A_{N-1}^{(N)}[k], Y_{\text{max}})$ theory with a regular maximal puncture on the sphere, and the $(A_{N-1},A_{k+N-1})$ theory with no regular puncture, via nilpotent Higgsing \cite{Giacomelli:2017ckh}. This flow can furthermore be understood purely from the Lagrangian perspective. It would be quite interesting to reproduce such an RG flow holographically, using the proposed holographic duals that we have now identified.


\section*{Acknowledgments}

We are grateful to 
Davide Gaiotto, Enoch Leung,
Kazunobu Maruyoshi, and Peter Weck
 for interesting
conversations. 
The work of IB and TW is supported in part by NSF grant PHY-2112699.
The work of IB and TW is also supported in part by the Simons
Collaboration on Global Categorical Symmetries.
FB is supported by STFC Consolidated Grant ST/T000864/1.
The research of EN is supported by World Premier International Research Center Initiative (WPI), MEXT, Japan.
Part of this work was completed
while IB was visiting the Institute for Advanced Study.
We also acknowledge Perimeter Institute, where
part of this work was completed.
Research at Perimeter Institute is supported in part by the Government of Canada through the Department of Innovation, Science and Economic Development Canada and by the Province of Ontario through the Ministry of Colleges and Universities.


\appendix




\section{Further details on separable solutions}\label{sec:fibr_change}

\subsection{Angular coordinates $\phi$, $z$}

As anticipated in the main text, it is convenient to perform a change of coordinates
from the Toda angles $\chi$, $\beta$ to a new pair of angles $\phi$, $z$.
The change of coordinates is engineered in such a way that, along the segment
$\mathbf P_3 \mathbf P_4$ in each case in Figure \ref{fig_cases_new}, the linear
combination of Killing vectors $\partial_\chi$, $\partial_\beta$ that
has vanishing norm is simply $\partial_\phi$.
This can be achieved by setting
\beq
\chi = \frac{\tilde{v}}{\tilde{v}-1}\phi - z,\qquad \beta = -\frac{1}{\tilde{v}-1}\phi + z,
\eeq
where   $\tilde{v} \equiv v_\beta(t,u_2)$
is the constant value attained by $v_\beta$ along the $\mathbf P_3 \mathbf P_4$ segment,
\beq
\tilde v  = 1 - \frac \sigma 2 \bigg( 1-   \frac{u_1}{u_2} \bigg)\qquad \Rightarrow \qquad \cC = \tilde v   -1 =  - \frac \sigma 2 \bigg( 1-   \frac{u_1}{u_2} \bigg) \ ,
\eeq
where we have also given the value of the constant $\cC$ that enters the parametrization
\eqref{new_angles} used in the main text.
In terms of the Killing vectors and differential forms, we can write
\beq
\begin{pmatrix}
\partial_\chi \\
\partial_\beta
\end{pmatrix} = 
\begin{pmatrix}
1 & \frac{1}{\tilde{v}-1} \\
1 & \frac{\tilde{v}}{\tilde{v}-1}
\end{pmatrix}
\begin{pmatrix}
\partial_\phi \\
\partial_z
\end{pmatrix},\qquad 
\begin{pmatrix}
d\chi \\
d\beta
\end{pmatrix} = 
\begin{pmatrix}
\frac{\tilde{v}}{\tilde{v}-1} & -1 \\
-\frac{1}{\tilde{v}-1} & 1
\end{pmatrix}
\begin{pmatrix}
d\phi \\
dz
\end{pmatrix}.
\eeq
With this, the metric takes the form  
\begin{align} \label{eq_app_metric}
ds_{11}^2 =& \frac{e^{2\tilde{\lambda}}}{m^2}\Bigg[ds^2(AdS_5)+ \frac{t^2u^2 e^{-6\tilde{\lambda}}}{4}ds^2(S^2)\\
&+ R_z^2 Dz^2 + R_\phi^2 d\phi^2-\partial_y D \frac{K_1u^2 + K_2 t^2}{4t u}\left(\frac{dt^2}{K_1} + \frac{du^2}{K_2}\right)\bigg] \ ,\nonumber
\end{align}
where $Dz = dz - Ld\phi$ and we have made the definitions
\begin{align}
R_z^2 =&\;
 \frac{- \partial_y D K_1 K_2}{4 t u }
+ \frac{(v_\beta-1)^2}{1 - tu \partial_y D} \nn \\
&\;=
 \frac{-\sigma^2}{4tu(t_1t_2u^2 - u_1u_2t^2)}\bigg(ut_1t_2(t_1+t_2)(u-u_1)(u-u_2)\\
&+t^2(t_1+t_2)\Big(u^3-u_1u_2(3u - u_1-u_2)\Big) - tt_1t_2\Big(4u^3 - (u_1+u_2)(3u^2-u_1u_2)\Big)  \nn \\
&+t^3\Big(4uu_1u_2 - (u_1+u_2)(u^2 + u_1u_2)\Big)\bigg) \ , \nn  \\
L =&\; -\frac{2(u-u_2)}{P}\bigg(tuu_1\left(t(ut_2-tu_1)+t_1(tu-t_2(2u-u_1))\right) \nn \\
&-(u-u_1)u_2\big(ut_1t_2(2t-t_1-t_2)+tu_1(t-t_1)(t-t_2)\big)  \nn \\
&+tu_1u_2^2(t-t_1)(t-t_2)\bigg) \ ,  \nn \\
R_\phi^2 =&\; \frac{\sigma(u_1-u_2)}{P}(t-t_1)(t-t_2)(u-u_1)(u-u_2)\Big((t_1+t_2)u - (u_1+u_2)t\Big) \ , \nn \\
P =&\; \sigma(u_1-u_2)\bigg(ut_1t_2(t_1+t_2)(u-u_1)(u-u_2) + t^2(t_1+t_2)\Big(u^3-u_1u_2(3u-u_1-u_2)\Big) \nn \\
&- t^3\Big(u^2(u_1+u_2) - u_1u_2(4u - u_1 - u_2)\Big) - tt_1t_2\Big(4u^3-(u_1+u_2)(3u^2-u_1u_2)\Big)\bigg) \  .  \nn
\end{align}
From these definitions we can readily read off that the $\phi$ circle shrinks along the intervals where the $K_i$ vanish, and we can find that the $Dz$ fibration will only vanish at the locations of the monopoles.

In terms of the new angular coordinates, the flux reads
\beq
\overline G_4  = - \frac{G_4}{(2\pi \ell_p)^3} = 
 \frac{ {\rm vol}_{S^2}  }{4\pi}\wedge d \bigg[ 
Y \, \frac{d\phi}{2\pi}  - W \, \frac{Dz}{2\pi} 
\bigg] \ ,
\eeq
where the 0-forms $Y$, $W$ are given by
\begin{align}
W &= t^3 u^3e^{-6\tilde{\lambda}}(v_\beta - 1) - t u v_\beta -\frac{1}{2}\mathcal{F} \ ,  \nn \\
Y + WL &= - t^3 u^3 e^{-6\tilde{\lambda}}\, \frac{\tilde{v} - v_\beta}{\tilde{v}-1}  - \frac{ t u v_\beta}{\tilde{v}-1} - \frac{\mathcal{F}}{2(\tilde{v}-1)}  \ ,
\end{align}
with $v_\beta$, $\cF$ as in \eqref{metric_functions}.

Finally, let us record the expression of the calibration 2-form $Y'$ given in \eqref{calibration2form}
in terms of functions appearing in the line element \eqref{eq_app_metric},
\begin{align}
Y=& \;\frac{1}{4}(tu)^3e^{-9\tilde{\lambda}}\mathrm{vol}_{S^2} + \frac{1}{2}tu e^{-3\tilde{\lambda}}(1-(tu)^2 e^{-6\tilde{\lambda}})(v_\beta - 1)d\tau\wedge Dz\nonumber\\
& + \frac{1}{2}tu e^{-3\tilde{\lambda}}(1-(tu)^2 e^{-6\tilde{\lambda}})\left(\frac{\tilde{v}-v_\beta}{\tilde{v}-1} - (v_\beta -1)L\right)d\tau\wedge d\phi\nonumber\\
& + \Bigg[\frac{1}{2}\tau e^{-3\tilde{\lambda}}u\left(\frac{\tilde{v}-v_\beta}{\tilde{v}-1} - (v_\beta -1)L\right) +\frac{1}{4}\frac{K_2tue^{-9\tilde{\lambda}}\tau}{1-(tu)^2e^{-6\tilde{\lambda}}}\left(\frac{1}{\tilde{v}-1}+L\right) \Bigg]dt\wedge d\phi\nonumber\\
& + \Bigg[\frac{1}{2}\tau e^{-3\tilde{\lambda}}t\left(\frac{\tilde{v}-v_\beta}{\tilde{v}-1} - (v_\beta -1)L\right) -\frac{1}{4}\frac{K_1tue^{-9\tilde{\lambda}}\tau}{1-(tu)^2e^{-6\tilde{\lambda}}}\left(\frac{1}{\tilde{v}-1}+L\right) \Bigg]du\wedge d\phi\nonumber\\
& + \Bigg[\frac{1}{2}\tau e^{-3\tilde{\lambda}}y_2(v_\beta - 1) - \frac{1}{4}\frac{K_2tue^{-9\tilde{\lambda}}\tau}{1-(tu)^2e^{-6\tilde{\lambda}}} \Bigg]dt\wedge Dz\nonumber\\
& + \Bigg[\frac{1}{2}\tau e^{-3\tilde{\lambda}}y_1(v_\beta - 1) + \frac{1}{4}\frac{K_1tue^{-9\tilde{\lambda}}\tau}{1-(tu)^2e^{-6\tilde{\lambda}}} \Bigg]du\wedge Dz, \label{eq:Ycalib_zphi}
\end{align}

\subsection{Solutions in Case II in the notation of \cite{Bah:2021hei}}

\paragraph{Brief review.}

The $AdS_5$ solutions discussed in  \cite{Bah:2021hei} were obtained by uplift
from 7d gauged supergravity. The 11d metric and flux are given by
\begin{align} \label{11d_metric}
m^2 \, ds^2_{11} =&\;\frac{2 \, B\, w^{1/3} \,  \cH(w,\mu)^{1/3}}{\sqrt{    1-w^2 } }   \, \bigg[
 ds^2(AdS_5) 
+ \frac{dw^2}{2 \, w \, h(w) \,   (1-w^2)^{3/2}}
+ \frac{\cC^2 \, h(w) \,dz^2}{B} 
\nn \\
& +  \frac{\sqrt{ 1-w^2 }}{2 \, B} \, \bigg(  \frac{d\mu^2}{w\,(1-\mu^2)}  + \frac{(1-\mu^2)\, D\phi^2}{w\, \cH(w,\mu)}\, 
+ \frac{w \, \mu^2 \, ds^2(S^2)}{\cH(w,\mu)}  \bigg)
\bigg] \ .
\end{align}
The quantities $B$, $\cC$ are parameters specifying the solution
and satisfy
\beq \label{eq_params}
0 < B < 1  \ , \qquad \cC = \frac{1}{\ell \, \sqrt{1-B^2}} \ , \qquad \ell \in \mathbb Z_{>0} \ .
\eeq
The coordinates $z$, $\phi$ are angles with period $2\pi$,
while the coordinates $\mu$, $w$ have ranges
\beq
0 \le \mu \le 1 \ , \qquad 0 \le w \le w_1 :=  \tfrac 12 \, \Big( \sqrt{1+B} -  \sqrt{1-B}\Big)
\eeq
We have introduced the shorthand notation
\beq \label{shorthand}
\cH(w,\mu) =  \mu^2 + w^2 \, (1-\mu^2)     \ , \qquad 
h(w) = B - 2 \, w \, \sqrt{ 1-w^2 } \ .
\eeq
The quantity $ds^2(S^2)$ is the metric on the round unit 2-sphere,
while the 1-form $D\phi$ reads
\beq \label{Dphi_def}
D\phi = d\phi + \cC \, (2 \, w^2 -1) \, dz \ .
\eeq
The expression for $G_4$ is
\beq \label{G4flux}
G_4 =- \frac{1}{ m^3} \, {\rm vol}_{S^2} \, d\bigg[
\frac{ \mu^3}{\mu^2 + w^2 \, (1-\mu^2)} \, D\phi
\bigg] \ ,
\eeq
where ${\rm vol}_{S^2}$ is the volume form on the 2-sphere of unit radius.

When these solutions are cast in canonical LLM form,
the Toda potential $D$ reads
\beq \label{expD}
e^D =
\frac{16\, B  \,  \cC^2 \,  \left(1-\mu ^2\right)^{1 +1 / \cC} \,  \left [ B - 2  \,  w \,  \sqrt{ 1  - 
   w^2  }\right] }{      \left(1 - w^2\right)   \, \cG(w)^2 }  \ .
\eeq
The LLM coordinates $r$, $y$ are related to the coordinates $w$, $\mu$
by the relations
\begin{align} \label{y_and_r}
y = \frac{4\, B \, w\, \mu}{  \sqrt{ 1-w^2  }  } \ , \qquad
r = (1-\mu^2)^{-\frac{1}{2\cC} } \, \cG(w) \ .
\end{align}
In the previous expressions, the quantity $\cG(w)$ is a function
of $w$ only, satisfying the ODE
\beq \label{cG_ODE}
\frac{\cG'(w)}{\cG(w)} = \frac{-B \, w }{\cC \, 
\left(1-w^2 \right)   \left[ B-2 \,  w \,  \sqrt{ 1 -   w^2 }\right] } \ .
\eeq
Let us define a new variable $\widehat t$ via
\beq
\widehat t = \frac{w}{\sqrt {1 - w^2}} \ .
\eeq
The above ODE can be written in the form
\beq
\frac{d}{d \widehat t} \log \cG (\widehat t )=  \frac{\alpha_1}{ \widehat t_1 - \widehat t}
+ \frac{\alpha_2}{ \widehat t_2 - \widehat t} \ ,
\eeq
where
\beq
\widehat t_{1,2} = \frac{1 \mp \sqrt{1-B^2}}{B}  \ , \qquad
\alpha_{1,2} = -\frac{B^2}{2 \cC [1-B^2 \pm \sqrt{1-B^2}]}  \ .
\eeq
The solution can be written as
\beq\label{correct_G}
\log \cG(\widehat t) =  - \alpha_1 \, \log(\widehat t_1 - \widehat t)
 - \alpha_2 \, \log(\widehat t_2 - \widehat t)  + \text{const} \ .
\eeq

%
%
%
%
%
%

\paragraph{Dictionary with Case II solutions.}
The $\mu$, $w$ coordinates are related to the $t$, $u$ coordinates of Case II via
\beq
u = u_2 \mu \ , \qquad 
t  = \frac{4 B w}{u_2 \sqrt{1-w^2}} = \frac{4B}{u_2} \widehat t \ .
\eeq
We also record the identifications
\beq
\sigma = - \cC \ , \qquad \frac{t_2}{t_1} = \frac{1 + \sqrt{1-B^2}}{1 - \sqrt{1-B^2}} \ . 
\eeq



\section{Formulae for the electrostatic potential}  \label{app_Vs}

In this appendix we discuss the electrostatic potential generated by
a piecewise linear charge density profile with an arbitrary number of monopoles.
The electrostatic potential is computed from the charge density using the 
standard Green's function for the Laplace operator on $\mathbb R^3$,
see \eqref{nice_V}.

We consider a total of $n$ monopoles, located at $\eta_i$, $i = 1,\dots, n$.
We start by computing the contribution of the linear charge density between
two consecutive monopoles, which we parametrize as
\beq
\lambda(\eta) = m_i \eta + q_i \qquad \text{for $\eta_i < \eta < \eta_{i+1}$} \  , \qquad i = 1,\dots, n-1 \ .
\eeq
This segment of charge density gives the electrostatic potential
\begin{align}
V(\eta_i,\eta_{i+1},m_i,q_i) =&\; \frac{1}{2}\Bigg[ m_i\sqrt{\rho^2 + (\eta-\eta_i)^2} -(m_i\eta +q_i){\rm arctanh}\left(\frac{\eta-\eta_i}{\sqrt{\rho^2 + (\eta-\eta_i)^2}}\right) \\
&- m_i\sqrt{\rho^2 + (\eta-\eta_{i+1})^2} + (m_i\eta +q_i){\rm arctanh}\left(\frac{\eta-\eta_{i+1}}{\sqrt{\rho^2 + (\eta-\eta_{i+1})^2}}\right)\Bigg] \ . \nn
\end{align}

Next, we study the semi-infinite line at the right of the last monopole. 
The charge density is written as
\beq
\lambda(\eta) = m_n \eta + q_n \qquad 
\text{for $\eta > \eta_n$} \ .
\eeq
This contribution suffers from divergences, which are treated in the same way
as for Case I in the main text.
We introduce a regulator $\eta_R$ and we subtract 
the divergences as $\eta_R \rightarrow \infty$, with the result
\begin{align}
V_{\rm right} =&\; \lim_{\eta_R \rightarrow \infty}
\bigg[ V(\eta_n, \eta_R  ,m_n, q_n)  + \frac{m_n \eta  + q_n}{2}\log(2\eta_R)
  + \frac{m_n}{2} \eta_R    
\bigg]
\nn \\
=&\; \frac{1}{2}\Bigg[m_n \sqrt{\rho^2 + (\eta-\eta_n)^2} - (m_n\eta + q_n){\rm arctanh}\left(\frac{\eta-\eta_{n}}{\sqrt{\rho^2 + (\eta-\eta_{n})^2}}\right)\nonumber\\
& +(m_n\eta + q_n   )\log\rho + m_n \eta\Bigg] \  .
\end{align}

The semi-infinite line at the left of the first monopole is treated 
in an analogous way. We parametrize the charge density as
\beq
\lambda(\eta) = m_0 \eta + q_0 \qquad 
\text{for $\eta < \eta_1$} \ .
\eeq
We then compute
\begin{align}
V_{\rm left} =&\; \lim_{\eta_L \rightarrow \infty}
\bigg[ V(- \eta_L, \eta_1  ,m_0, q_0)  + \frac{m_0 \eta  + q_0}{2}\log(2\eta_L)
  - \frac{m_0}{2} \eta_L    
\bigg]
\nn \\
=&\; \frac{1}{2}\Bigg[-m_0 \sqrt{\rho^2 + (\eta-\eta_1)^2} + (m_0\eta + q_0){\rm arctanh}\left(\frac{\eta-\eta_{1}}{\sqrt{\rho^2 + (\eta-\eta_{1})^2}}\right)\nonumber\\
& +(m_0\eta + q_0   )\log\rho + m_0 \eta\Bigg] \  .
\end{align}

Let us remark that continuity of the charge density imposes
the following constraints on the monopole locations $\eta_i$,
the slope parameters $m_i$, and the intercept 
parameters $q_i$,
\beq
m_{i-1 } \eta_{i} + q_{i-1}  = m_{i} \eta_{i} + q_{i} \ , \qquad
i = 1,\dots, n \ .
\eeq
The total electrostatic potential for the full
charge density      takes the form
\beq
V_{\text{tot}} = \sum_{i=1}^{n-1}V(\eta_i,\eta_{i+1},m_i,q_i) + V_{\rm left} + V_{\rm right} \ .
\eeq

\section{Detailed analysis of generalized Case II}  \label{app_caseII}

In this appendix we present a more detailed analysis of the
generalized Case II solutions discussed in section \ref{sec:gen_caseII}.
Before addressing these solutions, however, we describe
the electrostatic picture for Case II solutions in
the notation of \cite{Bah:2021hei}.

\subsection{Electrostatic interpretation of Case II, revisited}
\label{sec:old_solus}

\paragraph{Determination of $\rho$, $\eta$, $V$.}

The expression of $\rho$ as a function of $(w,\mu)$ is obtained directly
by combining 
\eqref{Backlund_def}, \eqref{expD}, and \eqref{y_and_r},
\beq
\rho = \frac{4 \,  \cC  \, \sqrt B \, \sqrt{1-\mu^2} \, \sqrt{B - 2  \, w \, \sqrt{ 1-w^2 }} }{ \sqrt{ 1-w^2 }} \ .
\eeq
The function $\eta = \eta(\mu,w)$ can be determined as follows.
Let us regard $V$ as a function of $\mu$, $w$. Its derivatives with respect to $w$, $\mu$ can
be computed from $\partial_\rho V$, $\partial_\eta V$ with the help of
the chain rule, in terms of an unspecified $\eta = \eta(\mu,w)$.
The result reads
\begin{align} \label{V_derivatives}
\partial_\mu V(\mu,w) & = 
 \partial_\mu \eta(\mu,w) \, \log \Big[
(1-\mu^2)^{- \frac{1}{2  \cC }} \, \cG(w)
 \Big] 
 - \frac{4\, B \, w \, \mu^2}{(1 - \mu^2) \, \sqrt{ 1-w^2 }}
\ ,   \\
\partial_w V (\mu,w)& = 
\partial_w \eta(\mu,w) \, \log \Big[
(1-\mu^2)^{- \frac{1}{2  \cC }} \, \cG(w)
 \Big] 
 +  \frac{4\, B \, w \, \mu \, (2w^2-1)}{(1-w^2) \, \Big[ B  -2 \, w\, \sqrt{ 1-w^2 } \Big]}
+ \frac{4\, B \,   w^2 \, \mu}{ ( 1 - w^2  )^{3/2}} \ . \nn
\end{align}
The integrability condition $\partial_w \partial_\mu V(\mu,w) = \partial_\mu \partial_w V(\mu,w)$ yields
\begin{align} \label{eq_integr}
0 =&\; B \,w \, (1-\mu^2) \, \partial_\mu \eta(\mu,w)
+ \mu \, (1-w^2) \, \bigg[ B   - 2 \,    w  \, \sqrt{ 1-w^2 } \bigg] \, \partial_w \eta(\mu,w)
\nn \\
& - 4 \, B \, \cC \, w \, (1+\mu^2) + \frac{4\, B^2 \, \cC \, \Big[ \mu^2 + w^2 \, (1-\mu^2) \Big]}{  \sqrt{ 1-w^2 }} \ .
\end{align}
We have used the expression for $\cG'(w)/\cG(w)$.
On the other hand, the function $V$ must satisfy the Laplace equation \eqref{Laplace_eq},
up to localized sources. 
In the first term of \eqref{Laplace_eq}, we make use of $\partial_\rho (\rho \, \partial_\rho V) =\partial_\rho y$, and in the second term
we write
$\partial_\eta^2V = \partial_\eta \log r$. The quantities $\partial_\rho y$, $\partial_\eta \log r$
are then expressed as functions of $\mu$, $w$ with the help of the chain rule,
in terms of   $\eta = \eta(\mu,w)$. We get
\beq \label{eq_other}
\mu \, \partial_\mu \eta(\mu,w) - w \, (1-w^2) \, \partial_w \eta(\mu,w) - 4 \,  \cC  \, \mu = 0   \ .
\eeq
Combining \eqref{eq_integr} and \eqref{eq_other}, 
we have two linear equations in $\partial_\mu \eta(\mu,w)$, $\partial_w \eta(\mu,w)$, with solution
\beq
\partial_\mu \eta (\mu, \eta) = 4 \,  \cC  \, \bigg( 1 - \frac{B \, w}{\sqrt{ 1-w^2 }} \bigg) \ , \qquad
\partial_w \eta (\mu,\eta)= - \frac{4 \, B \,  \cC  \, \mu \, \sqrt{  1-w^2   }}{    (1-w^2)^2 } \ .
\eeq
We can now solve these PDEs for the function $\eta = \eta(\mu,w)$,
\beq \label{eta_expr}
\eta = 4 \,  \cC  \, \mu\, \bigg( 1 - \frac{B \, w}{\sqrt{ 1-w^2 }} \bigg) \ .
\eeq
We have fixed an integration constant by requiring $\eta = 0$ for $\mu = 0$.

\begin{figure}
\centering
\includegraphics[width = 11 cm]{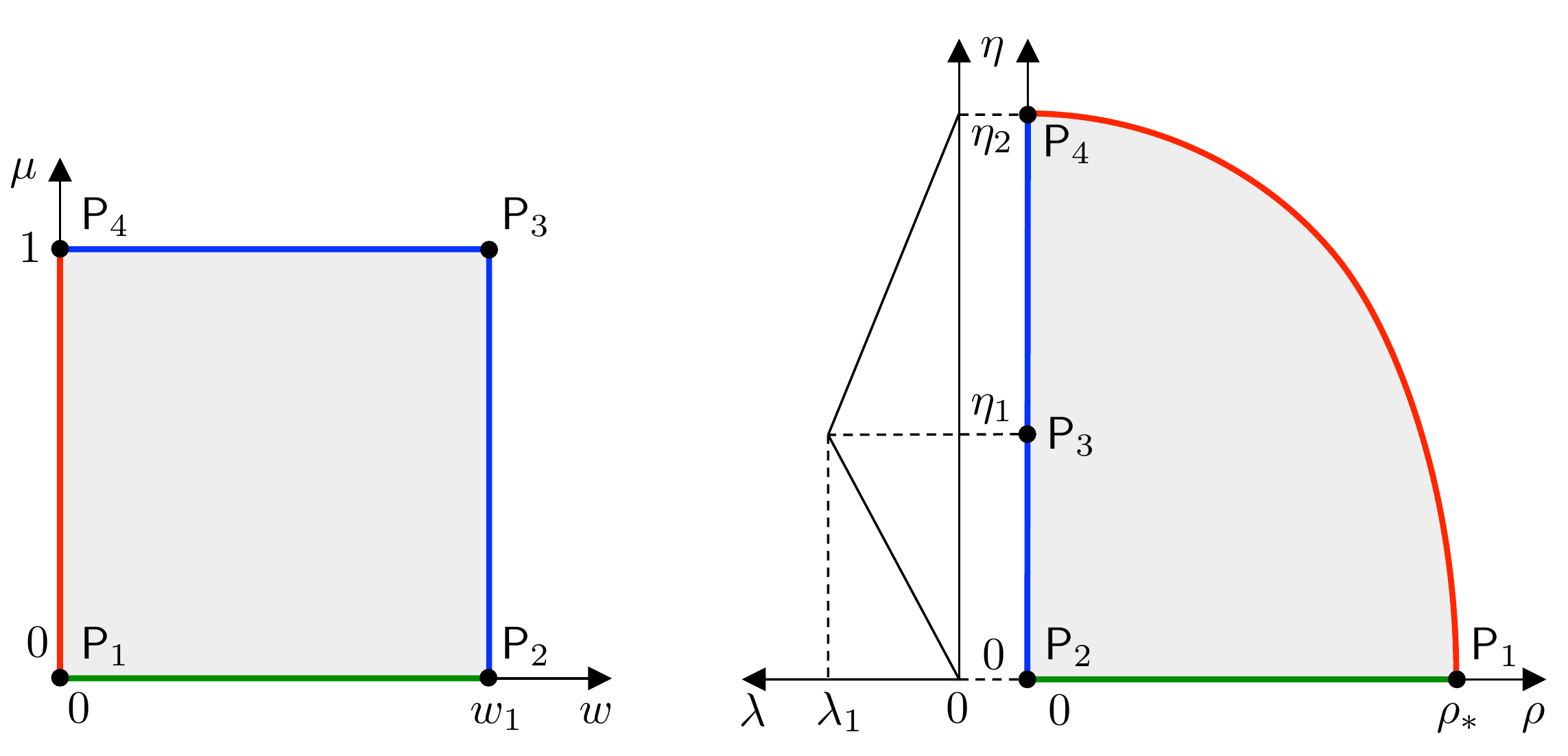}
\caption{
A schematic depiction of the relation between
the $(w,\mu)$ coordinates and $(\rho,\eta)$ coordinates.
The shaded regions on both sides correspond to the allowed values
of the $(w,\mu)$, $(\rho,\eta)$ coordinates.
On the right, we also include the plot of the charge density $\lambda(\eta)$.
}
\label{backlund}
\end{figure}

In Figure \ref{backlund} we depict schematically the change of coordinates
from $(w,\mu)$ to $(\rho,\eta)$.
It is convenient to define
\begin{align} \label{param_dictionary}
\eta_1  = 4 \,  \cC  \, \sqrt{1-B^2} = \frac{4}{\ell} \ , \qquad 
\eta_2  = 4 \,  \cC 
= \frac{4\, (N+k)}{N} \ , \qquad
\rho_*  = 4 \,B\,   \cC    \ .
\end{align}
The locus $w = 0$ is mapped to an arc of the ellipse defined by the equation
\beq \label{my_ellipse}
\frac{\rho^2}{16 \, B^2 \, \cC^2} + \frac{\eta^2}{16 \,  \cC^2} =1 \ .
\eeq 

Using \eqref{eta_expr}, we can find the explicit solution
to the PDEs \eqref{V_derivatives}.
We may start with the equation for $\partial_\mu V$ and integrate it in $\mu$.
To do so, we do not need the explicit form of $\cG(w)$.
The resulting expression for $V$, up to an arbitrary function of $w$,
is then plugged back in the equation for $\partial_w V$.
Making use of the expression for $\cG'(w)/\cG(w)$, we complete the determination
of $V$, up to an overall constant. The result reads
\begin{align} \label{actual_V}
V_{\rm T} & = 4 \, \mu - 2 \, \log \frac{1+\mu}{1-\mu} - 2 \, \bigg[
1 - \frac{B \, w}{\sqrt{ 1-w^2 }} 
\bigg] \, \mu \, \log(1-\mu^2)
+ 4 \, \cC \, \bigg[
1 - \frac{B \, w}{\sqrt{ 1-w^2 }} 
\bigg] \, \mu \, \log \cG(w) \ .
\end{align}
We have added a subscript `T' as a reminder that this  the electrostatic potential
that is inferred from the Toda form of the solution, via the B\"acklund transform.
We have fixed an arbitrary additive shift in $V_{\rm T}$ by demanding $V_{\rm T} = 0$ for $\mu = 0$.
We can also extract the charge density along the $\eta$ axis from $V_{\rm T}$ using \eqref{get_the_charge}.
The result, written as a function of $\eta$, takes the form of a continuous piecewise linear function
defined on the interval $[0,\eta_2]$,
\beq \label{eq_small_charge}
\lambda_{\rm T}(\eta) = \left\{
\begin{array}{ll}
\displaystyle \frac{\lambda_1}{\eta_1} \, \eta & \quad \text{for $ 0 \le \eta \le \eta_1$} \ ,
\\[4mm]
\displaystyle 
- \frac{\lambda_1}{\eta_2- \eta_1} \, (\eta - \eta_2)
 & \quad \text{for $  \eta_1 < \eta \le  \eta_2$} \ ,
\end{array}
\right.
\eeq
where we have introduced
\beq \label{eq_lambda1}
\lambda_1 = 4\, (1-\sqrt{1-B^2})   \ .
\eeq
Finally, we notice that the function $\cG(w)$ is determined only up to a multiplicative constant.
This ambiguity, however, translates into an ambiguity in the potential $V_{\rm T}$ of the form
\eqref{eq_eta_shift}, and therefore has no effect on the metric and flux.

\paragraph{Improved form for $V$.} The expression \eqref{actual_V}
for the electrostatic potential has two drawbacks:
it is not given in closed form as a function of $\rho$, $\eta$;
it is only determined in the interior of the shaded region
in Figure \ref{actual_V} in the $(\rho,\eta)$ plane.
Correspondingly, the charge density $\lambda_{\rm T}$ is known on the interval
$[0,\eta_2]$, but not on the entire $\eta$ axis.
We now discuss an improved potential $V$, which is
given explicitly as a function on the entire $(\rho,\eta)$ plane.

To write $V$, we start by prescribing a charge density along the $\eta$ axis,
which is continuous and piecewise linear, and extends \eqref{eq_small_charge} beyond
the interval $[0,\eta_2]$,
\beq \label{eq_charge}
\lambda(\eta) = \left\{
\begin{array}{ll}
\displaystyle \frac{\lambda_1}{\eta_1} \, \eta & \quad \text{for $ -\eta_1 \le \eta \le \eta_1$} \ ,
\\[4mm]
\displaystyle 
- \frac{\lambda_1}{\eta_2- \eta_1} \, (\eta - \eta_2)
 & \quad \text{for $  \eta > \eta_1$} \ ,
\\[4mm]
\displaystyle 
- \frac{\lambda_1}{\eta_2- \eta_1} \, (\eta + \eta_2)
 & \quad \text{for $  \eta < - \eta_1$} \ .
\end{array}
\right.
\eeq
Given this charge density, the electrostatic potential 
is determined using the standard Green's function for the Laplacian in $\mathbb R^3$.
The naive expression for $V$ would be
\beq\label{eq:V_Greens}
- \frac 12 \, \int_{-\infty}^{+\infty}  \frac{\lambda(\eta')}{\sqrt{\rho^2 + (\eta - \eta')^2}} \, d\eta' \ ,
\eeq
but this quantity suffers from logarithmic divergences from the 
large $|\eta'|$ region in the domain of integration.
We regularize the divergence by integrating in $\eta'$
in the range $[- \eta_*, \eta_*]$, with $\eta_*$ large and positive.
We perform a ``minimal subtraction'' of the 
divergence, and we send the regulator $\eta_*$ to infinity.
This prescription yields
\beq \label{eq_prescription}
V = \lim_{\eta_* \rightarrow + \infty} \bigg[
- \frac{\lambda_1}{\eta_2 - \eta_1} \, \eta \, \log \eta_* 
 -\frac 12 \, \int_{- \eta_*}^{+ \eta_*} \frac{\lambda(\eta')}{\sqrt{\rho^2 + (\eta - \eta')^2}} \, d\eta' 
\bigg] \ .
\eeq
The first term implements the minimal subtraction and it corresponds to a shift
in the electrostatic potential of the form   \eqref{eq_eta_shift}.
Computing the $\eta'$ integral and taking the limit,
we find 
\begin{align} \label{V_impr}
V  =&\;
\frac{\lambda_1}{2 \eta _1 \left(\eta _1-\eta _2\right)} \,  \bigg[
\eta _2 \sqrt{\left(\eta -\eta _1\right){}^2+\rho ^2}-\eta _2
   \sqrt{\left(\eta +\eta _1\right){}^2+\rho ^2}
   \nn \\
   & -\left(\eta -\eta
   _1\right) \eta _2  \, {\rm arctanh}  \left(\frac{\eta -\eta
   _1}{\sqrt{\left(\eta -\eta _1\right){}^2+\rho ^2}}\right)+\left(\eta
   +\eta _1\right) \eta _2  \,  {\rm arctanh}\left(\frac{\eta +\eta
   _1}{\sqrt{\left(\eta +\eta _1\right){}^2+\rho ^2}}\right)
   \nn \\
   & +2 \eta 
   \eta _1 \log (\rho )+2 \eta  \eta _1-\eta  \eta _1 \log (4) \bigg] \ .
\end{align}
The charge density \eqref{eq_charge} satisfies $\lambda(-\eta) = - \lambda(\eta)$,
implying that $V$ is equal to zero along the $\rho$ axis at $\eta = 0$:
this is a standard application of the method of images.

We may now compare $V$ with the potential $V_{\rm T}$
given in \eqref{actual_V}.
The values of $\eta_1$, $\eta_2$, $\lambda_1$ in terms of $B$, $\cC$ were given in \eqref{param_dictionary}, \eqref{eq_lambda1}.
After a lengthy but straightforward computation, using the 
  expression for $\cG(w)$
in \eqref{correct_G},
one verifies that
\beq
V_{\rm T} - V =  \frac{\cK}{ 4 \,  \cB  \, \cC} \eta \ ,
\eeq
where $\cK$ is a constant, given by
\beq
\cK = 4\, \cB + 4 \, \cB \, \cC \, \log \cG_*
+ 2 \, {\rm arctanh} \cB + \cB \, \log(16 \, B^2)  + 4 \, \cB \, \log \cC \ .
\eeq
We see that the difference $V_{\rm T} - V$ is of the form
 \eqref{eq_eta_shift}. It follows that, for the purposes of computing the 11d metric and flux,
we can use $V$ in \eqref{V_impr} instead of $V_{\rm T}$.

\subsection{Generalization of the charge density profile}

We now turn to the generalization of Case II solutions.
The metric and flux are as in 
\eqref{Backlund_form_BIS}.
Our task is to specify $\cC$ and $V$.
The main idea is to compute $V$ using the standard Green's function
starting from a given charge density profile $\lambda$ along the $\eta$ axis.
Building on solutions of Case II, we make the following working assumptions.
The charge density is taken to be continuous and piecewise linear.
Its slope changes at a finite number of points along the $\eta$ axis.
Moreover, we require
\beq
\lambda(-\eta) = - \lambda(\eta) \ ,
\eeq
so that it is sufficient to specify $\lambda$ for $\eta \ge 0$.

A generic continuous, piecewise linear profile for $\lambda$
can be parametrized as follows. 
Let $0 < w_1 < w_2 < \dots < w_p$
be the locations on the $\eta$ axis where the slope of $\lambda$ changes.
We may then write
\beq \label{eq_generalized_charge}
\lambda(\eta) = \left\{
\begin{array}{ll}
\displaystyle m_0 \, \eta  & \quad \text{for $  0  \le \eta <  w_1$} \ ,
\\[1mm] 
\displaystyle m_a \, \eta + y_a  & \quad \text{for $  w_{a}  \le \eta <  w_{a+1}$, $a=1,2,\dots,p-1$} \ ,
\\[1mm] 
\displaystyle m_p \, \eta + y_p  & \quad \text{for $\eta \ge  w_p$} \ ,
\end{array}
\right.
\eeq
where we have introduced the slope parameters $m_a$ ($a =0,1,\dots,p$)
and the intercepts $y_a$ ($a = 1,2,\dots p$).
We also define $q_0 := 0$.
 Continuity of $\lambda$ imposes
\beq \label{eq_charge_continuity}
(m_a - m_{a-1}) \, w_a + (y_a - y_{a-1}) = 0 \ , \qquad a = 1, 2, \dots, p \ .
\eeq
Further constraints on the slope and intercept parameters will be derived below
from metric regularity and flux quantization.
The outcome of our analysis will be the charge density \eqref{eq_decomposition}
discussed in the main text.

The electrostatic potential determined by the charge density \eqref{eq_generalized_charge}
can be computed as a sum of various contributions.
Firstly, we may consider the interval $[w_a, w_{a+1}]$, $a = 0,1,\dots,p-1$,
and its mirror image   $[- w_{a+1}, - w_a]$. (By definition, $w_0 :=0$.)
The corresponding contribution to the electrostatic potential reads
\begin{align} \label{finite_piece}
&V( w_{a}, w_{a+1}, m_a , y_a ) = 
- \frac 12 \, \int_{w_a}^{w_{a+1}}  \frac{m_a \, \eta' + y_a}{\sqrt{\rho^2 + (\eta - \eta')^2}} \, d\eta' 
- \frac 12 \, \int_{- w_{a+1}}^{w_{a}}  \frac{m_a \, \eta' - y_a}{\sqrt{\rho^2 + (\eta - \eta')^2}} \, d\eta' 
\nn \\
& =
\frac{1}{2} \left(  m_a \sqrt{\left(\eta - w _a\right){}^2+\rho
   ^2}
   -{\rm arctanh}\left(\frac{\eta -w_a }{\sqrt{\left(\eta -w_a\right){}^2+\rho ^2}}\right) \left(  
   m_a \eta +y_a\right)\right)
   \nn \\
   & +\frac{1}{2} \left( - m_a  \sqrt{\left(\eta
   + w_a \right){}^2+\rho ^2}
    + {\rm arctanh}\left(\frac{\eta
   + w_a }{\sqrt{\left(\eta + w _a \right){}^2+\rho ^2}}\right)
   \left(  m_a \eta - y_a\right)\right)
   \nn \\
   & +\frac{1}{2} \left( - m_a
   \sqrt{\left(\eta - w _{a+1}\right){}^2+\rho ^2}
   + {\rm arctanh}
   \left(\frac{\eta  - w _{a+1} }{\sqrt{\left(\eta -w
   _{a+1}\right){}^2+\rho ^2}}\right) \left(  
   m_a \eta +y_a\right)\right)
   \nn \\
   & +\frac{1}{2} \left(m_a \sqrt{\left(\eta+ w
   _{a+1} \right){}^2+\rho ^2}
   - {\rm arctanh}\left(\frac{\eta + w
   _{a+1} }{\sqrt{\left(\eta + w _{a+1} \right){}^2+\rho
   ^2}}\right) \left(   m_a \eta -y_a\right)\right) \ .
\end{align}
Next, we have the contribution of the semi-infinite interval $[ w_p,+ \infty)$
and its mirror image $( - \infty , - w_p]$. In this case, a na\"ive integration of the charge density
against the standard Green's function yields a logarithmic divergence. We 
regulate and subtract the divergence in the same way as in \eqref{eq_prescription}.
We thus obtain the quantity
\begin{align} \label{infinite_piece}
& V( w_p , \infty , m_p, y_p )  =  \nn \\
& = \lim_{w_* \rightarrow + \infty} \bigg[
 -\frac 12 \, \int_{w_p}^{ w_*} \frac{m_p \eta' + y_p}{\sqrt{\rho^2 + (\eta - \eta')^2}} \, d\eta' 
  -\frac 12 \, \int_{- w_*}^{ -w_p} \frac{m_p \eta' - y_p}{\sqrt{\rho^2 + (\eta - \eta')^2}} \, d\eta' 
+ m_p \, \eta \, \log \eta_*
\bigg] 
\nn \\
& = 
\frac{1}{2} m_p \sqrt{\left(\eta -w _p\right){}^2+\rho
   ^2}-\frac{1}{2} m_p \sqrt{\left(\eta +w _p\right){}^2+\rho
   ^2}
    -\frac{1}{2}    m_p \eta  (-2 \log  \rho  -2+\log   4 )
     \\
   & +\frac{1}{2}
   \left(   m_p \eta -y_p\right)  {\rm arctanh} \left(\frac{\eta +w
   _p}{\sqrt{\left(\eta +w _p\right){}^2+\rho ^2}}\right)
    - \frac{1}{2}
   \left(   m_p \eta +y_p\right)  {\rm arctanh} \left(\frac{\eta -w _p
   }{\sqrt{\left(\eta -w _p\right){}^2+\rho ^2}}\right) \ . \nn
\end{align}
The final expression for $V$ is
\beq \label{eq_tot_V}
V = \sum_{a=0}^{p-1} V({w_a, w_{a+1}, m_a, y_a}) + V(w_p , \infty , m_p, y_p ) \ .
\eeq
By virtue of the method of images,
$V$ is an odd function of $\eta$, and thus in particular
it is zero at $\eta = 0$.

\subsubsection{Metric regularity}

\paragraph{Monopole sources.}

The 11d metric functions entering \eqref{Backlund_form_BIS} may now be computed in closed form
by plugging \eqref{eq_tot_V} into   \eqref{stuff_from_V}.
In particular, we observe that
\begin{itemize}
\item The quantity $R_\phi^2$ vanishes along the $\eta$ axis in the $(\rho,\eta)$ plane.
\item The quantity $R_z^2$ has isolated zeros in the $(\rho,\eta)$ plane, situated
along the $\eta$ axis at the locations $w_a$ ($a = 1,2,\dots,p$) where
the slope in $\lambda(\eta)$ changes.  
\item The quantity $L$ is piecewise constant along the $\eta$ axis,
\beq
\text{if $w_a < \eta < w_{a+1}$,} \qquad \lim_{\rho \rightarrow 0^+} L(\rho,\eta) := \ell_{a+1} = m_a + \frac 1 \cC \ .
\eeq
\end{itemize}
We infer that the internal geometry admits the following description.
The total 6d space is a fibration of $S^2$ and $S^1_z$ over a 3d base space,
spanned by $\rho$, $\eta$, $\phi$.
The quantity $R_\phi$ is the radius of $S^1_\phi$ in the 3d base space.
The fact that it vanishes along the $\eta$ axis implies that $S^1_\phi$ shrinks smoothly
there inside the 3d base space.
The locations $w_a$ ($a = 1,2,\dots,p$) on the $\eta$ axis
are monopoles for the $S^1_z$ fibration over the 3d base space.
Indeed, the radius $R_z$ of $S^1_z$ goes to zero at $w_a$,
and the function $L$ which governs the fibration of $S^1_z$ over $S^1_\phi$
jumps at $w_a$.
The discontinuity of $L$ at $w_a$ is identified with the monopole charge $k_a$ of the $a$th monopole,
which must be a positive integer,
\beq
\ell_a - \ell_{a+1}   = k_a \in \mathbb Z_{>0} \ , \qquad a = 1,2,\dots,p \ .
\eeq
It follows that all the slope parameters $m_a$ can be deterimined recursively
in terms of the monopole charges $k_a$ and the outermost slope parameter $m_p$.
Based on analogy with the original solution given by the charge density \eqref{eq_charge},
we set
\beq \label{eq_last_slope}
m_p = - \frac 1 \cC \ , \qquad \text{or equivalently} \qquad \ell_{p+1} =0 \ .
\eeq
All slope parameters are thus fixed,
\beq \label{eq_slopes}
m_a = - \frac 1 \cC + \sum_{b=a+1}^p k_b \ , \qquad a = 0,1,\dots, p-1 \ .
\eeq
The intercepts   $y_a$ are then also fixed, using $y_0 = 0$
and the continuity condition \eqref{eq_charge_continuity},
\beq \label{eq_qs_are_fixed}
y_a = \sum_{b=1}^a k_b \, w_b \ , \qquad a = 1,\dots,p \ .
\eeq

The original solution based on the charge density \eqref{eq_charge}
corresponds to the case of one monopole, $p=1$.
In that case, the value of $\lambda$ at the location $w_1$ is positive.
By analogy, we now require that the value of $\lambda$
at the location $w_p$ of the last monopole be positive,
\beq
\lambda(w_p) = - \frac{w_p}{\cC} + \sum_{b=1}^{p} k_b \, w_b > 0 \  .
\eeq
It follows that the charge density profile has a zero at a point $w_{\rm m} > w_p$,
given by
\beq \label{eq_eta_m}
w_{\rm m} = \cC \, \sum_{b=1}^{p} k_b \, w_b \ .
\eeq
We observe that the charge density $\lambda$ is positive 
and concave in the interval $[0, w_{\rm m}]$.

\paragraph{Allowed region in the $(\rho,\eta)$ plane.}
As already explained in the main text, it is determined by
the inequalities \eqref{eq_allowed_region}.
The arc $\partial_\rho V = 0$ intersects the $\eta$ axis
at the value $w_{\rm m}$ where the positive zero of $\lambda$ is located,
see \eqref{eq_eta_m}.

\subsubsection{Flux quantization}
 
In our normalization conventions, the quantity $\overline G_4$ 
in \eqref{Backlund_form_BIS} has integral periods on any 4-cycle in the internal space.
The computation of the periods of $\overline G_4$   is facilitated by the following properties
of the functions $Y$, $W$ in \eqref{Backlund_form_BIS}, which can be verified
by direct computation using \eqref{stuff_from_V}, \eqref{eq_tot_V}.
\begin{itemize}
\item The quantity $Y(\rho,\eta)$ is piecewise constant along the $\eta$ axis. Its values
are determined by the intercepts parameters $y_a$ in the charge density profile,
\begin{align} \label{eq_Y_property}
 \qquad Y(0,\eta) &= y_a \ , \qquad \text{for } w_a < \eta < w_{a+1} \ , \qquad a = 0, \dots, p-1 \ , \nn \\
  \qquad Y(0,\eta) &= y_p \ , \qquad \text{for } \eta > w_{p} \ .
\end{align}

\item The values of the quantity $W(\rho,\eta)$ at the monopole locations
are
\beq \label{eq_W_property}
W(0,\eta_a) = w_a \ , \qquad a = 1,\dots,p \ .
\eeq

\item Both $Y(\rho,\eta)$ and $W(\rho,\eta)$ vanish for $\eta = 0$, for arbitrary $\rho$.
\end{itemize}

We can now list the 4-cycles in the geometry and evaluate
the corresponding $G_4$-flux parameters. Our discussion follows closely the
approach and notation of \cite{Bah:2019jts}.

\begin{figure}
\centering
\includegraphics[width = 5. cm]{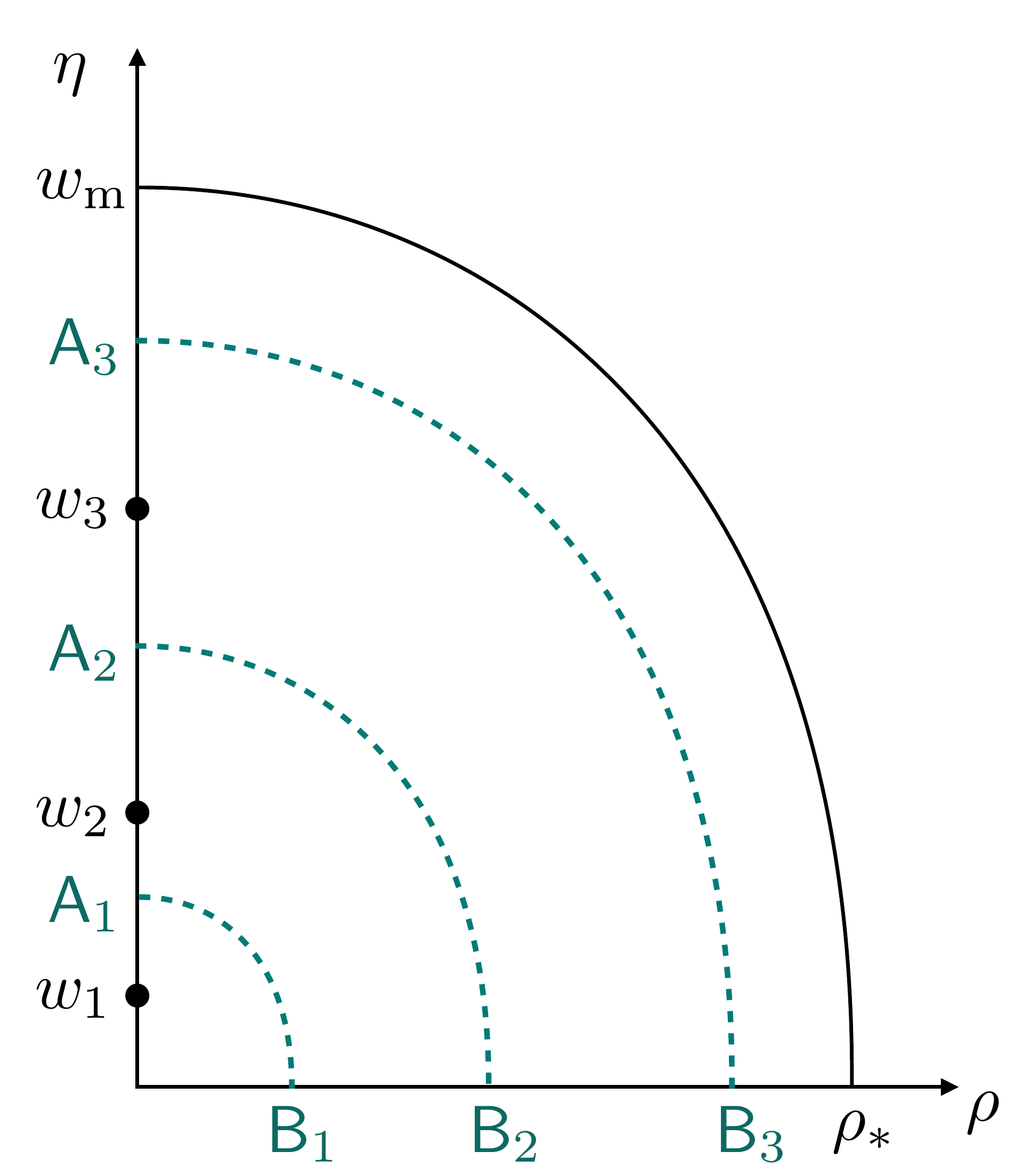}
\caption{
The 4-cycles of type $\mathbf C$ are obtained combining
the dashed arcs $\mathsf A_a \mathsf B_a$ with the $S^2$
and the $\phi$ circle in the base of the $Dz$ fibration.
The 4-cycles of type $\mathbf B$ are obtained
combining a segment $[w_a,w_{a+1}]$ on the $\eta$ axis
with the $S^2$ and the $Dz$ circle fiber.
}
\label{fig_some_cycles}
\end{figure}

\paragraph{Four-cycles of type $\mathbf C$.}
With reference to Figure \ref{fig_some_cycles}, 
let us consider 
the arc $\mathsf A_a \mathsf B_a$.
The $S^2$ shrinks at $\mathsf B_a$.
The $\phi$ circle in the base of the $Dz$ fibration
shrinks at $\mathsf A_a$, because $R_\phi^2$ goes to zero there.
We then have a four-cycle, which
we denote $\mathbf C_a$.

To identify the $\phi$ circle in the base of the $Dz$ fibration, we   use
that $L = \ell_{a+1}$ and we set $0 = Dz = dz - \ell_{a+1} \, d\phi$,
giving us $dz = \ell_{a+1} \, d\phi$ along the arc
$\mathsf A_a \mathsf B_a$.
As a result, the relevant terms in $\overline G_4$ are
\beq
\overline G_4 
= \frac{ {\rm vol}_{S^2} }{4\pi} \, d( Y + L \,W - \ell_{a+1} \, W )\, \frac{d\phi}{2\pi}
\eeq
The integral of this quantity over $\mathbf C_a$ yields the
value of the function $Y$ at the endpoint $\mathsf A_a$,
\beq
\int_{\mathbf C_a} \overline G_4 = Y(\mathsf A_a)   \ .
\eeq
But $Y(\mathsf A_a)  = y_a$ from \eqref{eq_Y_property}.
We conclude that the intercept parameters $y_a$ are all integrally quantized.
(Throughout this appendix, we fix the mass scale $m$
as in \eqref{eq_fix_m}).

We also notice that the outermost flux quantum $y_p$
($p=3$ in the example of Figure \ref{fig_some_cycles}) is identified with the number $N$ 
of M5-branes,
\beq \label{eq_the_last_q}
y_p = N \ .
\eeq


%
%
%
%
%
%
%

\paragraph{Four-cycles of type $\mathbf B$.}

We can consider the segment $[w _{a-1}, w_a]$ along the $\eta$ axis,
combined with the $S^2$ and the $Dz$ circle, to get a four-cycle denoted
$\mathbf B_a$. The relevant terms in $\overline G_4$ are
\begin{align}
\overline G_4 &\supset  \frac{ {\rm vol}_{S^2} }{4\pi} \, (- dW) \, \frac{Dz}{2\pi}  \ .
\end{align}
The corresponding flux quantization implies
that the following are integers, 
\beq
\int_{\mathbf B_a}\overline G_4 =  W(0,w_a) - W(0,w_{a-1})
= w_a - w_{a-1}   \ .
\eeq
In the second step we used \eqref{eq_W_property}. Since $w_0 = 0$, we conclude that
\beq \label{eq_etas_integer}
\text{ $\{ w_a \}_{a=1}^p$ is an increasing sequence of positive integers.} \ .
\eeq

We can also extend the arguments of the previous paragraphs
to the last segment $[w_p, w_{\rm m}]$. The interpretation is now different:
this is a four-cycle because the warp factor goes to zero
along the boundary of the $(\rho, \eta)$ region.
The lesson is that $w_{\rm m}$ must be also an integer.
The flux quantum $w_{\rm m}$ should be regarded as a property of the 
smeared M5-brane source
located at $\partial_\rho V = 0$, which is discussed in greater detail below.
We find it convenient to
parametrize it in terms of $N$ and another integer parameter $k$,  
\beq \label{this_is_etam}
w_{\rm m} = N + k \ , \qquad k \in \mathbb Z \ .
\eeq
The zero $w_{\rm m}$ of the charge density is located
at the right of the last monopole location, $w_{\rm m} > w_{p}$.
It follows that the integer $k$ must satisfy
\beq
N+k > w_p \ .
\eeq

%

\paragraph{Regularity of $\overline G_4$ near monopoles.}
We observed above that $Y$ and $L$ are piecewise constant along the $\eta$ axis.
This might potentially generated delta-function singularities
in $\overline G_4$, due to the presence of the derivatives $\partial_\eta Y$,
$\partial_\eta L$. One can verify, however, that these singularities
are absent by virtue of the conditions \eqref{eq_charge_continuity}, which guarantee the continuity of
the charge density profile.
The continuity condition implies \eqref{eq_qs_are_fixed},
which together with 
\eqref{eq_the_last_q} gives us
\beq
N  = \sum_{a=1}^p w_a \, k_a \ .
\eeq
We have thus verified the emergence of a partition of $N$
from   regularity and flux quantization.
 
\paragraph{The parameter $\cC$ is fixed by flux quanta.}
The slope of the charge density profile for $\eta > w_p$
is given in \eqref{eq_last_slope} in terms of $\cC$. It can be alternatively be computed
by connecting the points $(\eta,\lambda) = (w_p,  N - w_p/\cC)$ and $(\eta, \lambda) = (N+k,0)$.
The result is the following relation between $\cC$ and the flux
quanta $N$, $k$,
\beq \label{C_from_quanta}
\cC = \frac{N+k}{N} \ .
\eeq

\subsubsection{M5-brane source}

To clarify the behavior of the solution near the arc defined by $\partial_\rho V = 0$,
let us start from \eqref{Backlund_form_BIS} and   perform two operations:
\begin{itemize}
\item We break up $Dz$ and we complete the $d\phi$ square,
so that the line element is written in terms of $dz$ and $\mathcal D \phi = d\phi - \mathcal L \, dz$. (The quantity
  $\mathcal L$ is fixed requiring the absence of cross terms $dz \mathcal D \phi$).
\item We collect the leading terms in the limit $\dot V \rightarrow 0$.
\end{itemize}
All electric sources are localized along the $\eta$ axis. As a result,
at a generic point along the arc $\partial_\rho V = 0$, the potential $V$
satisfies the Laplace equation. Using this information,
we can write the resulting line element in the form
\begin{align} \label{eq_near_arc}
& \frac{ds^2_{11}}{(4\pi)^{2/3} \ell_p^2}   \approx  \dot V ^{1/3} \,  \bigg[  \frac{(\dot V')^2 - \ddot V \, V''}{2 \, V''} \bigg]^{1/3} \,  \bigg[  ds^2(AdS_5)
+ \cC^2  \, dz^2 \bigg]     \\
& \qquad  + \frac 14 \,  \dot V ^{-2/3} \,  \bigg[  \frac{(\dot V')^2 - \ddot V \, V''}{2 \, V''} \bigg]^{-2/3} \bigg[ 
\dot V^2 \, ds^2(S^2) + \frac{\rho^2}{\cC^2} \, \mathcal D \phi^2
+ \Big[ (\dot V')^2 - \ddot V V'' \Big] \, (d\rho^2 + d\eta^2)
\bigg]   \ . \nn
\end{align}
We observe that, at leading order near the locus $\partial_\rho V = 0$,
we can write $\mathcal L \approx \cC$, and hence $\mathcal D \phi \approx d\phi - \cC \, dz$.
This line element is compatible with an interpretation in terms of smeared M5-branes.
Near the arc $\partial_\rho V = 0$, we can parametrize the $(\rho,\eta)$ 2d space
in terms of a normal coordinate $n = \dot V$, and a tangential coordinate $t$,
which varies along the arc. From the term $\dot V^2 \, ds^2(S^2) = n^2 \, ds^2(S^2)$ inside the bracket
on the second line, we see that, at small $n$ near the arc, $n$ and $S^2$ combine
into a local $\mathbb R^3$. We also observe the appearance of overall $\dot V$ 
prefactors with powers 1/3, $-2/3$ in the first and second lines of \eqref{eq_near_arc},
respectively,
This structure implies that the M5-branes are: 
 extended in the $AdS_5$ and $z$ directions;
 smeared in the $t$ and $\phi$ directions;
localized at the origin of the local $\mathbb R^3$
parametrized by $n$ and $S^2$. 
These findings are directly analogous to the analysis of 
 \cite{Bah:2021hei}, which applies to the case $p=1$.
 
We can also analyze the form of the $G_4$-flux
in the vicinity of the  arc defined by $\partial_\rho V = 0$.
Making use of \eqref{Backlund_form_BIS}, \eqref{stuff_from_V}, we verify that,
as we approach the locus $\partial_\rho V = 0$, we have
\beq
L \approx \frac 1 \cC  , \qquad   W \approx \eta   \ , \qquad
  Y \approx  0 \ .
\eeq 
It follows that the $G_4$ is given at leading order by
\beq
\overline G_4 \approx     \frac{ {\rm vol}_{S^2}  }{4\pi}\wedge \frac{d\eta}{  \cC} \wedge \frac{\cD \phi}{2\pi} \ .
\eeq
We can integrate this quantity along the $S^2$ and $\mathcal D \phi$,
combined with an arc in the $(\rho,\eta)$ plane that approaches the
boundary component $\partial_\rho V =0$ from the inside of the allowed region.
We can use the coordinate $\eta$ to parametrize this arc,
regarding $\rho =\rho(\eta)$ as fixed by $\partial_\rho V = \epsilon$,
with small positive $\epsilon$. As $\epsilon$ goes to zero,
the integral of $\overline G_4$ approaches a  finite value,
\beq \label{eq_near_smeared}
\int \overline G_4 \approx    \frac{w_{\rm m}}{  \cC}  = \frac{N+k}{\cC} = N  \ .
\eeq
We have made use of \eqref{this_is_etam} and \eqref{C_from_quanta}.
We conclude that the smeared M5-brane source has a total charge equal to $N$.

\subsection{Inflow analysis}\label{sec:inflow}

In this section we derive the 't Hooft anomaly coefficients
quoted in \eqref{A_anom_coeff}, \eqref{flavor_from_inflow}. We first construct $E_4$, the equivariant completion
of the background flux $\overline G_4$, and we then compute the integral
of $E_4^3$ on the internal space $M_6$.

\subsubsection{Construction of $E_4$}

The background flux $\overline G_4$ is presented in \eqref{Backlund_form_BIS}
as the wedge product of ${\rm vol}_{S^2}$ with the total derivative of a locally defined 1-form.
This suggests a 
na\"ive 
candidate for $E_4$. Firstly, we replace 
 ${\rm vol}_{S^2}/(4\pi)$ with $e_2$, which is the standard global angular form of $SO(3)$,
normalized to integrate to $1$ on $S^2$.
(For more details, see for instance \cite{Harvey:1998bx,Bah:2019jts}.)
Secondly, we 
consider the local 1-form inside the total derivative,
and we perform the
 replacements
\beq
d\phi \rightarrow d\phi + A_\phi \ , \qquad
dz \rightarrow dz + A_z  \ .
\eeq 
Here $A_\phi$, $A_z$ denote the external background gauge fields
associated to the isometries $\partial_\phi$, $\partial_z$.
The $E_4$ resulting  is manifestly
closed and gauge invariant. It takes the form
\begin{align}
E_4^{\rm naive}  & = e_2 \, \bigg[
(dY + W \,  dL) \, \frac{\mathcal D \phi}{2\pi}
- dW \, \frac{\cD z}{2\pi}
 \bigg]
- e_2 \, W \, \frac{F_z}{2\pi}
+ e_2 \, (Y+W\,L) \, \frac{F_\phi}{2\pi} \ ,
\end{align}
where we have introduced
\beq
\mathcal D \phi = d\phi  + A_\phi \ , \qquad  \mathcal D z = dz + A_z - L \, D \phi \ , \qquad
F_\phi = dA_\phi \ , \qquad
F_z = dA_z \ . 
\eeq
Crucially, however, $E_4$ is not automatically guaranteed to
be globally defined. To clarify this point 
it is convenient to trade
$A_z$, $A_\phi$ for the external gauge fields
$A_\beta$, $A_\chi$, associated to
the canonical LLM angular variables,
making use of \eqref{angular_vars}.
We obtain
\begin{align}
E_4^{\rm naive}  =&\; e_2 \, \bigg[
(dY + W \,  dL) \, \frac{\mathcal D \phi}{2\pi}
- dW \, \frac{\cD z}{2\pi}
 \bigg]
\nn \\
& + e_2 \, \Big[ Y +(L -\cC^{-1}) \, W  \Big] \, \frac{F_\chi}{2\pi}
+ e_2 \, \Big[ Y +(L -\cC^{-1}-1) \, W  \Big] \, \frac{F_\beta}{2\pi} \ .
\end{align}
The coefficients of $F_\chi$, $F_\beta$
must be well-defined 2-forms in the internal space $M_6$.
Recall that the $S^2$ shrinks both along the $\rho$ axis,
and along the arc in the $(\rho,\eta)$ plane defined by the condition
$\partial_\rho V = 0$.
We have already notices that both $Y$ and $W$ vanish
along the $\rho$ axis, so regularity there is guaranteed
both for the $F_\chi$ and   the $F_\beta$ term.
Along the arc where $\partial_\rho V = 0$, on the other hand,
we have $  W \approx \eta$, $L \approx \cC^{-1}$ and $  Y \approx 0$.
As a result, we observe that the coefficient of $F_\chi$ goes to zero,
but the coefficient of $F_\beta$ does not.

Equivalently, we can observe that the 3-form
$\iota_\chi \overline G_4$ is exact,
while the 3-form $\iota_\beta \overline G_4$ is closed but not exact.
This is the same phenomenon encountered in \cite{Bah:2021hei}
for the solutions of Case II (before generalization).
Since $\iota_\beta \overline G_4$ is cohomologically non-trivial,
expansion of the M-theory 3-form on 
 $\iota_\beta \overline G_4$ yields an axion.
The would-be massless gauge field $A_\beta$
participates to a St\"uckelberg coupling with this axion,
and is thus massive. It does not correspond to a continuous
$U(1)$ global symmetry on the field theory side.
We refer the reader to \cite{Bah:2021hei} for further details
about this St\"uckelberg mechanism.

%
%
%
%
%


Since we are interested in studying anomalies for continuous symmetries,
we proceed setting $A_\beta = 0$. We can then write
\begin{align}
E_4   =&\; e_2 \, \bigg[
(dY + W \,  dL) \, \frac{\mathcal D \phi}{2\pi}
- dW \, \frac{\cD z}{2\pi}
 \bigg]
 + e_2 \, \Big[ Y +(L -\cC^{-1}) \, W  \Big] \, \frac{F_\chi}{2\pi} \nn \\
 & +  \sum_{a=1}^p \, \sum_{I=1}^{k_a-1} \, \frac{\widehat F_{a,I}}{2\pi} \, \widehat \omega_{a,I} \ .
\end{align}
On the second line we have introduced the contributions originating
from the Cartan generators of the non-Abelian $\mathfrak{su}(k_a)$ flavor algebra associated to the $a$th
monopole, of charge~$k_a$.
The 2-forms $\widehat \omega_{a,I}$ are dual
to the resolutions 2-cycles in the local $\mathbb C^2/\mathbb Z_{k_a}$.

\subsubsection{Integration of $E_4^3$}

Let us start from the contributions that do not involve the 
$\mathfrak{su}(k_a)$ flavor symmetries. By a standard application of 
the Bott-Cattaneo formula \cite{bott1999integral},
we arrive at
\begin{align} \label{E4cube_piece}
- I_6^{\rm inflow} = \int_{M_6} \frac 16 \, E_4^3 & \supset - \frac 18 \, 
c_1(U(1)_\chi) 
  \, p_1(SO(3)) \, \int_{\cB_2} dW \wedge d\Big[ Y + W \, (L-\cC^{-1}) \Big]^2 \ .
\end{align}
Here
$c_1(U(1)_\chi) =F_\chi/(2\pi)$ andf
 $p_1(SO(3))$ is the first Pontryagin class
of the $SO(3)$ background gauge field
associated to the isometries of the $S^2$ in the geometry.
We have assigned positive orientation to $\mathcal Dz \wedge \cD \phi$.
The symbol $\cB_2$ denotes the domain in the first $(\rho,\eta)$ quadrant,
see Figure \ref{fig_more_monopoles}.
The integral over $\cB_2$ can be written as
\beq
\int_{\cB_2} dW \wedge d\Big[ Y + W \, (L-\cC^{-1}) \Big]^2 = 
\int_{\partial \cB_2} W \, d\Big[ Y + W \, (L-\cC^{-1}) \Big]^2 \ .
\eeq	
Let us analyze in turn the components of $\partial \cB_2$:
\begin{itemize}
\item Along the $\rho$ axis, $W =0$, hence we get zero.
\item Along the arc defined by $\partial_\rho V = 0$,
we know that $  W = \eta$, $  Y = 0$, $L = \cC^{-1}$, hence we get zero.
\item Since $L$ and $Y$ are   piecewise constant along the $\eta$ axis, it is convenient to treat
each segment in turn. On a segment of the form $[w_a, w_{a+1}]$, 
with $a =0, \dots, p-1$, we know that $L = \ell_{a+1} = m_a + \cC^{-1}$. 
We also have $  Y =  y_a$.
We thus get a contribution
\begin{align} \label{term_in_E4cube}
\int_{\partial \cB_2} W \, &d\Big[ Y + W \, (L-\cC^{-1}) \Big]^2   \supset
- \int_{[w_a, w_{a+1}]} W \, d\Big[ y_a + W \, m_a \Big]^2
\nn \\
& = -   \int_{[w_a, w_{a+1}]} W \, d\Big[ y_a^2 + 2 \, m_a \, y_a \, W + m_a^2 \, W^2 \Big] \nn \\
& =  -   \int_{[w_a, w_{a+1}]}  d\Big[  \tfrac 23 \,    m_a ^2   \, W^3 + m_a \, y_a \, W^2 \Big]  \nn \\
& = - \frac 23 \, m_a^2 \, (w_{a+1}^3  -w_a^3 ) - m_a \, y_a \, (w_{a+1} ^2  - w_a^2 ) \ .
\end{align}
We have used $W(0,w_a) = w_a$.
The minus sign in front comes from the fact that we are taking $\partial \cB_2$
with a counterclockwise orientation in the $(\rho, \eta)$ plane.
Finally, we have the final segment $[w_p, w_{\rm m}]$.
It gives a contribution of the same form,
formally obtained taking $a = p$ with the convention
$w_{p+1} = N+k$, $\ell_{p+1} =0$.

\end{itemize}

Making use of \eqref{E4cube_piece} and  \eqref{term_in_E4cube},
as well as \eqref{background_identifications},
we recover the expression \eqref{A_anom_coeff} for $\cA_{r,R}$ quoted in the main text.
The comparison between \eqref{A_anom_coeff} and \eqref{conjecture}  relies on the following identity,
\begin{align}
 & \sum_{a=0}^p \bigg[ 
\frac 13 \, m_a^2 \, (w_{a+1}^3 - w_a^3)
+ m_a \, y_a \, (w_{a+1}^2 - w_{a}^2)
+ y_a^2 \, (w_{a+1} - w_a)
\bigg]   \nn \\
& =  -  \sum_{a=0}^p \bigg[ 
\frac 23 \, m_a^2 \, (w_{a+1}^3 - w_a^3)
+ m_a \, y_a \, (w_{a+1}^2 - w_{a}^2)
\bigg]  \ .
\end{align}

%

%

Let us now turn to the contributions associated to the $\mathfrak{su}(k_a)$ factor
of the symmetry associated to the regular puncture.
The relevant terms in the inflow anomaly polynomial are
\beq
- I_6^{\rm inflow}    \supset \frac 12 \, \frac{F_\chi}{2\pi} \, 
 \sum_{a=1}^p \, \sum_{I,J=1}^{k_a-1} \, \frac{\widehat F_{a,I}}{2\pi} \, \frac{\widehat F_{a,J}}{2\pi} \, \int \widehat \omega_{a,I} \, \widehat \omega_{a,J}\, \Big[  Y + (L- \cC^{-1}) \, W \Big] \ .
\eeq
The 2-forms dual to the resolution cycles are localized at the monopole locations.
Even though $Y$ and $L$ have jumps at the monopoles, the quantity $Y + (L- \cC^{-1}) \, W$ is continuous at each monopole location,
as may be verified using the continuity conditions \eqref{eq_charge_continuity} for the
charge density profile $\lambda$.
We can write
\beq
\Big[ Y + (L- \cC^{-1}) \, W\Big]_{(\rho,\eta) = (0,w_a)} = y_a + m_a \, w_a \ .
\eeq
It follows that the relevant term in the inflow anomaly polynomial reads
\beq
- I_6^{\rm inflow}    \supset \frac 12 \, \frac{F_\chi}{2\pi} \, 
 \sum_{a=1}^p \, \sum_{I,J=1}^{k_a-1} \, \frac{\widehat F_{a,I}}{2\pi} \, \frac{\widehat F_{a,J}}{2\pi} \, (y_a + m_a \, w_a) \, (- C_{IJ}^{\mathfrak{su}(k_a)})   \ .
\eeq
We have used the fact that the intersection pairing
among the 2-forms $\widehat \omega_{a,I}$ reproduces minus
the Cartan matrix $C_{IJ}^{\mathfrak{su}(k_a)}$ of $\mathfrak{su}(k_a)$.
This result implies the expression \eqref{flavor_from_inflow} for the flavor central
charge quoted in the main text.

%
%
%
%
%


\bibliographystyle{./ytphys}
\bibliography{./refs}

\end{document}